\documentclass[%
 reprint,
superscriptaddress,
 amsmath,amssymb,
 aps,
]{revtex4-2}

\usepackage{graphicx}
\usepackage{dcolumn}
\usepackage{bm}
\usepackage{color}
\usepackage{comment}
\usepackage{siunitx}
\usepackage{booktabs}
\usepackage{multirow}
\usepackage{upgreek}

\newcommand{\goodchi}{\protect\raisebox{2pt}{$\chi$}}

\begin{document}

\preprint{APS/123-QED}

\title{A saga on the $\gamma$-decay branching ratio of the Hoyle state}

\author{W.~Paulsen}
\email{wanja.paulsen@fys.uio.no}
\affiliation{Department of Physics, University of Oslo, N-0316 Oslo, Norway}%
\affiliation{Norwegian Nuclear Research Centre, Norway}
\author{K.~C.~W.~Li}%
\affiliation{Department of Physics, University of Oslo, N-0316 Oslo, Norway}%
\affiliation{Norwegian Nuclear Research Centre, Norway}
\author{S.~Siem}%
\affiliation{Department of Physics, University of Oslo, N-0316 Oslo, Norway}%
\affiliation{Norwegian Nuclear Research Centre, Norway}
\author{V.~W.~Ingeberg}%
\affiliation{Department of Physics, University of Oslo, N-0316 Oslo, Norway}%
\affiliation{Norwegian Nuclear Research Centre, Norway}
\author{A.~C.~Larsen}%
\affiliation{Department of Physics, University of Oslo, N-0316 Oslo, Norway}%
\affiliation{Norwegian Nuclear Research Centre, Norway}
\author{T.~K.~Eriksen}%
\affiliation{Department of Physics, University of Oslo, N-0316 Oslo, Norway}%
\affiliation{Norwegian Nuclear Research Centre, Norway}
\affiliation{Department of Radiation protection and Physics, Sector NUK Kjeller, Institute for Energy Technology, 2007 Kjeller, Norway}
\author{\\H.~C.~Berg}%
\email{Current address: Facility for Rare Isotope Beams, 640 S Shaw Ln, East Lansing, MI 48824, USA}
\affiliation{Department of Physics, University of Oslo, N-0316 Oslo, Norway}%





\author{F.~L.~B.~Garrote}%
\affiliation{Department of Physics, University of Oslo, N-0316 Oslo, Norway}%

\author{D.~Gjestvang}%
\affiliation{Department of Physics, University of Oslo, N-0316 Oslo, Norway}%
\affiliation{Norwegian Nuclear Research Centre, Norway}
\author{A.~G\"orgen}%
\affiliation{Department of Physics, University of Oslo, N-0316 Oslo, Norway}%
\affiliation{Norwegian Nuclear Research Centre, Norway}

\author{M.~Markova}%
\affiliation{Department of Physics, University of Oslo, N-0316 Oslo, Norway}%
\affiliation{Norwegian Nuclear Research Centre, Norway}
\author{V.~Modamio}%
\affiliation{Department of Physics, University of Oslo, N-0316 Oslo, Norway}%
\affiliation{Norwegian Nuclear Research Centre, Norway}
\author{E.~Sahin}%
\affiliation{Department of Physics, University of Oslo, N-0316 Oslo, Norway}%
\affiliation{Norwegian Nuclear Research Centre, Norway}

\author{G.~M.~Tveten}%
\affiliation{Department of Physics, University of Oslo, N-0316 Oslo, Norway}%

\author{V.~M.~Valsdòttir}%
\affiliation{Department of Physics, University of Oslo, N-0316 Oslo, Norway}%
\affiliation{Norwegian Nuclear Research Centre, Norway}
\date{\today}

\begin{abstract}
\begin{description}
\item[Background] 
The radiative branching ratio of the Hoyle state is crucial to estimate the triple-$\alpha$ reaction rate in stellar environments at medium temperatures of $T=0.1$ to 2 GK.
Knowledge of the $\gamma$-decay channel is critical as this is the dominant radiative decay channel for the Hoyle state.
A recent study by Kib\'edi \textit{et al.} [Phys. Rev. Lett. \textbf{125}, 182701 (2020)] has challenged our understanding of this astrophysically significant branching ratio and its constraints.

\item[Purpose] 
The main purpose was to perform a new measurement of the $\gamma$-decay branching ratio of the Hoyle state to deduce the radiative branching ratio of the Hoyle state.
An additional objective was to independently verify aspects of the aforementioned measurement conducted by Kib\'edi \textit{et al.}

\item[Method] 
For the primary experiment of this work the Hoyle state was populated by the $^{12}\textrm{C}(p,p')$ reaction at 10.8 MeV at the Oslo Cyclotron Laboratory.
The $\gamma$-decay branching ratio was deduced through triple-coincidence events, each consisting of a proton-ejectile energy corresponding to population of the $0_{2}^{+}$ Hoyle state, and the subsequent cascade of 3.21 MeV and 4.44 MeV $\gamma$-rays.
To verify the analysis, a surrogate $\gamma$-ray cascade from the $0_{2}^{+}$ state in $^{28}\textrm{Si}$ was also studied.
Following the same methodology, an independent analysis of the 2014 data published by Kib\'edi \textit{et al.} [Phys. Rev. Lett. \textbf{125}, 182701 (2020)] has been carried out.

\item[Results] 
From the main experiment of this work, a $\gamma$-decay branching ratio of the Hoyle state was determined as \mbox{$\Gamma_{\gamma}^{7.65}/\Gamma^{7.65}=4.0(3)\times 10^{-4}$}, yielding a radiative branching ratio of \mbox{$\Gamma_{\textrm{rad}}/\Gamma=4.1(4) \times 10^{-4}$}.
The independent reanalysis of the 2014 experiment published by Kib\'edi \textit{et al.} [Phys. Rev. Lett. \textbf{125}, 182701 (2020)] in this work yielded \mbox{$\Gamma_{\gamma}^{7.65}/\Gamma^{7.65}=4.5(6)\times 10^{-4}$}, with a corresponding radiative branching ratio of \mbox{$\Gamma_{\textrm{rad}}/\Gamma=4.6(6) \times 10^{-4}$}.

\item[Conclusions] The radiative branching ratio of the Hoyle state reported in this work is in excellent agreement with several recent studies, as well as the previously adopted ENSDF average of \mbox{$\Gamma_{\textrm{rad}}/\Gamma=4.16(11)\times 10^{-4}$}.
In this work, several issues were found in the analysis of [Phys. Rev. Lett. \textbf{125}, 182701 (2020)], with the corrected values no longer being discrepant with the ENSDF average.

\end{description}

\end{abstract}

\maketitle





\section{Introduction}
\label{sec:Introduction}
The production of $^{12}\textrm{C}$ through the triple-$\alpha$ reaction is a crucial process to our understanding of stellar nucleosynthesis (see Refs.~\cite{BF2H, FYNBO_NATURE_TRIPLE_ALPHA, FREER20141} for further discussions).
At medium temperatures between 0.1--2 GK, the triple-$\alpha$ reaction is predominantly mediated by the Hoyle state: the $0_{2}^{+}$ resonance of $^{12}\textrm{C}$ at $E_{\textrm{x}} = 7.65407(19)$ MeV \cite{KELLEY201771}.
The existence of this resonance was first proposed by Fred Hoyle~\cite{hoyle1953state} and was first experimentally observed by Dunbar \textit{et al.}~\cite{DUNBAR_PhysRev.92.649}.
The Hoyle state also plays a fundamental role in our understanding of nuclear clustering and its theoretical calculation remains an indispensable test for the development of nuclear models \cite{FREER20141, RevModPhys.90.035004}.
As such, the properties of the Hoyle state continue to be of tremendous interest for the scientific community, both theoretically and experimentally.

Accurate evaluations for the triple-$\alpha$ reaction rate are important to correctly model the subsequent stellar nucleosynthesis.
At low temperatures below $0.1$ GK, the direct mechanism dominates the triple-$\alpha$ reaction \cite{PhysRevLett.109.141101,Garrido2011,PhysRevC.92.022801}.
At higher temperatures beyond $2$ GK, there is significant uncertainty in the contributions to the triple-$\alpha$ reaction due to the presence of complex, broad resonance structures \cite{LI2022136928, PhysRevC.105.024308} as well as extremely rare radiative decay branches \cite{TSUMURA2021136283, PhysRevC.104.064315}.
In contrast to the low- and high-temperature regimes, the medium-temperature regime between 0.1 and 2 GK has generally been understood to be well constrained \cite{FYNBONATURE, KELLEY201771}.
Specifically, it is the nuclear properties of the Hoyle state which directly determine the triple-$\alpha$ reaction rate at medium-temperatures and as such, accurate knowledge of these properties is essential.

A recent measurement by Kib\'edi \textit{et al.}~\cite{PhysRevLett.125.182701} has challenged our understanding of the crucial radiative branching ratio of the Hoyle state, with reported value of \mbox{$\Gamma_{\textrm{rad}}/\Gamma=6.2(6) \times 10^{-4}$}.
This value is highly discrepant with the adopted value of \mbox{$\Gamma_{\textrm{rad}}/\Gamma=4.16(11)\times 10^{-4}$} given by Kelley \textit{et al.}~\cite{KELLEY201771}.
Given the astrophysical importance of the Hoyle state, the result of Ref.~\cite{PhysRevLett.125.182701} has triggered several further investigations measuring the recoiling $^{12}\textrm{C}$ from alpha-scattering \cite{TSUMURA2021136283, PhysRevC.109.025801, dellAquila_hoyleState,RANA2024139083}.
These measurements support the adopted value by Kelley \textit{et al.}~\cite{KELLEY201771}.
This work presents an effort to remeasure the $\gamma$-decay and radiative branching ratios of the Hoyle state, complementing the measurements performed by Kib\'edi \textit{et al.}~\cite{PhysRevLett.125.182701} and Obst \textit{et al.}~\cite{PhysRevC.13.2033}.
The radiative width of the Hoyle state cannot be measured directly, but it can be deduced indirectly with three independently measured quantities as
\begin{equation} \label{eq: gamma_rad}
    \Gamma_{\textrm{rad}}=\left[\frac{\Gamma_{\textrm{rad}}}{\Gamma^{7.65}}\right] \times\left[\frac{\Gamma^{7.65}}{\Gamma_{\pi}^{E 0}}\right] \times\left[\Gamma_{\pi}^{E 0}\right].
\end{equation}
The current recommended value for the radiative width of the Hoyle state is $\Gamma_{\textrm{rad}} = 3.81(39)$ meV \cite{KELLEY201771}, with an uncertainty of about $10\%$. The most precise term in Eq.~\ref{eq: gamma_rad} is $\Gamma_{\textrm{rad}}/\Gamma$, which can be expressed as
\begin{equation}\label{eq: gamma_rad/gamma}
    \frac{\Gamma_{\textrm{rad}}}{\Gamma^{7.65}}=\frac{\Gamma_{\gamma}^{E2}\left(1+\alpha_{\textrm{tot}}\right) + \Gamma_{\pi}^{E0}}{\Gamma^{7.65}},
\end{equation}
where $\alpha_{\textrm{tot}}$ is the theoretical total $E2$ conversion coefficient and $\Gamma_{\pi}^{E0}/\Gamma$ is the partial $E0$ pair decay width. 
This work reports a new measurement of $\Gamma_{\gamma}^{E2}/\Gamma^{7.65}$, which is deduced by measuring the $\gamma$-decay branching ratio of the Hoyle state.
This rare decay mode from the Hoyle state corresponds to an $E2$-$E2$ $\gamma$-ray cascade that proceeds through the first-excited $2_{1}^{+}$ state to the ground state.
Such events were observed as proton-$\gamma$--$\gamma$ coincidences, corresponding to proton ejectiles having populated the Hoyle state, as well as the emission of 3.21 and 4.44 MeV $\gamma$ rays of the following $E2$-$E2$ $\gamma$-ray cascade.
The $\Gamma_{\gamma}^{E2}/\Gamma^{7.65}$ branching ratio can be expressed as
\begin{equation}\label{eq: gamma_gamma/gamma}
    \frac{\Gamma_{\gamma}^{E2}}{\Gamma^{7.65}} = \frac{N_{020}^{7.65}}{N_{\text{inclusive}}^{7.65} \times \epsilon_{3.21} \times \epsilon_{4.44} \times c_{\textrm{det}} \times W_{020}^{7.65}},
\end{equation}
\noindent where $N_{020}^{7.65}$ is the number of \mbox{$p$-$\gamma$--$\gamma$} triple-coincidence events, each corresponding to a proton exciting the $0_{2}^{+}$ Hoyle state and two photopeak signals corresponding to $\gamma$ rays with energies of $E_{\gamma} = 3.21$ and 4.44 MeV.
$N_{\text{inclusive}}^{7.65}$ is the inclusive amount of protons populating the Hoyle state, while $\epsilon_{3.21}$ and $\epsilon_{4.44}$ are the absolute photopeak efficiencies per detector, corresponding to the $E_{\gamma}= 3.21$ and $E_{\gamma}=4.44$ MeV $E2$-$E2$ $\gamma$-ray cascade from the Hoyle state.
In this work, we define these absolute photopeak efficiencies to correspond to events within the narrow photopeak, which are separated from any smooth underlying contributions (e.g., the Compton continuum or background events) by means of a fit.
$W_{020}^{7.65}$ is the angular correlation correction factor for the two $\gamma$ rays and $c_{\textrm{det}}=n_{\textrm{det}}(n_{\textrm{det}}-1)$ is a combinatorial factor, where $n_{\textrm{det}}$ is the total number of $\gamma$-ray detectors in the setup.

Practically, $N_{020}^{7.65}$ is determined from the fit of a 1-dimensional proton spectrum, generated by either gating on the energy of a single photopeak in a $\gamma$--$\gamma$ matrix, or the summed energy of both photopeaks in a summed-$\gamma$ matrix.
In Sec.~\ref{subsec:Data_Analysis_main_measurement_12C_2019}, we detail how the gating procedure can bias the number of observed triple-$\alpha$ coincidences.
This effect was not accounted for in the study by Kib\'edi \textit{et al.}~\cite{PhysRevLett.125.182701}.
In this work, we detail how this is the main source of the discrepant $\gamma$-decay branching ratio reported in Ref.~\cite{PhysRevLett.125.182701}.
\section{Experimental apparatus}
\begin{table*}[btp]
\caption{
\label{tab:Experimental_apparatus_OSCAR}%
A summary of the experimental conditions of the three measurements studied in this work using the OSCAR detector array.
The primary experiment of this work was the $^{12}\textrm{C}(p,p')$ measurement performed in 2019.
}
\begin{ruledtabular}
\begin{tabular}{l D{,}{}{4.0} D{,}{}{4.0} D{,}{}{4.0} D{,}{}{4.0} } 
 & \multicolumn{1}{c}{$^{12}\textrm{C}(p,p')$ with $E_p=10.8$ MeV} &  \multicolumn{1}{c}{$^{28}\textrm{Si}(p,p')$ with $E_p=10.8$ MeV} &   \multicolumn{1}{c}{$^{28}\textrm{Si}(p,p')$ with $E_p=16.0$ MeV}   \\ [0.5ex]
 & \multicolumn{1}{c}{performed in 2019} &  \multicolumn{1}{c}{performed in 2019} &   \multicolumn{1}{c}{performed in 2020}\\ 
\midrule
Ejectile detector          & \multicolumn{1}{c}{SiRi ($\theta=126.0\text{--}140.0^{\circ}$) \cite{GUTTORMSEN2011168}} & \multicolumn{1}{c}{SiRi ($\theta=126.0\text{--}140.0^{\circ}$) \cite{GUTTORMSEN2011168}} & \multicolumn{1}{c}{SiRi ($\theta=126.0\text{--}140.0^{\circ}$) \cite{GUTTORMSEN2011168}}  \\ [0.8ex]
$\gamma$-ray detector     & \multicolumn{1}{c}{LaBr$_3$(Ce) ($n_{\textrm{det}}=30$) \cite{ZEISER2021164678}}  & \multicolumn{1}{c}{LaBr$_3$(Ce) ($n_{\textrm{det}}=30$) \cite{ZEISER2021164678}} & \multicolumn{1}{c}{LaBr$_3$(Ce) ($n_{\textrm{det}}=30$) \cite{ZEISER2021164678}}  \\ [0.8ex]
Total $\gamma$-ray efficiency (1.3 MeV) & \approx, 56,.0\% & \approx 56,.0\% & \approx 56,.0\% \\ [0.8ex]
Detector distance from target & 16,.3\textrm{ cm} & 16,.3\textrm{ cm} & 16,.3\textrm{ cm} \\ [0.8ex]
\multirow{2}{*}{Target}     & \multicolumn{1}{c}{$^{\mathrm{nat}}\textrm{C}$ (180 $\upmu$g/cm$^2$)} & \multicolumn{1}{c}{SiO$_{2}$ (140 $\upmu$g/cm$^2$) with} & \multicolumn{1}{c}{SiO$_{2}$ (140 $\upmu$g/cm$^2$) with} \\ [0.8ex]
     &  & \multicolumn{1}{c}{$^{\mathrm{nat}}\textrm{C}$ (30 $\upmu$g/cm$^2$) backing} & \multicolumn{1}{c}{$^{\mathrm{nat}}\textrm{C}$ (30 $\upmu$g/cm$^2$) backing}\\ [0.8ex]
\end{tabular}
\end{ruledtabular}
\end{table*}

\begin{table*}[btp]
\caption{\label{tab:Experimental_apparatus_CACTUS}%
A summary of the experimental conditions of the three measurements studied in this work using the CACTUS detector array.
}
\begin{ruledtabular}
\begin{tabular}{l D{,}{}{4.0} D{,}{}{4.0} D{,}{}{4.0} } 
 & \multicolumn{1}{c}{$^{12}\textrm{C}(p,p')$ and $^{28}\textrm{Si}(p,p')$ with $E_p=16.0$ MeV} &   \multicolumn{1}{c}{$^{12}\textrm{C}(p,p')$ and $^{28}\textrm{Si}(p,p')$ with $E_p=10.7$ MeV}   \\ [0.8ex]
 & \multicolumn{1}{c}{performed in 2012 } &   \multicolumn{1}{c}{performed in 2014 \cite{PhysRevLett.125.182701}}\\ [0.8ex]
\midrule
Ejectile detector          & \multicolumn{1}{c}{SiRi ($\theta = 40.0\text{--}54.0^{\circ}$) \cite{GUTTORMSEN2011168}} & \multicolumn{1}{c}{SiRi ($\theta=126.0\text{--}140.0^{\circ}$) \cite{GUTTORMSEN2011168}}  \\ [0.8ex]
$\gamma$-ray detector     & \multicolumn{1}{c}{NaI(Tl) ($n_{\textrm{det}}=23$) \cite{M_Guttormsen_1990}} & \multicolumn{1}{c}{NaI(Tl) ($n_{\textrm{det}}=26$) \cite{M_Guttormsen_1990}}  \\ [0.8ex]
$\gamma$-ray efficiency (1.3 MeV)    & \approx12,.8\% & \approx14,.5\% \\ [0.8ex]
Detector distance from target    & 22,.0\textrm{ cm} & 22,.0\textrm{ cm} \\ [0.8ex]
\multirow{3}{*}{Targets}  & \multicolumn{1}{c}{$^{\mathrm{nat}}\textrm{C}$ (1 mg/cm$^2$)} & \multicolumn{1}{c}{$^{\mathrm{nat}}\textrm{C}$ (180 $\upmu$g/cm$^2$) and } \\ [0.8ex]
& \multicolumn{1}{c}{and $^{\mathrm{nat}}\textrm{Si}$ (4 mg/cm$^2$)} & \multicolumn{1}{c}{SiO$_{2}$ (140 $\upmu$g/cm$^2$) with} \\ [0.8ex]
&  & \multicolumn{1}{c}{$^{\mathrm{nat}}\textrm{C}$ (30 $\upmu$g/cm$^2$) backing} \\ [0.8ex]
\end{tabular}
\end{ruledtabular}
\end{table*}
Three experiments were performed in this study, with the experimental conditions summarized in Table~\ref{tab:Experimental_apparatus_OSCAR}.
Several measurements from previous experiments were also analyzed, with the experimental conditions summarized in Table~\ref{tab:Experimental_apparatus_CACTUS}.
In this work, the primary experiment to study the $0_{2}^{+}$ Hoyle state was performed in 2019 (Sec.~\ref{subsec:experimentalApparatus_12C_2019}).
Data from $^{28}\textrm{Si}$ and $^{12}\textrm{C}$ measurements was used to obtain efficiencies and validate the analysis methodology employed in this work.
These $^{28}\textrm{Si}$ measurements were also used to study the decay of the $0_{2}^{+}$ state in $^{28}\textrm{Si}$ as a surrogate for that of the Hoyle state (one of these measurements is described in Sec.~\ref{subsec:experimentalApparatus_28Si_2020}).
Finally, an independent analysis of the data published by Kib\'edi \textit{et al.}~\cite{PhysRevLett.125.182701} was performed in this work (Sec.~\ref{subsec:experimentalApparatus_12C_2014}). 
\subsection{$\mathbf{^{12}\textrm{C}(p,p')}$ and $\mathbf{^{28}\textrm{Si}(p,p')}$ with $\mathbf{E_{p}=10.8} \text{ MeV}$ performed in 2019}
\label{subsec:experimentalApparatus_12C_2019}
The Hoyle state was populated through inelastic proton scattering on a $^{\mathrm{nat}}\textrm{C}$ target with an areal density of 180 $\mu\textrm{g} / \textrm{cm}^{2}$.
A 140 $\mu\textrm{g} / \textrm{cm}^{2}$-thick SiO$_{2}$ target with a 30 $\mu\textrm{g} / \textrm{cm}^{2}$-thick $^{\mathrm{nat}}\textrm{C}$ backing was also employed to study relevant states in $^{28}\textrm{Si}$.
The beam energy was $E_{p}\approx 10.8$ MeV and was delivered with a current of 2--6 nA by the MC-35 Scanditronix cyclotron at the Oslo Cyclotron Laboratory (OCL).
Ejectiles were detected using the Silicon Ring (SiRi) particle-telescope system, consisting of eight trapezoidal modules mounted at a distance of $\approx5$ cm from the target \cite{GUTTORMSEN2011168}.
These modules covered a polar-angle range of $126^{\circ}$--$140^{\circ}$, with $\approx 2^{\circ}$ being subtended by each of the eight rings acting as the front $\Delta E$-layer. 
The thickness of the $\Delta E$ and $E$ detectors are approximately 130 and 1550 \si{\micro\metre}, respectively \cite{GUTTORMSEN2011168}. 
The coincident $\gamma$-rays decays were detected with the OSCAR multidetector system \cite{ZEISER2021164678}.
The large-volume LaBr$_{3}$(Ce) detectors of OSCAR were configured at closest possible distance of $\approx 16.3$ cm from the target, each detector subtending a solid angle of $\approx1.9\%$ of $4\pi$.
Events were triggered by signals in the $\Delta E$ detectors, with the time-to-digital (TDC) values from coincidentally triggered LaBr$_{3}$(Ce) detectors being recorded relative to that of the detected proton.

\subsection{$\mathbf{^{28}\textrm{Si}(p,p')}$ with $\mathbf{E_{p}=16.0 \text{ MeV}}$ performed in 2020}
\label{subsec:experimentalApparatus_28Si_2020}
In 2020, a 140 $\mu\textrm{g} / \textrm{cm}^{2}$-thick SiO$_{2}$ target with a 30 $\mu\textrm{g} / \textrm{cm}^{2}$-thick $^{\mathrm{nat}}\textrm{C}$ backing was employed to study states in $^{28}\textrm{Si}$ to test the background subtraction method and obtain in-beam efficiencies relevant for Eq.~\ref{eq: gamma_gamma/gamma}.
Events were triggered by signals in the $\Delta E$ detectors of the SiRi particle-telescope system, configured to a polar-angle range of $126^{\circ}$--$140^{\circ}$.
The coincident $\gamma$ rays were detected with the OSCAR multidetector system.

\subsection{$\mathbf{^{12}\textrm{C}(p,p')}$ and $\mathbf{^{28}\textrm{Si}(p,p')}$ with $\mathbf{E_{p}=10.7 \text{ MeV}}$ performed in 2014}
\label{subsec:experimentalApparatus_12C_2014}

The data from this experiment were previously published by Kib\'edi \textit{et al.}~\cite{PhysRevLett.125.182701}. 
The Hoyle state was populated through inelastic proton scattering on a $^{\mathrm{nat}}\textrm{C}$ target with an areal density of 180 $\mu\textrm{g} / \textrm{cm}^{2}$ using a beam energy of 10.7 MeV.
A 140 $\mu\textrm{g} / \textrm{cm}^{2}$-thick SiO$_{2}$ target with a 30 $\mu\textrm{g} / \textrm{cm}^{2}$-thick $^{\mathrm{nat}}\textrm{C}$ backing was also employed to study relevant states in $^{28}\textrm{Si}$.
Events were triggered by signals in the $\Delta E$ detectors of the SiRi particle-telescope system, configured to a polar-angle range of $126^{\circ}$--$140^{\circ}$ and detected by the analog data-aquisition system.
The coincident $\gamma$ rays were detected with the CACTUS multidetector system \cite{M_Guttormsen_1990}.
The large-volume NaI(Tl) detectors of CACTUS were configured at $\approx 22$ cm from the target, collimated with 10 cm of lead, with each detector subtending a solid angle of $\approx0.63\%$ of $4\pi$.

\section{Data Analysis: New measurements performed in this work}\label{sec:Data_Analysis_main_measurement}
For the primary experiment of this work, the data analysis methods are explained in detail (Sec.~\ref{subsec:Data_Analysis_main_measurement_12C_2019}).
The same methodology is employed in the analyses of the 2020, 2014 and 2012 experiments, which are presented in Secs.~\ref{subsec:Data_Analysis_main_measurement_28Si_2020}, \ref{subsec:Data_Analysis_main_measurement_28Si_2019}, \ref{subsec:Data_Analysis_reanalysis_12C_2014} and \ref{subsec:Data_Analysis_reanalysis_28Si_2014}, respectively.

\begin{figure*}[htbp]
\centering
\includegraphics[width=2\columnwidth]{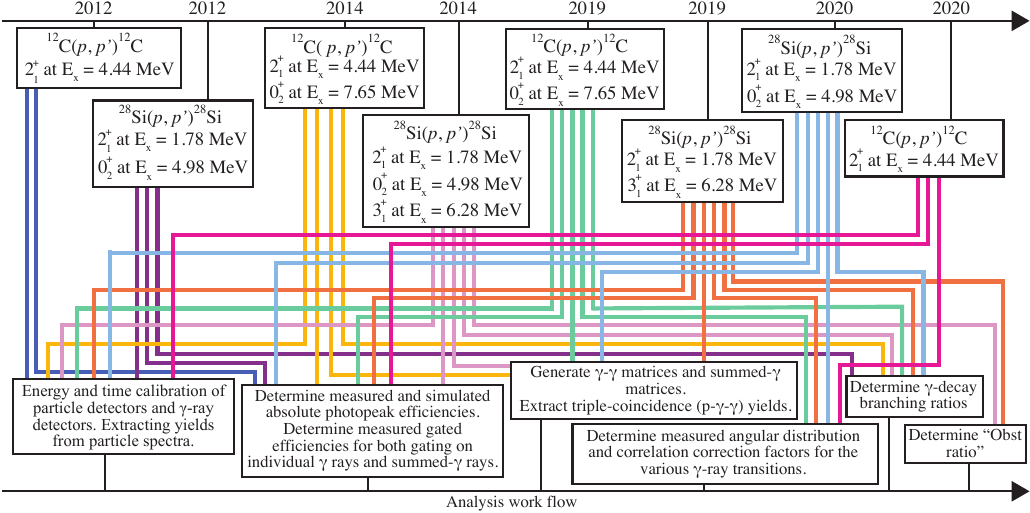}
\caption{\label{fig:experimentalOverviewDiagram}
Diagram illustrating the various measurements utilized in this work (refer to Tables~\ref{tab:Experimental_apparatus_OSCAR} and \ref{tab:Experimental_apparatus_CACTUS} for detailed information), along with the analysis pipeline.
The diagram includes the years in which the measurements were performed, the excited states of interest for each measurement, and the connections between the measurements and their corresponding points of use in the analysis pipeline.
}
\end{figure*}

Given the complexity and extensive scope of this study, Fig.~\ref{fig:experimentalOverviewDiagram} presents a diagram that illustrates the various measurements utilized in this work, along with the corresponding analysis pipeline.
This diagram details the years in which the measurements were performed, highlights the excited states of interest for each measurement, and outlines the connections between the measurements and their relevant stages in the analysis pipeline.
\subsection{$\mathbf{^{12}\textrm{C}(p,p')}$ with $\mathbf{E_{p}=10.8}$ MeV performed in 2019}\label{subsec:Data_Analysis_main_measurement_12C_2019}
\begin{figure}[htbp]
\centering
\includegraphics[width=\columnwidth]{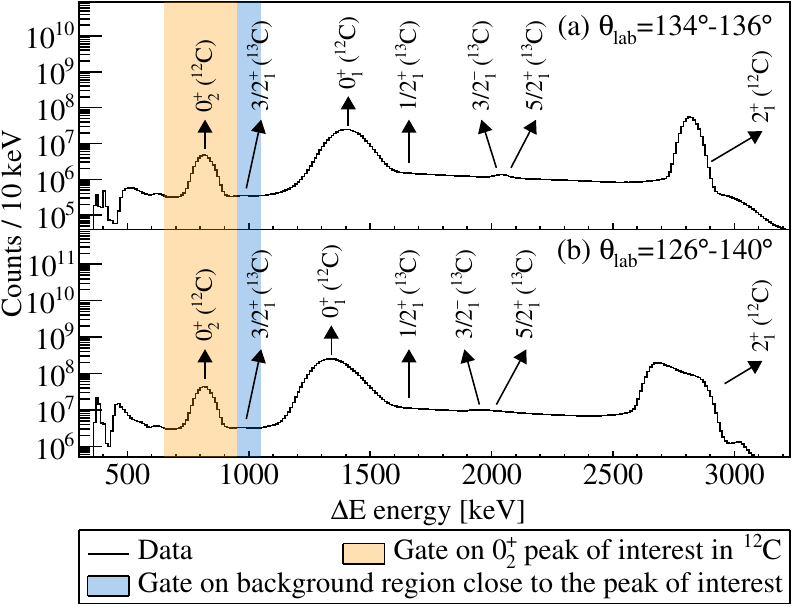}
\caption{\label{fig:InclusiveProtonSpectrum_2019}
(a) $\Delta E$ energy spectrum of inelastically scattered proton ejectiles, detected at $\theta_{\textrm{lab}} = 134^{\circ}$--$136^{\circ}$ with SiRi from the measurement performed in 2019.
The orange shaded area denotes the energy gate employed for the $0_{2}^{+}$ Hoyle state.
The blue shaded area denotes the energy gate employed for the background gated analysis, as explained in Appendix~\ref{subsec:investigationOfBackgroundInclusiveSpectra}.
The total projection of all angles of SiRi ($\theta_{\textrm{lab}} = 126^{\circ}$--$140^{\circ}$) is shown in panel (b).
The energy calibration for this data was conducted under the assumption of no dead layer in the particle detector (see text for details).
}
\end{figure}

For the $^{\mathrm{nat}}\textrm{C}(\textit{p},\textit{p}')$ reaction, the energy depositions in the $\Delta E$ detector of SiRi for angles $\theta_{\textrm{lab}}=126^{\circ}$--$140^{\circ}$ from scattered protons are shown in Fig.~\ref{fig:InclusiveProtonSpectrum_2019}.
Only the $\Delta E$ signal from SiRi was employed to gate on the Hoyle state as the corresponding proton ejectiles were stopped in the $\Delta E$ detector.
The Hoyle/$2_{1}^{+}$ population ratio of this 2019 experiment is half of the 2014 experiment due to the change in beam energy of about $\Delta E_{p}=100$ keV, due to the sharply changing excitation function for the Hoyle state in this energy range \cite{PhysRevC.104.024620}.
For the 2014 experiment \cite{PhysRevLett.125.182701}, the Hoyle/$2_{1}^{+}$ population ratio is in agreement with the study by Cook \textit{et al.}~\cite{PhysRevC.104.024620} (measured between \mbox{$E_{p}\approx$ 10.2--10.7} MeV).
A linear extrapolation of the Hoyle/$2_{1}^{+}$ population ratios in Ref.~\cite{PhysRevC.104.024620} up to $E_{p}=10.8$ MeV is in good agreement with that observed in the 2019 measurement (the primary experiment of this work).
Due to the reaction kinematics, the kinetic energy of the proton is highly dependent on the polar angle between the detector ring and the beam detection axis.
Because of the resultant kinematic smearing across each subtended angle, the data from each ring is fitted separately.
The ordering of the $0_{1}^{+}$, $2_{1}^{+}$ and $0_{2}^{+}$ peaks does not follow that of the corresponding excitation energies as the proton ejectiles corresponding to the Hoyle state are stopped in the $\Delta E$ layer of SiRi.
In contrast, the ejectiles corresponding to the lower-lying $2_{1}^{+}$ and $0_{1}^{+}$ states do penetrate the $\Delta E$ layer and are stopped in the $E$ detector.
However, to reproduce the ordering of the $0_{1}^{+}$, $2_{1}^{+}$ and $0_{2}^{+}$ peaks in kinematic simulations, including a dead layer is a necessity.
The dead layer of each individual detector is nearly impossible to estimate; it is also not necessary as the Hoyle state peak is well separated from other peaks in the inclusive spectrum.
Due to this complex kinematic behavior in the $\Delta E$ detectors of SiRi, and variations in the thickness of the $\Delta E$ dead layer, the calibration was performed under the assumption that no dead layer exists.
This assumption is justified because the absolute energies of the proton ejectiles do not affect the analysis; the $\Delta E$ detector spectra are simply used to select for events from the state of interest, such as Hoyle-state events and to estimate the background underneath the peak of interest.
A few weakly populated states from $^{13}\textrm{C}$ are also observed in the inclusive spectra, however, the decays from these states do not interfere with the analysis of the Hoyle state.

In this work, the triple-coincidence yield is obtained from a 1-dimensional spectrum by gating on the state of interest in the inclusive proton spectrum and then either a gate on on the energy of a single photopeak in the $\gamma$--$\gamma$ matrix, or the summed energy of both photopeaks in the summed-$\gamma$ matrix; the latter method being used by Kib\'edi \textit{et al.} (see Fig.~5 in Ref.~\cite{PhysRevLett.125.182701}).
An example of a $\gamma$--$\gamma$ matrix can be seen in Fig.~\ref{fig:gammaGamma_2019}(a), where the energies of two $\gamma$ rays in coincidence are plotted against one another, with the larger energy being on the y-axis.
Similarly a summed-$\gamma$ matrix is shown in Fig.~\ref{fig:summedGamma_2019}(a), where pairs of coincident $\gamma$-ray energies are each plotted against their sum.
When the $\gamma$-ray detector resolution is sufficiently high, the photopeak is well separated from the smooth Compton continuum response of the detector and the gate contains almost only the photopeak yield as previously defined for the absolute photopeak efficiencies.
However, with spectra of poorer resolution, a $3\sigma$ gate on a photopeak can also contain correlated events which form part of the smooth Compton continuum and first-escape peak in the detector response.
Since these correlated events beneath the narrow photopeak are not accounted for in the aforementioned absolute photopeak efficiencies extracted with fits, the effect of gating on the $\gamma$--$\gamma$ and summed-$\gamma$ matrices must be accounted for.
This is achieved by using a gated efficiency for the $\gamma$ ray or sum of $\gamma$ rays utilized for gating.
This gated efficiency consists of all correlated events in both the narrow photopeak and the smooth contribution underneath the peak.
This effect was not accounted for in the study by Kib\'edi \textit{et al.}~\cite{PhysRevLett.125.182701}.
\begin{figure}[htbp]
\centering
\includegraphics[width=\columnwidth]{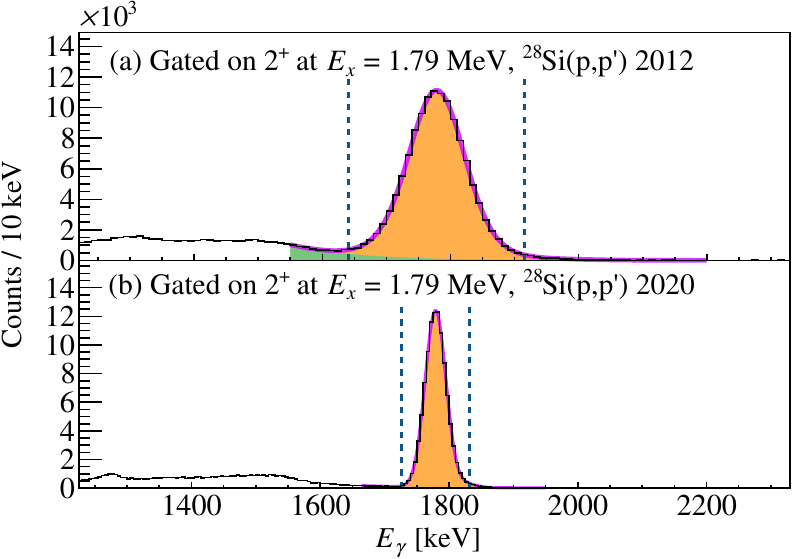}
\includegraphics[width=\columnwidth]{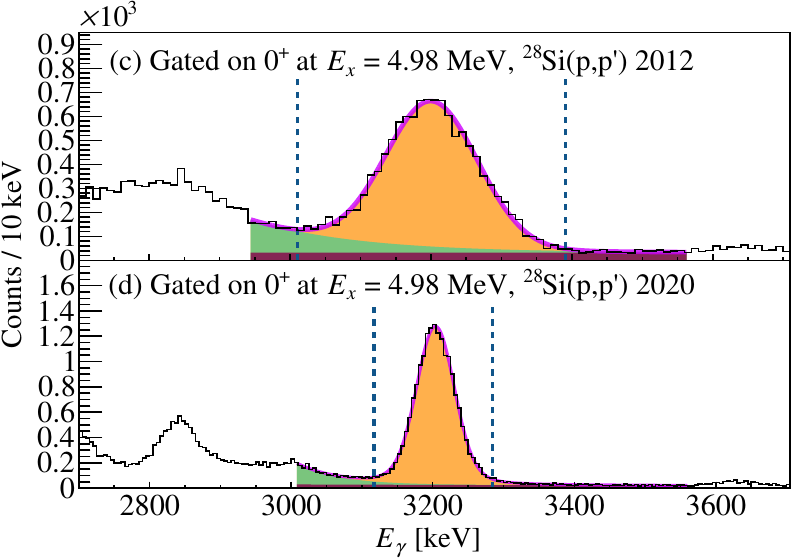}
\includegraphics[width=\columnwidth]{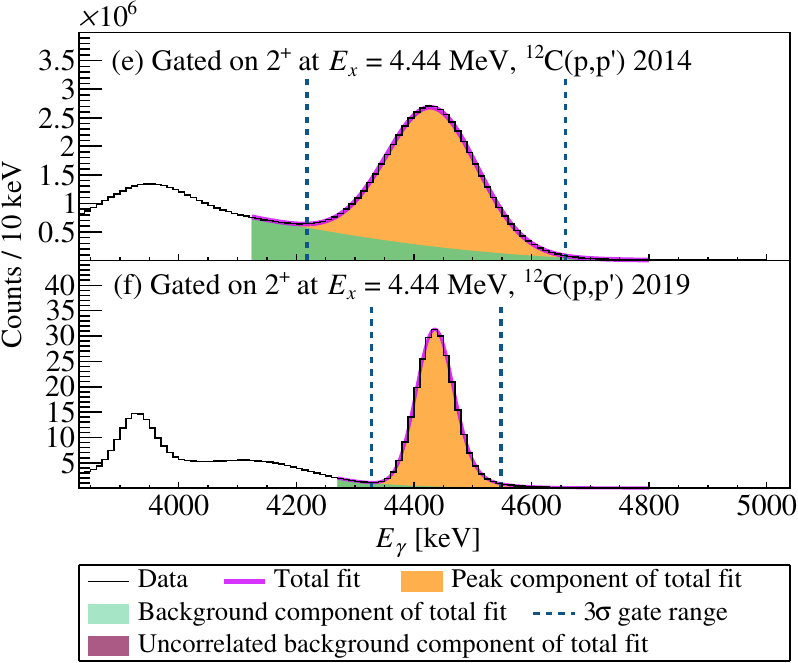}
\caption{\label{fig:gammaGatingCorrectionFactor_all}
{
$\gamma$-ray spectra background-subtracted in regards to time obtained by gating on proton ejectiles, where the various components of the spectra are indicated by different colors.
The orange filled area denotes the absolute photopeak yield of the spectrum, the green filled area denotes the smooth varying and correlated background underneath the peak of interest, the dark pink filled area denotes the smooth uncorrelated background component of the total background, the violet line denotes the total fit to the spectrum and the black dashed lines indicate the valid fit range for this decomposition.
Figures (a), (c) and (e) are obtained by measurements performed with the CACTUS array, while figures (b), (d) and (f) are obtained through measurements performed with the OSCAR array (see text for details).
}
}
\end{figure}
The discrepancy between the fitted peak yield (corresponding to a fitted peak efficiency), and the total yield selected by a direct gate, is shown in Figs.~\ref{fig:gammaGatingCorrectionFactor_all} (a)-(f).
In this figure, the $\gamma$-ray spectra (background subtracted with respect to time) for various transitions are shown.
For the $E_{\gamma}=4.44$ MeV $\gamma$ ray shown in panels (e) and (f), the corresponding photopeak detected with the LaBr$_{3}$(Ce) detectors of OSCAR is relatively well separated from the Compton continuum, as shown in panel (f).
In contrast, the analogous spectrum detected with the NaI(Tl) detectors of the CACTUS array yields a photopeak which overlaps with a significantly large, smoothly varying component, as shown in panel (e).

As mentioned in Sec.~\ref{sec:Introduction}, the absolute photopeak efficiencies in Eq.~\ref{eq: gamma_gamma/gamma} are defined to correspond to events within the narrow photopeak, which are separated from any smooth underlying contributions (e.g., the Compton continuum or background events) by means of a fit.
Depending on the case, these peaks were parameterized with either a single Gaussian, several Gaussians or a single skewed Gaussian.
For the smooth underlying contribution, a polynomial (up to second order) and/or an exponential was employed.
In panels (c) and (d) of Fig.~\ref{fig:gammaGatingCorrectionFactor_all}, there is a non-negligible amount of events underneath the $E_{\gamma}=3.20$ MeV peak of interest.
This stems from the gate on the $0_{2}^{+}$ state in $^{28}\textrm{Si}$ yielding a cascade of $E_{\gamma}=3.20$ MeV and $E_{\gamma}=1.79$ MeV $\gamma$ rays, yielding a summed $\gamma$-ray coincidence within a single detector.
This summing yields background events above the $E_{\gamma}=3.20$ MeV peak of interest, which are assumed to be approximately linear in this region.
The blue shaded area in panels (c) and (d) is a parametrization of this uncorrelated background as a zeroth order polynomial originating from the summed $\gamma$-ray photopeak in a single detector for this transition.
Due to this summing effect being present for both the $0_{2}^{+}$ state at $E_{\mathrm{x}}=4.98$ MeV in $^{28}\textrm{Si}$ and the $E_{\mathrm{x}}=7.65$ MeV $0_{2}^{+}$ in $^{12}\textrm{C}$ it is not possible to obtain a gated efficiency for the $\approx$ $E_{\gamma}=3.20$ MeV $\gamma$ ray without the presence of this summing effect, altough the effect is significantly smaller in panel (d) due to the improved resolution of the OSCAR detectors.
To mitigate systematic error from this uncorrelated background, the $\approx$ $E_{\gamma}=3.20$ MeV $\gamma$ ray in the $0_{2}^{+}$ $E2$--$E2$ cascades of interest from $^{12}\textrm{C}$ and $^{28}\textrm{Si}$ are not utilized in any gating to obtain a triple-coincidence yield.
Instead, it is the complementary transition that was gated on ($2_{1}^{+}$ state at $E_{\textrm{x}}=1.79$ and $2_{1}^{+}$ at $E_{\textrm{x}}=4.44$ for $^{12}\textrm{C}$ and $^{28}\textrm{Si}$, respectively) and the triple-coincidence yield extraction was then performed on the $\approx E_{\gamma}=3.20$ MeV peak.
With this tactic the gated efficiency is obtained from the transitions without the summing effect present and the response from the $\approx E_{\gamma}=3.20$ MeV $\gamma$ ray is obtained with a fitted efficiency.
The $E_{\mathrm{x}}=4.98$ MeV $0_{2}^{+}$ transition from $^{28}\textrm{Si}$ is illustrated in this figure to explain its exclusion in gating processes.
Additionally, it demonstrates how the number of correlated events within a smooth continuum beneath the peak rises with increasing $\gamma$-ray energy.

To ensure self consistency for how the triple-coincidence yield is parameterized with respect to smooth underlying contributions, the spectra of interest were simultaneously fitted (i.e., shared parameters for the peak, etc.) with the high-statistics in-beam spectra which were used to determine the absolute photopeak efficiency.
For each discussed case, the specific efficiency spectrum that was simultaneously fitted is mentioned.
The efficiencies employed for the OSCAR and CACTUS detectors are summarized in Tables~\ref{tab:Efficiency_OSCAR} and \ref{tab:Efficiency_CACTUS}, respectively.
See Sec.~\ref{sec:angularCorrelation_efficiency} for more information regarding efficiencies and angular correlation correction factors.
\begin{figure}[htbp]
\centering
\includegraphics[width=\columnwidth]{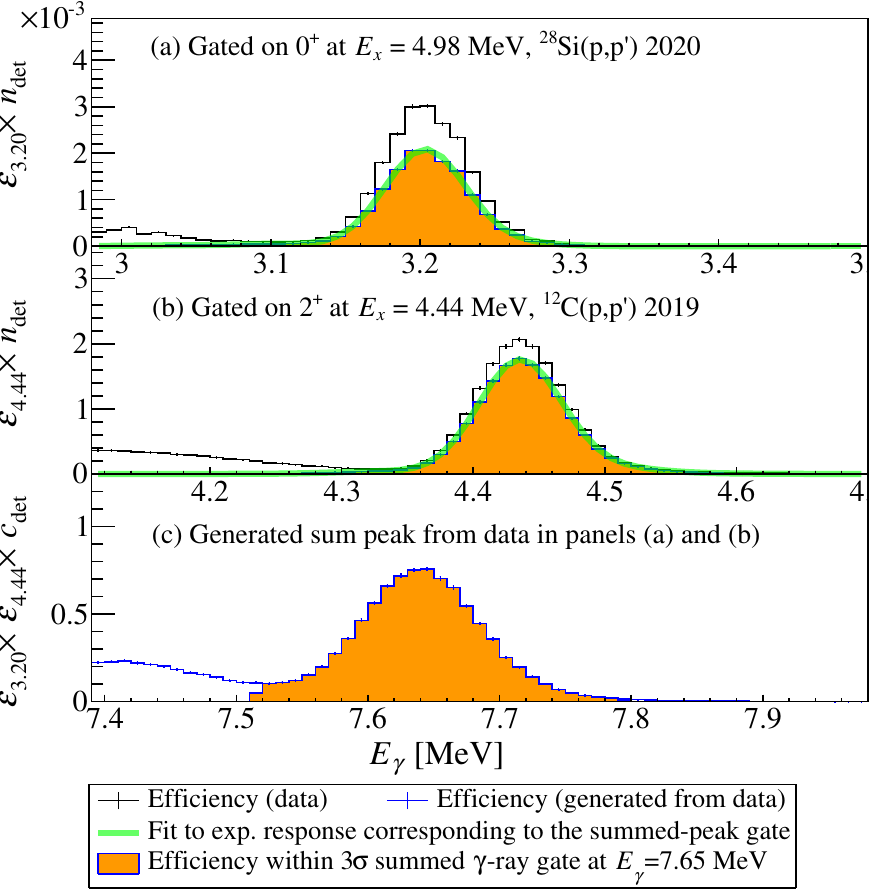}
\caption{\label{fig:summedGammaGatingCorrectionFactor_12C_2019}
{
Detector response at $E_{\gamma}=3.20$ MeV, $E_{\gamma}=4.44$ MeV and $E_{\gamma}=7.65$ MeV using the OSCAR array.
The black points are efficiency per 10 keV, where (a) is the efficiency around $E_{\gamma}=3.21$ MeV and (b) is the efficiency at $E_{\gamma}=4.44$ MeV.
The blue points in (c) are the summed detector response per 10 keV at $E_{\gamma}=7.65$ MeV from the convoluted efficiency spectra in (a) and (b).
The orange filled area is the $3\sigma$ gate of the sum peak at $E_{\gamma}=7.65$ MeV in (c), consisting of the convoluted response from the orange filled area in (a) and (b).
The green line is the fit to the experimental response corresponding to the $3\sigma$ gate at $E_{\gamma}=7.65$ MeV.
}
}
\end{figure}

To extract the triple-coincidence yield from a $\gamma$--$\gamma$ matrix, one may gate on the state of interest in the inclusive particle spectrum, place a gate on one of the $\gamma$-rays in the cascade and fit the complementary photopeak of the cascade.
With such a gating method, it is therefore appropriate to employ a gated efficiency for the gated transition as previously mentioned, which corresponds to the probability for a $\gamma$ ray with energy $E_{\gamma}$ to induce a correlated signal in a detector within the gate range of $E_{\gamma} \pm 3\sigma$.
This is in contrast to the previously defined fitted absolute photopeak efficiency in Eq.~\ref{eq: gamma_gamma/gamma}, where the narrow photopeak is extracted from the spectrum and employed in the efficiency.
For the subsequent fit on the complementary $\gamma$-ray photopeak energy, the corresponding fitted absolute photopeak efficiency should be employed.
For example, for the Hoyle state, a $3\sigma$ gate is employed on the $E_{\gamma} = 4.44$ MeV photopeak, and on the resultant projection, the yield from the $E_{\gamma} = 3.21$ MeV transition is extracted with a fit.
As such, when applying Eq.~\ref{eq: gamma_gamma/gamma}, the efficiency used for the $E_{\gamma} = 4.44$ MeV corresponds to $\epsilon_{4.44}$ (data, gated) in Tables~\ref{tab:Efficiency_OSCAR} and \ref{tab:Efficiency_CACTUS} (for OSCAR and CACTUS, respectively), whilst $\epsilon_{3.21}$ (data, fitted) is used to account for the fitted $E_{\gamma} = 3.21$ MeV photopeak.

\begin{figure}[htbp]
\centering
\includegraphics[width=\columnwidth]{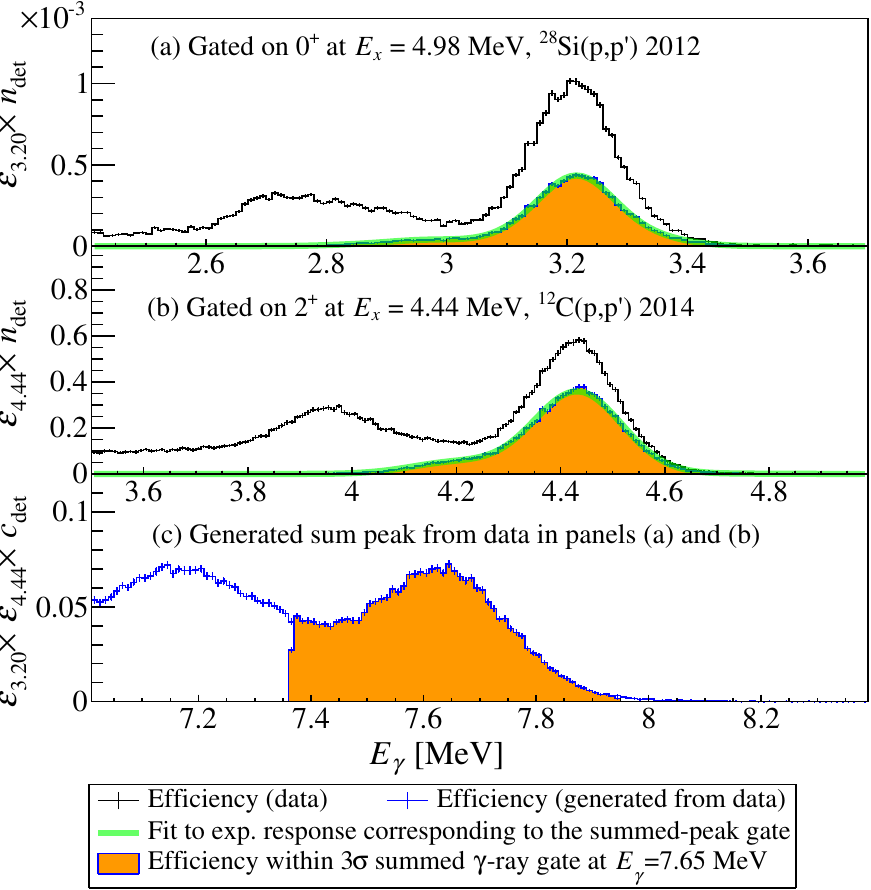}
\caption{\label{fig:summedGammaGatingCorrectionFactor_12C_2014}
{
Detector response at $E_{\gamma}=3.20$ MeV, $E_{\gamma}=4.44$ MeV and $E_{\gamma}=7.65$ MeV using the CACTUS array.
The black points are efficiency per 10 keV, where (a) is the efficiency around $E_{\gamma}=3.20$ MeV and (b) is the efficiency at $E_{\gamma}=4.44$ MeV.
The blue points in (c) are the summed detector response per 10 keV at $E_{\gamma}=7.65$ MeV from the convoluted efficiencies in (a) and (b).
The orange filled area is the $3\sigma$ gate of the sum peak at $E_{\gamma}=7.65$ MeV in (c), consisting of the convoluted response from the orange filled area in (a) and (b).
The green line is the fit to the experimental response corresponding to the $3\sigma$ gate at $E_{\gamma}=7.65$ MeV.
}
}
\end{figure}

\begin{figure}[htbp]
\centering
\includegraphics[width=\columnwidth]{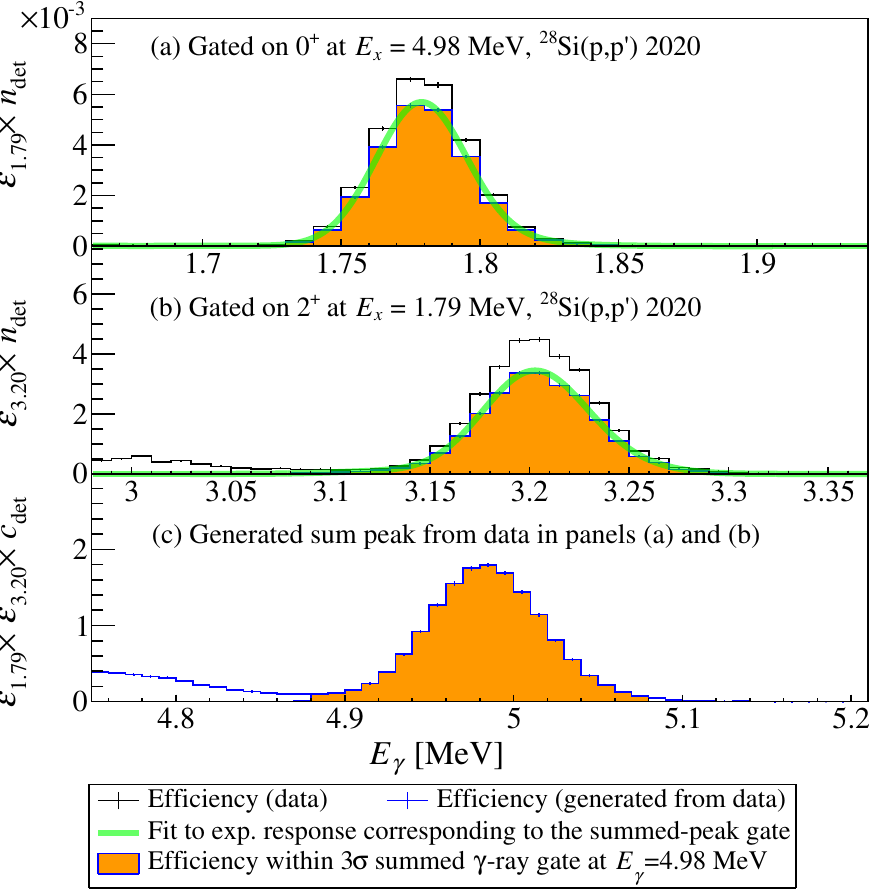}
\caption{\label{fig:summedGammaGatingCorrectionFactor_28Si_2020}
{
Detector response at $E_{\gamma}=1.78$ MeV, $E_{\gamma}=3.20$ MeV and $E_{\gamma}=4.98$ MeV using the OSCAR array.
The black points are efficiency per 10 keV, where (a) is the efficiency around $E_{\gamma}=1.78$ MeV and (b) is the efficiency at $E_{\gamma}=3.20$ MeV.
The blue points in (c) are the summed detector response per 10 keV at $E_{\gamma}=4.98$ MeV from the convoluted efficiencies in (a) and (b).
The orange filled area is the $3\sigma$ gate of the sum peak at $E_{\gamma}=4.98$ MeV in (c), consisting of the convoluted response from the orange filled area in (a) and (b).
The green line is the fit to the experimental response corresponding to the $3\sigma$ gate at $E_{\gamma}=4.98$ MeV.
}
}
\end{figure}

To extract the triple-coincidence yield from a summed-$\gamma$ matrix, one may gate on the state of interest in the inclusive particle spectrum, place a gate on the summed photopeak of the cascade and fit the constituent photopeaks to extract the triple-coincidence yield, where both constituent photopeaks should contain equal amount of counts.
By gating on a sum of two $\gamma$-ray energies in the summed-$\gamma$ matrix the correlated and smooth underlying Compton background becomes non-trivial and a decomposition of the spectrum itself (photopeak and smooth underlying correlated background) becomes nearly impossible to perform.
For this method a summed $\gamma$-ray gated efficiency must be applied, meaning that it is the convolved detector response of the two constituent $\gamma$ rays that must be obtained.
Extracting this experimental summed $\gamma$-ray gated efficiency is possible as long as the statistics are high enough.
In the case of the Hoyle state a different approach must be employed.
To generate a summed $\gamma$-ray detector response of a certain energy, the experimental spectra of the constituent transitions must be convolved.
This method of generating a summed $\gamma$-ray detector response is used to extract two quantities of interest from the experimental spectra: Firstly, the gated efficiency for the summed $\gamma$-ray peak corresponding to the entire array at the summed $\gamma$-ray energy of interest.
Secondly, this analysis provides us with the response for the individual transitions which, when summed, passed within the $3\sigma$ gate of the sum peak.
This is important because the peak shapes of the individual transitions are non-trivial when a gate is performed on the summed-$\gamma$ rays.
One can observe how this effect is very dependent on the resolution of the detectors, much similar to how the $\gamma$--$\gamma$ gating method is very dependent on the detector resolution. 
To generate a summed-$\gamma$ detector response of $E_{\gamma}=7.65$ MeV using OSCAR, the experimental spectra from the $E_{\gamma}=3.20$ MeV transition in Fig.~\ref{fig:summedGammaGatingCorrectionFactor_12C_2019} (a) and $E_{\gamma}=4.44$ MeV transition in Fig.~\ref{fig:summedGammaGatingCorrectionFactor_12C_2019} (b) were convolved to create the sum peak shown in Fig.~\ref{fig:summedGammaGatingCorrectionFactor_12C_2019} (c).
To generate a summed-$\gamma$ detector response of $E_{\gamma}=4.98$ MeV using OSCAR, the experimental spectra from the $E_{\gamma}=1.78$ MeV transition in Fig.~\ref{fig:summedGammaGatingCorrectionFactor_28Si_2020} (a) and $E_{\gamma}=3.20$ MeV transition in Fig.~\ref{fig:summedGammaGatingCorrectionFactor_28Si_2020} (b) were convolved to create the sum peak shown in Fig.~\ref{fig:summedGammaGatingCorrectionFactor_28Si_2020} (c).
The effect that detector resolution has on the peak shapes is very obvious when observing the corresponding summed-$\gamma$ detector response of $E_{\gamma}=7.65$ MeV using CACTUS shown in Fig.~\ref{fig:summedGammaGatingCorrectionFactor_12C_2014} (a)-(c), where the peak shapes are clearly distorted towards the lower-energy tail.

\begin{figure}[htbp]
\centering
\includegraphics[width=\columnwidth]{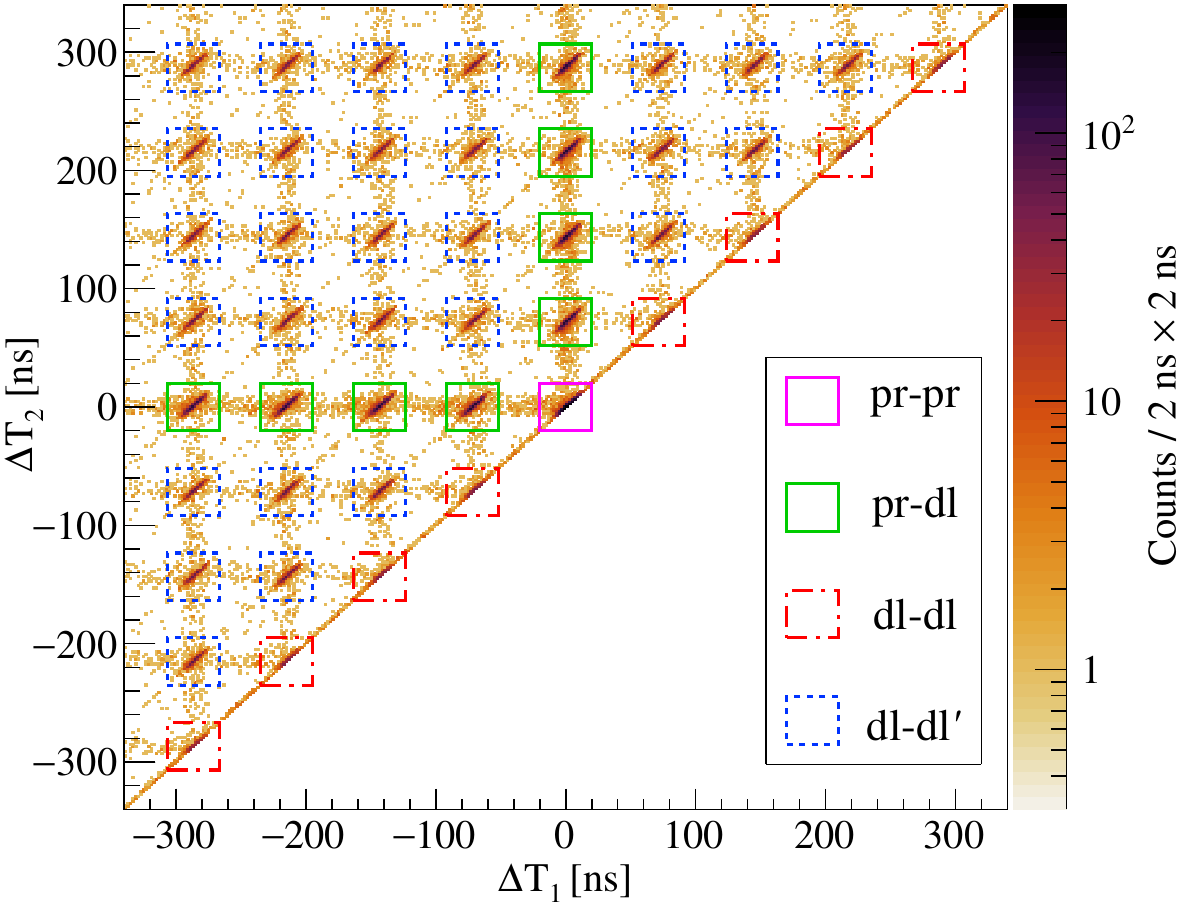}
\caption{\label{fig:labr3_time_matrix_2019}
Time differences between protons populating the Hoyle state and two $\gamma$ rays with $E_{\gamma}=3.21$ MeV and $E_{\gamma}=4.44$ MeV, from the experiment performed in 2019.
The matrix is sorted so that the largest of the two measured times is along the y-axis.
The colored boxes denote the different timing gates used for the analysis. 
See text for an explanation of coincidence events.
}
\end{figure}

\begin{figure}[htbp]
\centering
\includegraphics[width=\columnwidth]{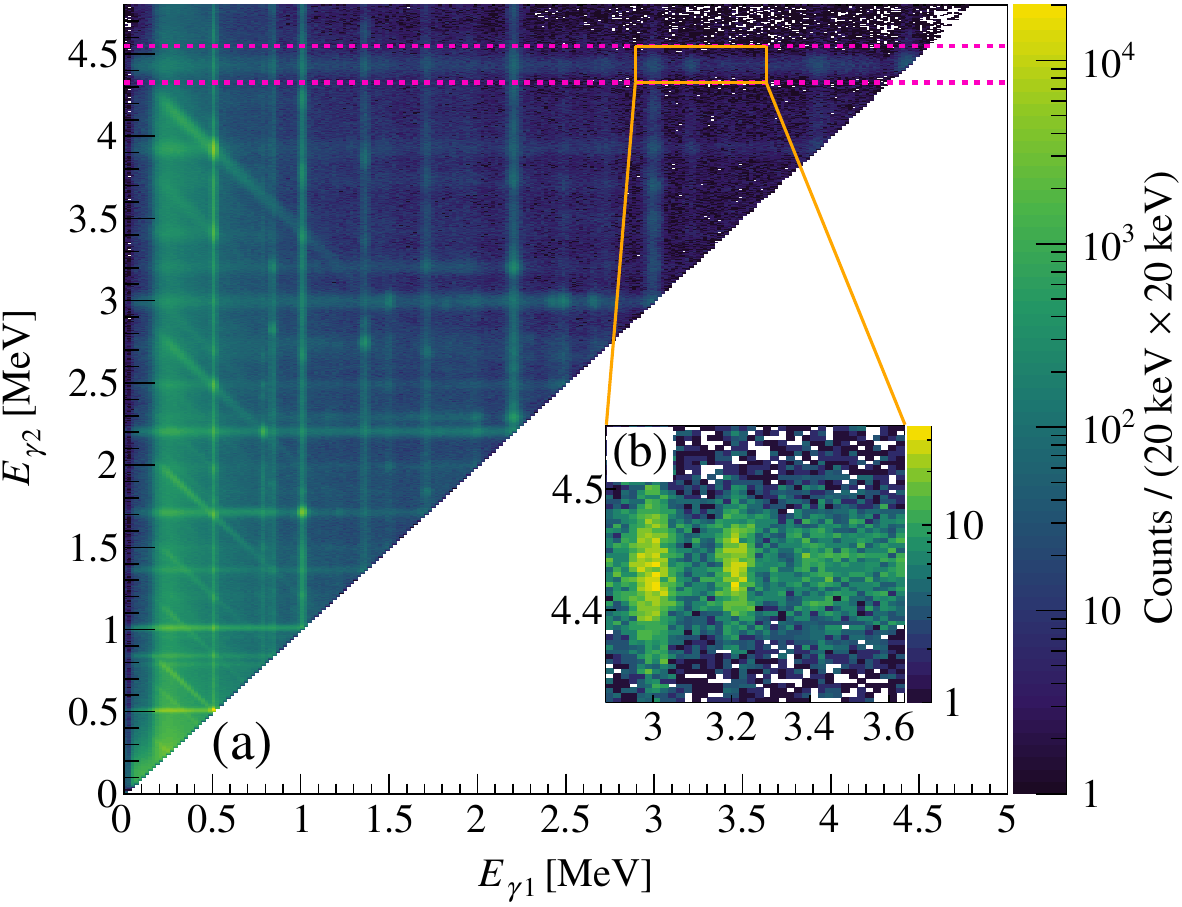}\\
\vspace{4pt}
\includegraphics[width=\columnwidth]{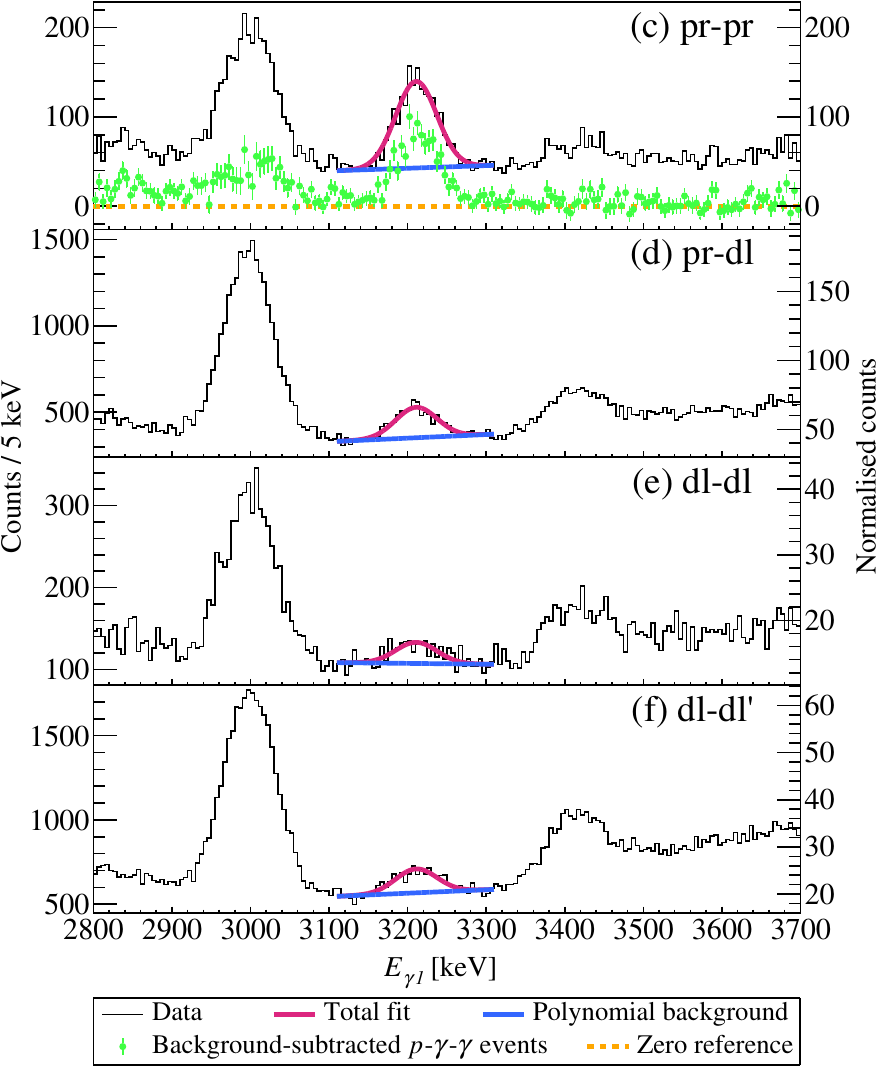}
\caption{\label{fig:gammaGamma_2019}
(a) The $\gamma$--$\gamma$ matrix (largest energy on the y-axis) gated on the \mbox{pr-pr} time locus (see Fig.~\ref{fig:labr3_time_matrix_2019}) from the $^{12}\textrm{C}(p,p')$ measurement performed in 2019.
A zoomed-in region, indicated with solid orange lines, is shown in panel (b).
The vertical, violet dashed lines correspond to the $3\sigma$ gate on the $E_{\gamma}=4.44$ MeV $\gamma$ ray from the Hoyle state cascade.
The events of interest are those within the violet dashed region, with the projected spectra shown in panels (c)--(f) for the different time loci \mbox{pr-pr}, \mbox{pr-dl}, \mbox{dl-dl} and \mbox{dl-dl'} (see Fig.~\ref{fig:labr3_time_matrix_2019} for details on time loci), respectively.
The y-axis on the right side shows the normalized counts, accounting for the number of time loci for each spectrum.
The solid violet line shows the simultaneous peak fit in panels (c)--(f), where the peak shape has been constrained by the data in panel (c).
Individual first-order polynomials are fitted in each panel (c)--(f).
This simultaneous fit was performed with maximum likelihood estimation due to the low statistics of the various background components.
}
\end{figure}

\begin{figure}[htbp]
\centering
\includegraphics[width=\columnwidth]{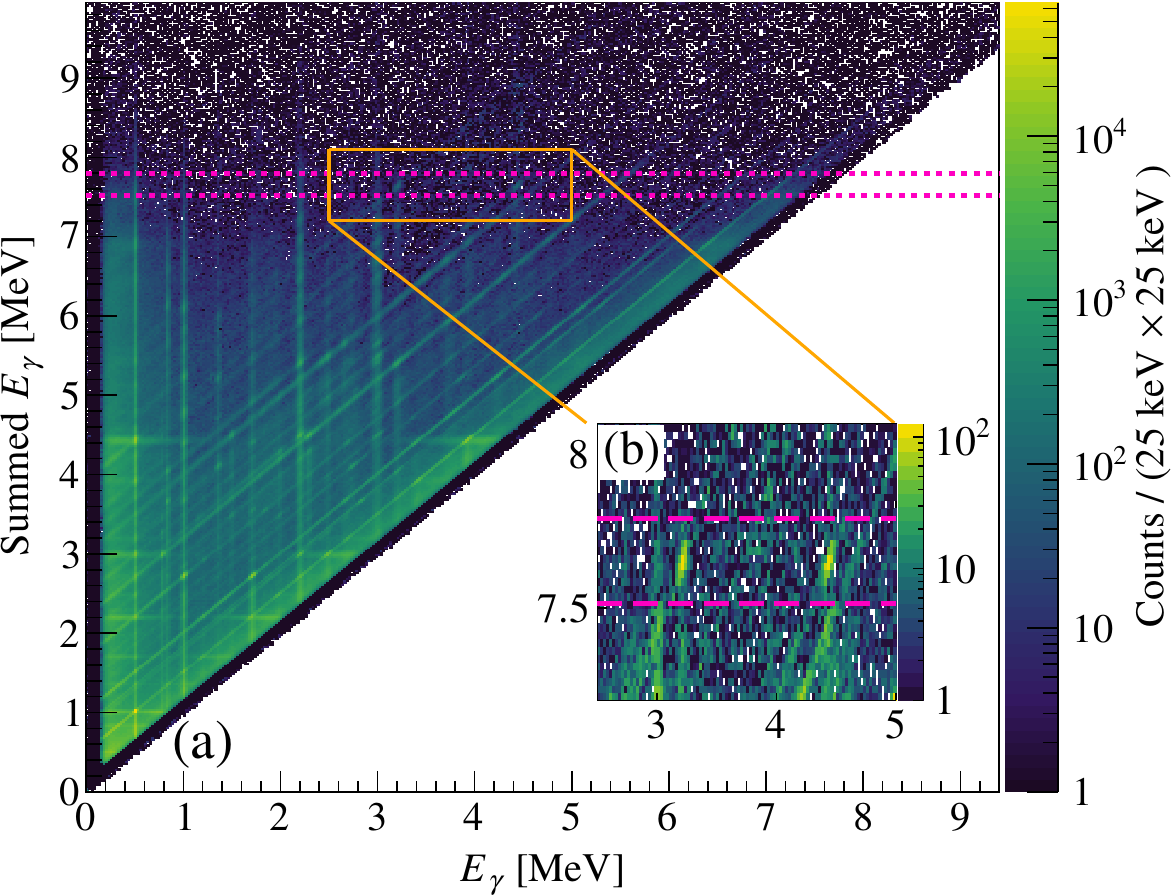}\\
\vspace{4pt}
\includegraphics[width=\columnwidth]{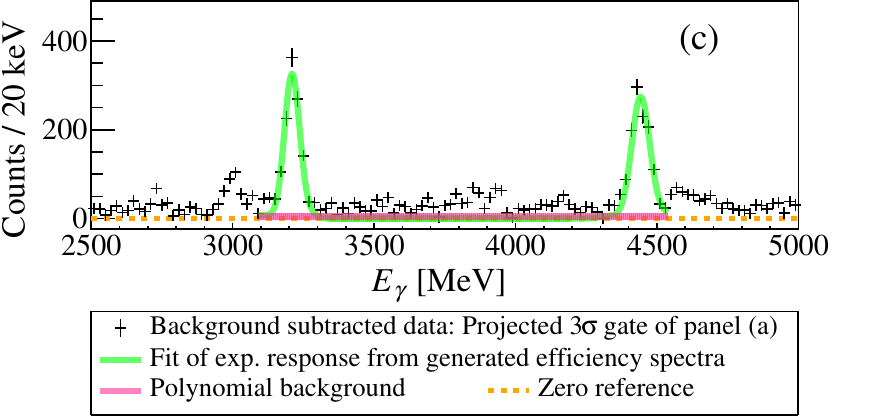}
\caption{\label{fig:summedGamma_2019}
(a) The summed-$\gamma$ matrix background subtracted with respect to time (see Fig.~\ref{fig:labr3_time_matrix_2019}) according to Eq.~\ref{eq:TripleCoincidence_backgroundSubtraction}) from the $^{12}\textrm{C}(p,p')$ measurement performed in 2019.
A zoomed-in region, indicated with solid orange lines, is shown in panel (b).
The horizontal, violet dashed lines correspond to the $3\sigma$ gate on the summed-$\gamma$ energies from the Hoyle state in $^{12}\textrm{C}$, similar to that employed in Ref.~\cite{PhysRevLett.125.182701}.
In panel (c) the background subtracted (according to Eq.~\ref{eq:TripleCoincidence_backgroundSubtraction}) projection of the violet dashed lines in panel (a) is presented.
The triple-coincidence yield extraction is performed on a single spectrum, this is to utilize the extracted non-trivial peak shape from the generated summed $\gamma$-ray detector response in Fig.~\ref{fig:summedGammaGatingCorrectionFactor_12C_2019}.
The fit is done with two parameters, one for the shared amplitude of the peaks and one for a zeroth-order polynomial of negligible size.
}
\end{figure}

To select the time-correlated $\gamma$-ray decays, a timing matrix is employed.
Fig.~\ref{fig:labr3_time_matrix_2019} shows the time difference between detected $\Delta E$ proton ejectiles corresponding to the Hoyle state, and the LaBr$_{3}$(Ce) timing signals corresponding to $E_{\gamma}=3.21$ MeV and $E_{\gamma}=4.44$ MeV.
The separation between the different time loci is $\approx 72$ ns, corresponding to the separation of beam pulses from the cyclotron.
The time locus at $\Delta T_{1} \approx \Delta T_{2} \approx 0$ ns, denoted \mbox{pr-pr} (\mbox{prompt-prompt}), corresponds to \mbox{$p$-$\gamma$--$\gamma$} coincidences where both $\gamma$ rays are detected in prompt coincidence with the proton ejectile corresponding to the Hoyle state.
To select for correlated \mbox{$p$-$\gamma$--$\gamma$} events corresponding to the Hoyle state, a gate was placed around the \mbox{pr-pr} time locus.
However, within this \mbox{pr-pr} time locus, uncorrelated events are also present and these background components can be determined by gating on other timing loci.

The first of these uncorrelated components within the \mbox{$p$-$\gamma$--$\gamma$} events is where only one of the two detected $\gamma$ rays is coincident with the detected ejectile corresponding to the Hoyle state.
To estimate this component, which we denote \mbox{pr-dl} (\mbox{prompt-delayed}), we gate on events where only one $\gamma$ ray is in prompt coincidence, whilst the other is delayed by at least one beam pulse.
The second uncorrelated component, where both $\gamma$ rays are correlated with each other, but are both uncorrelated with the proton (e.g., from a delayed beam pulse), is denoted \mbox{dl-dl} (\mbox{delayed-delayed}).
Finally, the component for the completely random background, where none of the particles are correlated in time, is denoted \mbox{dl-dl'} (\mbox{delayed-delayed'}).
The square (40 ns-wide) gates for these \mbox{pr-dl}, \mbox{dl-dl} and \mbox{dl-dl'} components are presented in Fig.~\ref{fig:labr3_time_matrix_2019}.
In order to increase the statistics for the estimation of these various background yield components, multiple \mbox{pr-dl}, \mbox{dl-dl} and \mbox{dl-dl'} loci are selected up to four beam pulses away from the \mbox{pr-pr} time locus (see Fig.~\ref{fig:labr3_time_matrix_2019}).
These yields are then normalized to the single \mbox{pr-pr} time locus.
The background-subtracted yield for triple-coincidence events corresponding to the Hoyle state is thus determined as 
\begin{equation} \label{eq:TripleCoincidence_backgroundSubtraction}
    N_{\textrm{020}}^{7.65} = N_{\textrm{\mbox{pr-pr}}} - N_{\textrm{\mbox{pr-dl}}} - N_{\textrm{\mbox{dl-dl}}} + N_{\textrm{\mbox{dl-dl'}}},
\end{equation}
where $N_{020}^{7.65}$ is employed in Eq.~\ref{eq: gamma_gamma/gamma}.
The reason why the $N_{\textrm{\mbox{dl-dl'}}}$ is added rather than subtracted, is because this completely random background component is subtracted twice through the $N_{\textrm{\mbox{pr-dl}}}$ and $N_{\textrm{\mbox{dl-dl}}}$ components, which each contain this completely random background.

Fig.~\ref{fig:gammaGamma_2019}(a) presents the $\gamma\textrm{-}\gamma$ matrix gated on the \mbox{pr-pr} time locus.
The violet, dashed horizontal lines correspond to the $3\sigma$ gate employed on the $E_{\gamma} = 4.44$ MeV $\gamma$-ray in the Hoyle-state cascade.
To obtain the background-subtracted triple-coincidence yield ($N_{020}^{7.65}$), the various components in Eq.~\ref{eq:TripleCoincidence_backgroundSubtraction} must be determined.
The corresponding $\gamma$-ray spectra for the \mbox{pr-pr}, \mbox{pr-dl}, \mbox{dl-dl} and \mbox{dl-dl'} components are generated by gating on the $\gamma$--$\gamma$ matrix (Fig.~\ref{fig:gammaGamma_2019}(a)) as well as on the corresponding time loci in Fig.~\ref{fig:labr3_time_matrix_2019}.
To accurately fit the data, the spectra for the \mbox{pr-pr}, \mbox{pr-dl}, \mbox{dl-dl}, \mbox{dl-dl'} and the spectrum employed to extract the efficiency of the $E_{\gamma}=3.21\approx 3.20$ MeV $\gamma$-ray were simultaneously fitted with the mean and experimental width of the peaks being shared parameters.
This simultaneous fit was performed with maximum likelihood estimation due to the low statistics of some of the components.
The independent backgrounds for each spectrum were simultaneously optimized in the fit analysis with a first-order polynomial for the spectra for the \mbox{pr-pr}, \mbox{pr-dl}, \mbox{dl-dl}, \mbox{dl-dl'} and a first-order polynomial and an exponential for the spectrum employed to extract the efficiency of the $E_{\gamma}=3.21\approx 3.20$ MeV $\gamma$-ray.
This technique enables the fit parameters to be better constrained, particularly for the spectra with low statistics.
Fig.~\ref{fig:gammaGamma_2019}(c)--(f) present the fitted histograms corresponding to the \mbox{pr-pr}, \mbox{pr-dl}, \mbox{dl-dl} and \mbox{dl-dl'} components.

An equivalent analysis to obtain the background-subtracted triple-coincidence yield $N_{020}^{7.65}$ was employed by gating on the summed-$\gamma$ matrices, shown in Fig.~\ref{fig:summedGamma_2019}(a).
The gating technique employed here is similar to that in Ref.~\cite{PhysRevLett.125.182701}, however, there are some notable differences.
The horizontal, violet dashed lines correspond to the $3\sigma$ gate on the summed $\gamma$-ray energies, similar to that employed in Ref.~\cite{PhysRevLett.125.182701}.
In Fig.~\ref{fig:summedGamma_2019}(c) the background subtracted (according to Eq.~\ref{eq:TripleCoincidence_backgroundSubtraction}) projection of the violet dashed lines in Fig.~\ref{fig:summedGamma_2019}(a) is presented.
The triple-coincidence yield extraction is performed on a single spectrum, this is to utilize the extracted non-trivial peak shape from the generated summed $\gamma$-ray detector response in Fig.~\ref{fig:summedGammaGatingCorrectionFactor_12C_2019}.
The fit is done with two parameters, one for the shared amplitude of the peaks and one for a zeroth-order polynomial of negligible size.

The results of this analysis is presented in Sec.~\ref{subsec:results_12C_2019}.
\subsection{$\mathbf{^{28}\textrm{Si}(p,p')}$ with $\mathbf{E_{p}=16.0}$ MeV performed in 2020}\label{subsec:Data_Analysis_main_measurement_28Si_2020}
The primary objectives of the analysis of the data from 2020 are to test the validity of the analysis method employed in Sec.~\ref{subsec:Data_Analysis_main_measurement_12C_2019} and to obtain efficiencies for the LaBr$_3$(Ce) detectors which comprise OSCAR.
The $0_{2}^{+}$ state at $E_{\textrm{x}}=4.98$ MeV in $^{28}\textrm{Si}$ has a $\gamma$-decay branching ratio of $\Gamma_{\gamma}/\Gamma = 1$ and emits a $\gamma$ ray with $E_{\gamma}=3.20$ MeV.
This transition is followed by the $E_{\gamma}=1.79$ MeV $\gamma$ ray from the $2_{1}^{+}$ state at $E_{\textrm{x}}=1.79$ MeV to the ground state in $^{28}\textrm{Si}$.
The reason why the $\gamma$-ray cascade from the $0_{2}^{+}$ state in $^{28}\textrm{Si}$ is a good surrogate for that of the Hoyle state is because the spin and parities of the corresponding states are the same.
This results in a nearly identical angular correlation for the cascade.
In addition, the $E_{\gamma}=3.20$ ($0_{2}^{+} \rightarrow 2_{1}^{+}$) transition is almost identical in energy to the $\gamma$ ray emitted from the Hoyle state.
Finally, since $\Gamma_{\gamma}/\Gamma = 1$ for the $0_{2}^{+}$ in $^{28}\textrm{Si}$, the branching ratio for this state can be calculated as a validation of the overall analysis method.
The $\Gamma_\gamma^{E 2}/\Gamma$ branching ratio of the $0_{2}^{+}$ state at $E_{\textrm{x}}=4.98$ MeV in $^{28}\textrm{Si}$ can be expressed as
\begin{equation}\label{eq: gamma_gamma/gamma_28Si_4_98MeV}
    \frac{\Gamma_{\gamma}^{E2}}{\Gamma^{4.98}} = \frac{N_{020}^{4.98}}{N_{\text{inclusive}}^{4.98} \times \epsilon_{1.78} \times \epsilon_{3.20} \times c_{\textrm{det}} \times W_{020}^{4.98}} = 1.0,
\end{equation}
which is analogous to Eq.~\ref{eq: gamma_gamma/gamma}.

\begin{figure}[htbp]
\centering
\includegraphics[width=\columnwidth]{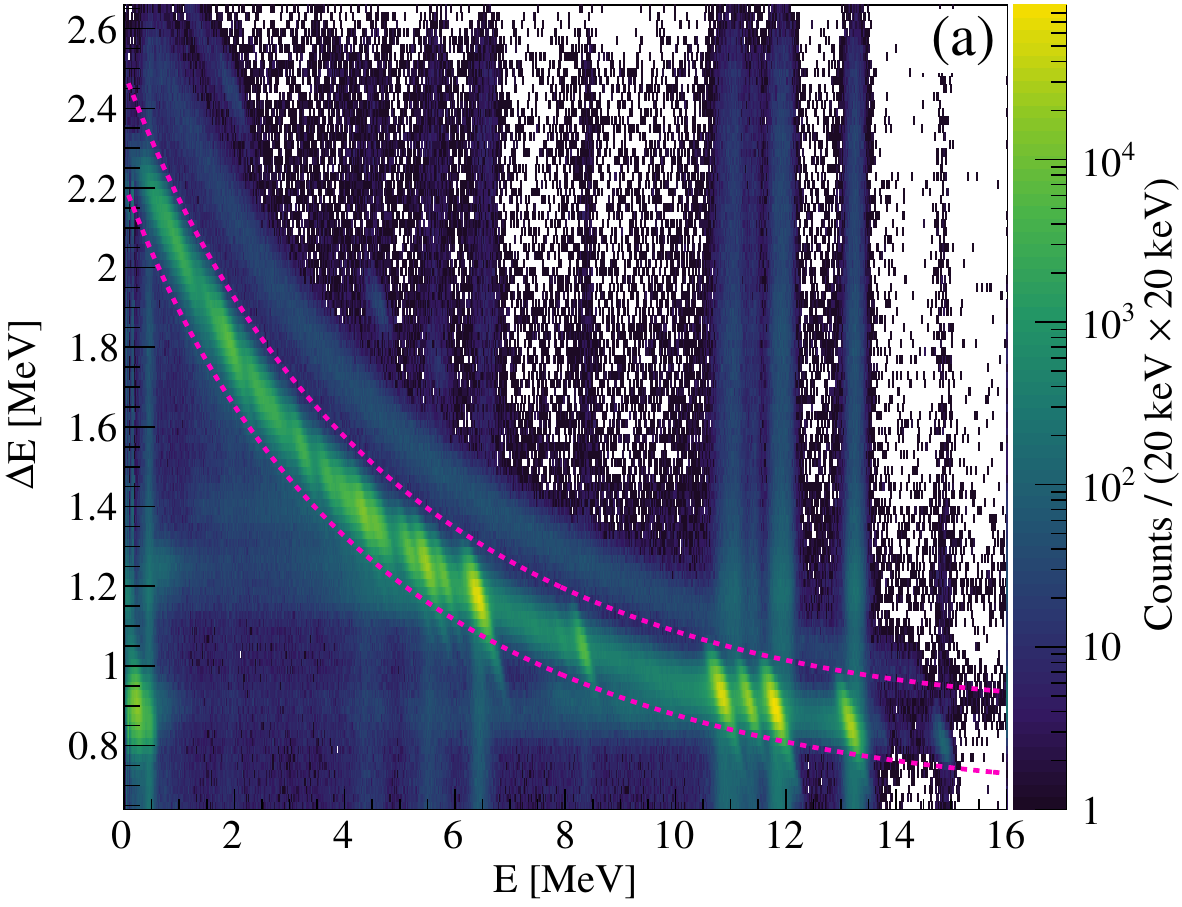}
\includegraphics[width=\columnwidth]{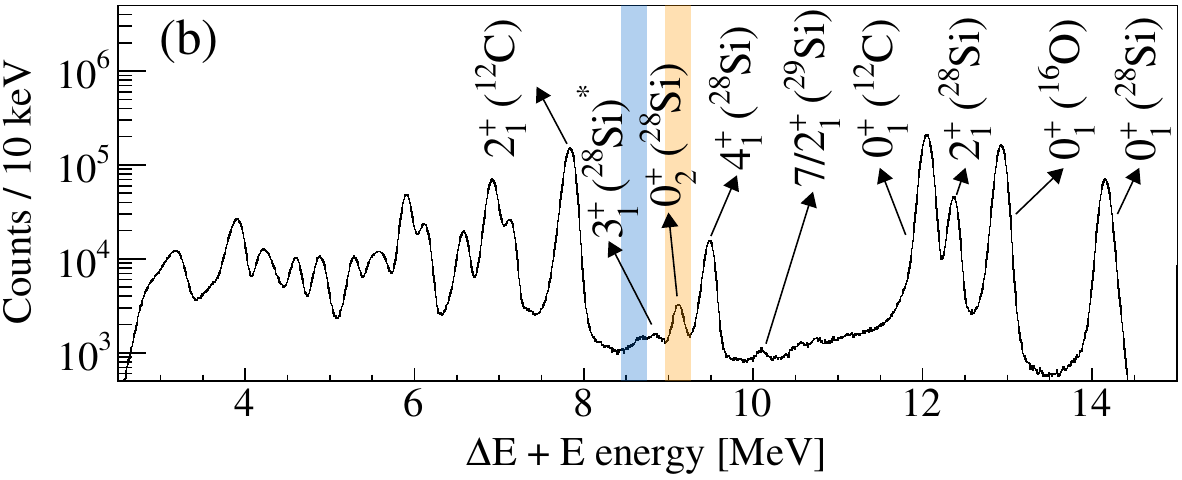}
\caption{\label{fig:InclusiveProtonSpectrum_2020}
(a) $\Delta E$-$E$ energy spectrum of ejectiles, detected at $\theta_{\textrm{lab}} = 128^{\circ}$--$130^{\circ}$ with SiRi from the measurement performed in 2020.
The violet dashed lines denote the energy gate employed for the inelastically scatted protons.
(b) The sum of \mbox{$\Delta E$-$E$} events in the violet gated area in (a). The orange filled area at $E_{p}\approx9$ MeV denotes the energy gate employed for the $0_{2}^{+}$ state in $^{28}\textrm{Si}$. 
The state denoted $3_{1}^{+}(^{28}\textrm{Si})^{*}$ also has $\gamma$-ray transitions consistent with states from $^{29}\textrm{Si}$ and $^{30}\textrm{Si}$.
The blue shaded area denotes the energy gate employed for the background gated analysis, as explained in Appendix~\ref{subsec:investigationOfBackgroundInclusiveSpectra}.
}
\end{figure}

For the $^{28}\textrm{Si}(p,p')$ reaction with a beam energy \mbox{$E_{p}=16.0$} MeV performed in 2020, both layers of SiRi could be used.
This is in contrast to the $^{28}\textrm{Si}(p,p')$ reaction with a beam energy of \mbox{$E_{p}=10.8$} MeV performed in 2019 (see Sec.~\ref{subsec:Data_Analysis_main_measurement_28Si_2019}), where the states of interest in $^{28}\textrm{Si}$ were only accessible through the front layer of SiRi.
The $0_{2}^{+}$ state at $E_{\textrm{x}}=4.98$ MeV was inaccessible in the $^{28}\textrm{Si}(p,p')$ measurement from 2019 due to overlap between the $0_{2}^{+}$ state at $E_{\textrm{x}}=4.98$ MeV and the $4_{1}^{+}$ state at $E_{\textrm{x}}=4.62$ MeV, therefore a $^{28}\textrm{Si}$ measurement from 2020 was used.
The energy depositions in SiRi are shown in Fig.~\ref{fig:InclusiveProtonSpectrum_2020} (a), and the sum of the relevant $\Delta E-E$ energies are shown in Fig.~\ref{fig:InclusiveProtonSpectrum_2020} (b), with the blue colored area denoting the energy gate around the $0_{2}^{+}$ state at $E_{\textrm{x}}=4.98$ MeV in $^{28}\textrm{Si}$.
Due to the kinematic smearing, the data from each ring are fitted separately.

\begin{figure}[htbp]
\centering
\includegraphics[width=\columnwidth]{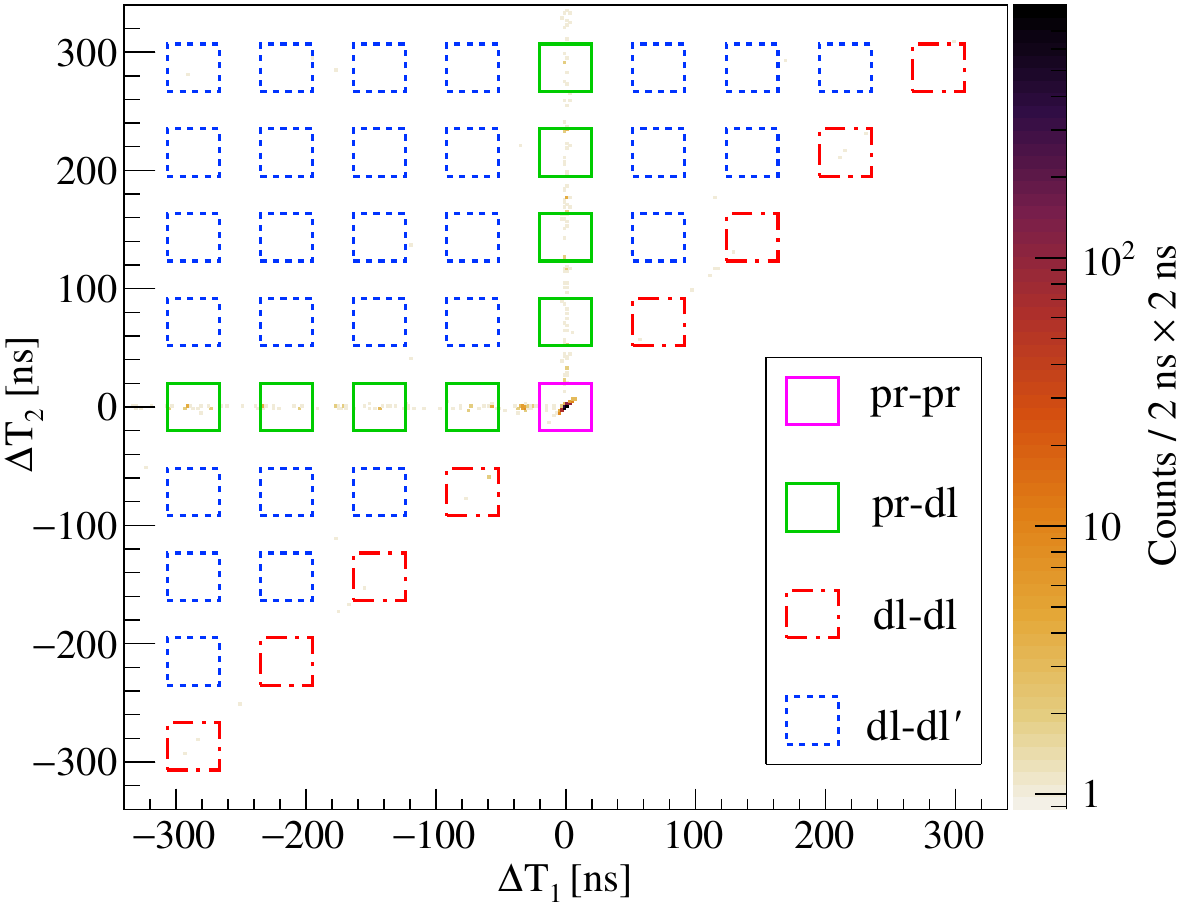}
\caption{\label{fig:labr3_time_matrix_2020}
Time differences between protons populating the $0_{2}^{+}$ state at $E_{\textrm{x}}=4.98$ MeV in $^{28}\textrm{Si}$ and two $\gamma$ rays with $E_{\gamma}=3.20$ MeV and $E_{\gamma}=1.79$ MeV, from the experiment performed in 2020.
The matrix is sorted so the largest value of the two timing values is along the y-axis.
The colored boxes denote the different background gates used for the analysis.
}
\end{figure}

Fig.~\ref{fig:labr3_time_matrix_2020} shows the time difference between detected proton ejectiles corresponding to the Hoyle state, and the LaBr$_{3}$(Ce) timing signals corresponding to the $\gamma$-ray cascade from the $0_{2}^{+}$ state at $E_{\textrm{x}}=4.98$ MeV.
As the total statistics is much lower in this experiment compared to the other experiments (2019 and 2014, see Sec.~\ref{subsec:Data_Analysis_main_measurement_12C_2019} and Sec.~\ref{subsec:Data_Analysis_reanalysis_12C_2014}) combined with the much cleaner signal, the total yield and spread of events in the timing matrix are also lower.

\begin{figure}[htbp]
\centering
\includegraphics[width=\columnwidth]{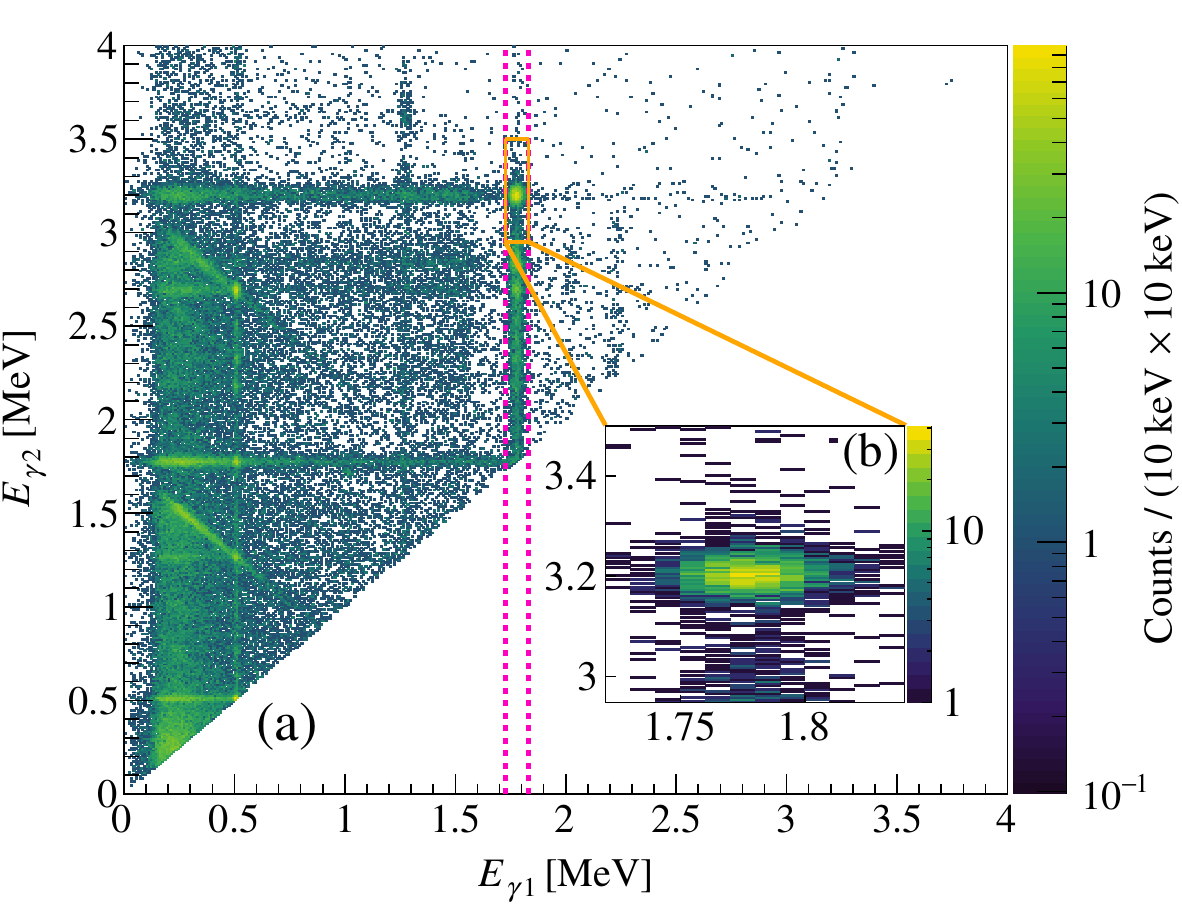}\\
\vspace{2pt}
\includegraphics[width=\columnwidth]{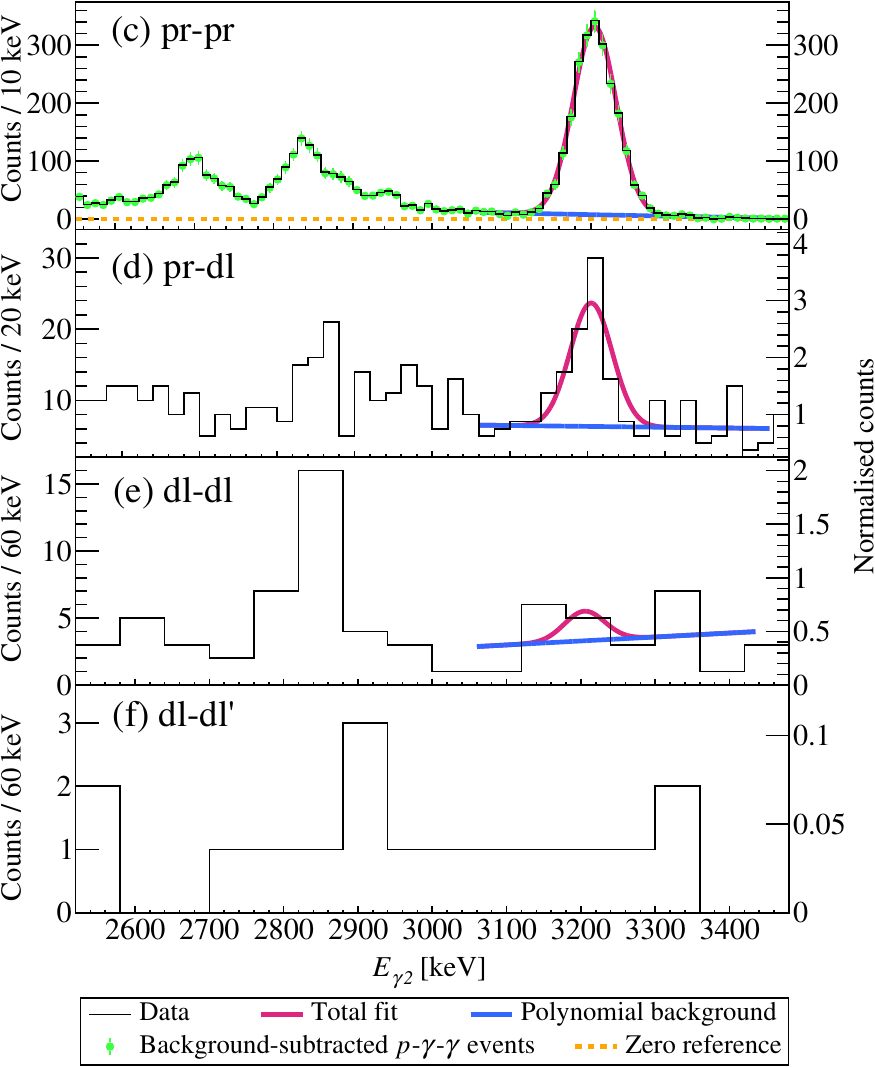}
\caption{\label{fig:gammaGamma_2020}
(a) The $\gamma$--$\gamma$ matrix (largest energy on the y-axis) gated on the \mbox{pr-pr} time locus (see Fig.~\ref{fig:labr3_time_matrix_2020}) from the $^{28}\textrm{Si}(p,p')$ measurement performed in 2020.
A zoomed-in region, indicated with solid orange lines, is shown in panel (b).
The vertical, violet dashed lines correspond to the $3\sigma$ gate on the $E_{\gamma}=1.78$ MeV $\gamma$ ray from the $0_{2}^{+}$ state at $E_{x}=4.98$ MeV.
The events of interest are those within the violet dashed region, with the projected spectra shown in panels (c)--(f) for the different time loci \mbox{pr-pr}, \mbox{pr-dl}, \mbox{dl-dl} and \mbox{dl-dl'} (see Fig.~\ref{fig:labr3_time_matrix_2020} for details on time loci), respectively.
The y-axis on the right side shows the normalized counts, accounting for the number of time loci for each spectrum.
The solid violet line shows the simultaneous peak fit in panels (c)--(e), where the peak shape has been constrained by the data in panel (c).
Individual first-order polynomials are fitted in each panel (c)--(e).
For panel (f) data from time locus \mbox{dl-dl'}, an integral over the $3\sigma$ of the $E_{\gamma}=3.2$ MeV peak contributes to the uncertainty in the final triple-coincidence yield.
This simultaneous fit was performed with maximum likelihood estimation due to the very low statistics of the various background components.
}
\end{figure}

In Fig.~\ref{fig:gammaGamma_2020}(a), the $\gamma$--$\gamma$ matrix gated on the \mbox{pr-pr} time locus is shown. 
The violet, dashed vertical lines correspond to the $3\sigma$ gate employed on the $E_{\gamma}=1.78$ MeV $\gamma$-ray in the $0_{2}^{+}$ cascade at $E_{\textrm{x}}=4.98$ MeV in $^{28}\textrm{Si}$, with a zoomed-in region focused on the $E_{\gamma}=1.78$ MeV and $E_{\gamma}=3.20$ coincidence locus shown in Fig.~\ref{fig:gammaGamma_2020}(b).
For the $\gamma$-ray spectra corresponding to these gated $\gamma$--$\gamma$ matrices (for the various time loci \mbox{pr-pr}, \mbox{pr-dl}, \mbox{dl-dl} and \mbox{dl-dl'}), the simultaneous fit is shown in Fig.~\ref{fig:gammaGamma_2020} (c)--(f).
To accurately fit the data, the spectra for the \mbox{pr-pr}, \mbox{pr-dl}, \mbox{dl-dl}, \mbox{dl-dl'} and the spectrum employed to extract the efficiency of the $E_{\gamma}=3.20$ MeV $\gamma$-ray were simultaneously fitted with the mean and experimental width of the peaks being shared parameters.
This simultaneous fit was performed with maximum likelihood estimation due to the very low statistics of some of the components.
The independent backgrounds for each spectrum were simultaneously optimized in the fit analysis with a first-order polynomial for the spectra for the \mbox{pr-pr}, \mbox{pr-dl}, \mbox{dl-dl}, \mbox{dl-dl'} and a first-order polynomial and an exponential for the spectrum employed to extract the efficiency of the $E_{\gamma}=3.20$ MeV $\gamma$-ray.

\begin{figure}[htbp]
\centering
\includegraphics[width=\columnwidth]{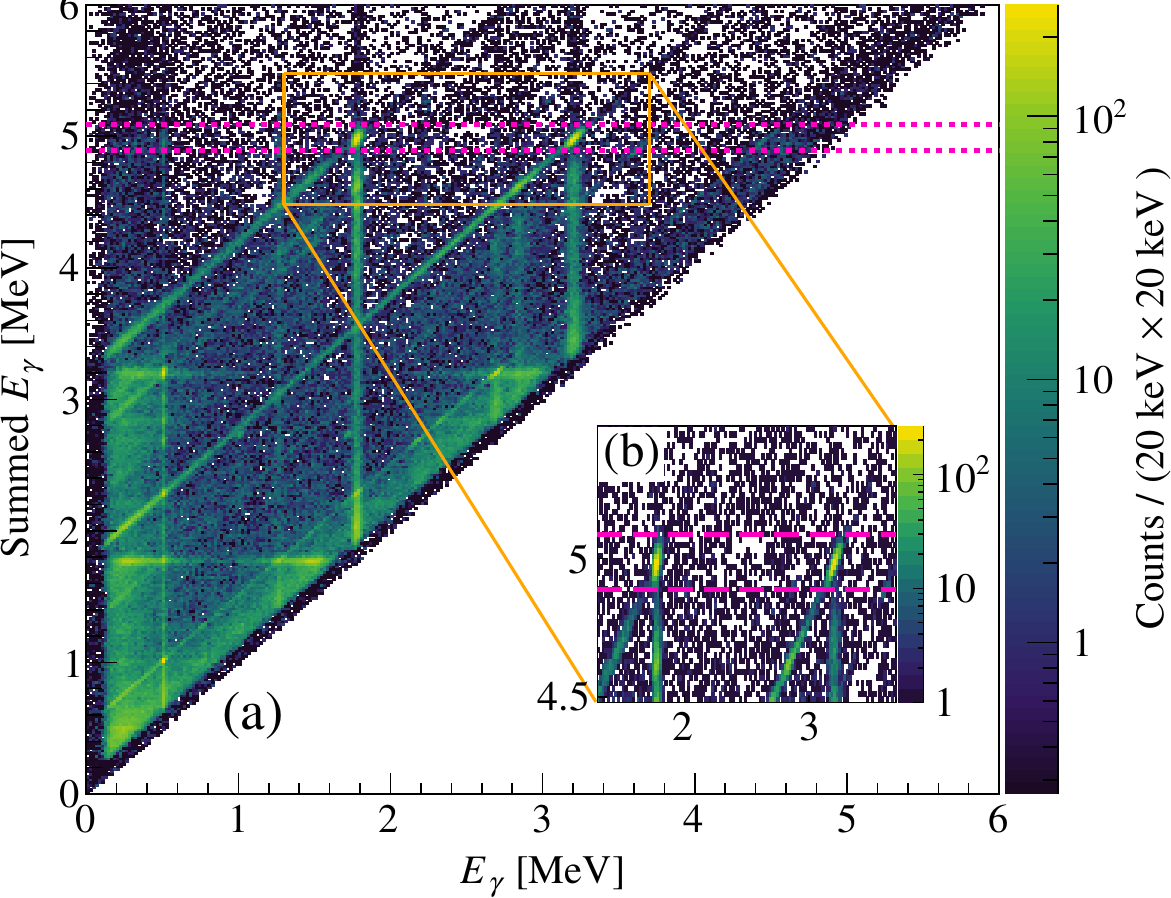}\\
\vspace{4pt}
\includegraphics[width=\columnwidth]{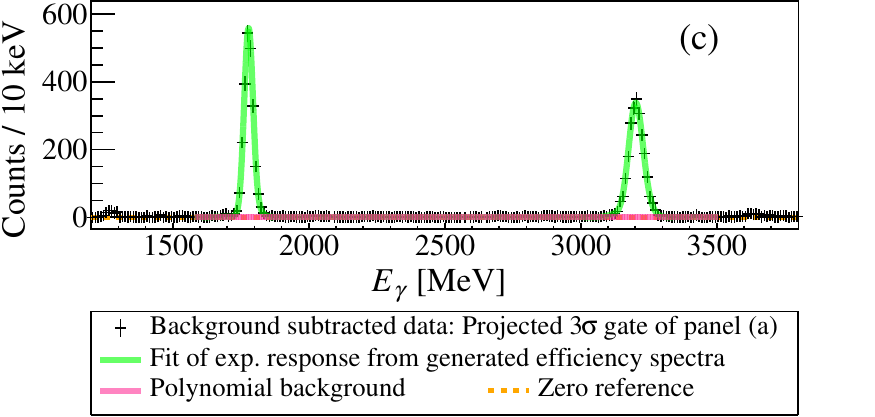}
\caption{\label{fig:summedGamma_2020}
(a) The summed-$\gamma$ matrix background subtracted with respect to time (see Fig.~\ref{fig:labr3_time_matrix_2020}) according to Eq.~\ref{eq:TripleCoincidence_backgroundSubtraction}) from the $^{28}\textrm{Si}(p,p')$ measurement performed in 2020.
A zoomed-in region, indicated with solid orange lines, is shown in panel (b).
The horizontal, violet dashed lines correspond to the $3\sigma$ gate on the summed-$\gamma$ energies from the $0_{2}^{+}$ state at $E_{x}=4.98$ MeV in $^{28}\textrm{Si}$, similar to that employed in Ref.~\cite{PhysRevLett.125.182701}.
In panel (c) the background subtracted (according to Eq.~\ref{eq:TripleCoincidence_backgroundSubtraction}) projection of the violet dashed lines in panel (a) is presented.
The triple-coincidence yield extraction is performed on a single spectrum, this is to utilize the extracted non-trivial peak shape from the generated summed $\gamma$-ray detector response in Fig.~\ref{fig:summedGammaGatingCorrectionFactor_28Si_2020}.
The fit is done with two parameters, one for the shared amplitude of the peaks and one for a zeroth-order polynomial of negligible size.
}
\end{figure}

The summed-$\gamma$ matrix is presented in Fig.~\ref{fig:summedGamma_2020} (a)-(c), background subtracted according to Eq.~\ref{eq:TripleCoincidence_backgroundSubtraction}.
In panel (c) the projection of the $3\sigma$ gate on the $E_{\gamma}=4.98$ MeV summed $\gamma$-ray, indicated by the violet lines in panel (a), is shown together with the fit obtained from the $E_{\gamma}=4.98$ MeV generated summed $\gamma$-ray efficiency (see Fig.~\ref{fig:summedGammaGatingCorrectionFactor_28Si_2020}).
The results of this analysis is presented in Sec.~\ref{subsec:results_28Si_2020}.
\subsection{$\mathbf{^{28}\textrm{Si}(p,p')}$ with $\mathbf{E_{p}=10.8}$ MeV performed in 2019}\label{subsec:Data_Analysis_main_measurement_28Si_2019}
The $\gamma$-decay branching ratio of the Hoyle state as measured by Obst \textit{et al.}~\cite{PhysRevC.13.2033} was deduced by incorporating the $\gamma$-decay branching ratio of the $0_{2}^{+}$ state in $^{28}\textrm{Si}$ at $E_{\textrm{x}}=4.98$ MeV (see Sec.~\ref{subsec:Data_Analysis_main_measurement_28Si_2020}) and the ratio between triple-coincidences and inclusive particles populating the higher lying $3_{1}^{+}$ state at $E_{\textrm{x}}=6.28$ MeV in $^{28}\textrm{Si}$.
Obst \textit{et al.} state that the $3_{1}^{+}$ state at $E_{\textrm{x}}=6.28$ MeV in $^{28}\textrm{Si}$ was used in the analysis since the $0_{2}^{+}$ state in $^{28}\textrm{Si}$ at $E_{\textrm{x}}=4.98$ MeV was unresolved in their particle spectrum.
As such, the $3_{1}^{+}$ state at $E_{\textrm{x}}=6.28$ MeV was used as a common monitor transition.
Eq.~14 of Obst \textit{et al.}~\cite{PhysRevC.13.2033} presents the $\gamma$-decay branching ratio of the Hoyle state as five ratios (A-E), where these quantities from $^{28}\textrm{Si}$ are incorporated such that the $\gamma$-decay branching ratio of the Hoyle state becomes
\begin{equation}\label{eq:ObstEquationGDBR}
\begin{split}
\frac{\Gamma_{\gamma}^{E2}}{\Gamma^{7.65}} &= \frac{N_{020}^{7.65}}{N_{320}^{6.28}} \times \frac{N_{320}^{6.28}}{N_{020}^{4.98}} \times \frac{N_{\text{inclusive}}^{6.28}}{N_{\text{inclusive}}^{7.65}} \times \frac{N_{\text{inclusive}}^{4.98}}{N_{\text{inclusive}}^{6.28}} \times \frac{\epsilon_{1.78}}{\epsilon_{4.44}} \\
&= ABCDE,
\end{split}
\end{equation}
where the terms $N_{020}^{4.98}$, $N_{320}^{6.28}$ and $N_{020}^{7.65}$ are the number of triple coincidences and the terms $N_{\text{inclusive}}^{4.98}$, $N_{\text{inclusive}}^{6.28}$ and $N_{\text{inclusive}}^{7.65}$ are the inclusive yield from the states of interest in $^{12}\textrm{C}$ and $^{28}\textrm{Si}$.
The terms $\epsilon_{1.78}$ and $\epsilon_{4.44}$ are the absolute photopeak efficiencies at $E_{\gamma}=1.78$ MeV and $E_{\gamma}=4.44$ MeV.
The missing term \mbox{$(\epsilon_{3.20}/\epsilon_{3.21})\times (W_{020}^{4.98}/W_{020}^{7.65})=0.99(5)$} in Eq.~\ref{eq:ObstEquationGDBR} (Eq.~14 in Ref.~\cite{PhysRevC.13.2033}) was separated into Eq.~13 in Ref.~\cite{PhysRevC.13.2033}.
In the work of Kib\'edi \textit{et al.}~\cite{PhysRevLett.125.182701} two of these ratios (B and D) were measured with the result being significantly different from Obst \textit{et al.}~\cite{PhysRevC.13.2033}.
Kib\'edi \textit{et al.}~\cite{PhysRevLett.125.182701} also state that ``Despite some differences between their experiment and ours, various combinations of these ratios ($A$-$E$) should agree within a few percent''.
The combination of the ratios $B\times D$, henceforth dubbed the ``Obst ratio'', can be expressed as:
\begin{equation}\label{eq:ObstRatio}
    B\times D = \frac{N_{320}^{6.28}}{N_{inc}^{6.28}} \times \frac{N_{inc}^{4.98}}{N_{020}^{4.98}},
\end{equation}
where $N_{320}^{6.28}$ and $N_{inc}^{6.28}$ are the respective triple-coincidence and inclusive yields for the $3_{1}^{+}$ state at $E_{\textrm{x}}=6.28$ MeV in $^{28}\textrm{Si}$, and $N_{020}^{4.98}$ and $N_{inc}^{4.98}$ are the respective triple-coincidence and inclusive yields for the $0_{2}^{+}$ state at $E_{\textrm{x}}=4.98$ MeV in $^{28}\textrm{Si}$ (see Sec.~\ref{subsec:Data_Analysis_main_measurement_28Si_2020}).

An analysis of the Obst ratio was performed in this work solely to test the validity of comparing the results of the Obst ratio between Kib\'edi \textit{et al.}~\cite{PhysRevLett.125.182701} and Obst \textit{et al.}~\cite{PhysRevC.13.2033}, as was originally conducted in Ref.~\cite{PhysRevLett.125.182701}.
The $\gamma$-decay branching ratio of the $3_{1}^{+}$ state at $E_{\textrm{x}}=6.28$ MeV in $^{28}\textrm{Si}$ was measured, with the corresponding $\Gamma_\gamma^{E 2}/\Gamma$ branching ratio determined as
\begin{equation}\label{eq: gamma_gamma/gamma_28Si_6_28MeV}
    \frac{\Gamma_{\gamma}^{E2/M1}}{\Gamma^{6.28}} = \frac{N_{320}^{6.28}}{N_{\text{inclusive}}^{6.28} \times \epsilon_{1.78} \times \epsilon_{4.49} \times c_{\textrm{det}} \times W_{320}^{6.28}},
\end{equation}
which is analogous to Eq.~\ref{eq: gamma_gamma/gamma}.
Substituting the terms in Eq.~\ref{eq:ObstRatio} with the equivalent terms in Eq.~\ref{eq: gamma_gamma/gamma_28Si_4_98MeV} and Eq.~\ref{eq: gamma_gamma/gamma_28Si_6_28MeV}, the Obst ratio can be expressed in terms of the efficiencies and angular correlation correction terms as
\begin{equation}\label{eq:ObstRatio_W}
    B\times D = \frac{\epsilon_{4.49}}{\epsilon_{3.20}} \times \frac{W_{320}^{6.28}}{W_{020}^{4.98}} \times \frac{\Gamma_{\gamma}^{E2}/\Gamma^{6.28}}{\Gamma_{\gamma}^{E2}/\Gamma^{4.98}}.
\end{equation}
From this equation one can see that the Obst ratio is highly dependent on the difference between efficiency and angular correlation correction factors in the experimental setup.

For the $^{28}\textrm{Si}(p,p')$ reaction with a beam energy of \mbox{$E_{p}=10.8$} performed in 2019, only the front layer of SiRi could be used to access the $3_{1}^{+}$ state at $E_{\textrm{x}}=6.28$ MeV.
Due to kinematic smearing across each subtended angle, the data from each individual $\Delta E$ detector were fitted separately (see Sec.~\ref{subsec:Data_Analysis_main_measurement_12C_2019} for a more in-depth explanation).
The extraction of the triple coincidences was performed using an identical analysis as in Secs.~\ref{subsec:Data_Analysis_main_measurement_12C_2019} and \ref{subsec:Data_Analysis_main_measurement_28Si_2020}.
\section{Data Analysis: Reanalysis of previous work}\label{sec:Data_Analysis_reanalysis}
The same analysis method conducted on the data obtained from the 2019 and 2020 experiments (see Sec.~\ref{sec:Data_Analysis_main_measurement}) was performed on the data obtained in 2014 (published in Ref.~\cite{PhysRevLett.125.182701}) and data obtained in 2012 (see Table~\ref{tab:Experimental_apparatus_CACTUS} for an overview), with the goal of independently verifying aspects of the analysis performed in Ref.~\cite{PhysRevLett.125.182701}.
\subsection{$\mathbf{^{12}\textrm{C}(p,p')}$ with $\mathbf{E_{p}=10.7}$ MeV performed in 2014}\label{subsec:Data_Analysis_reanalysis_12C_2014}
\begin{figure}[htbp]
\centering
\includegraphics[width=\columnwidth]{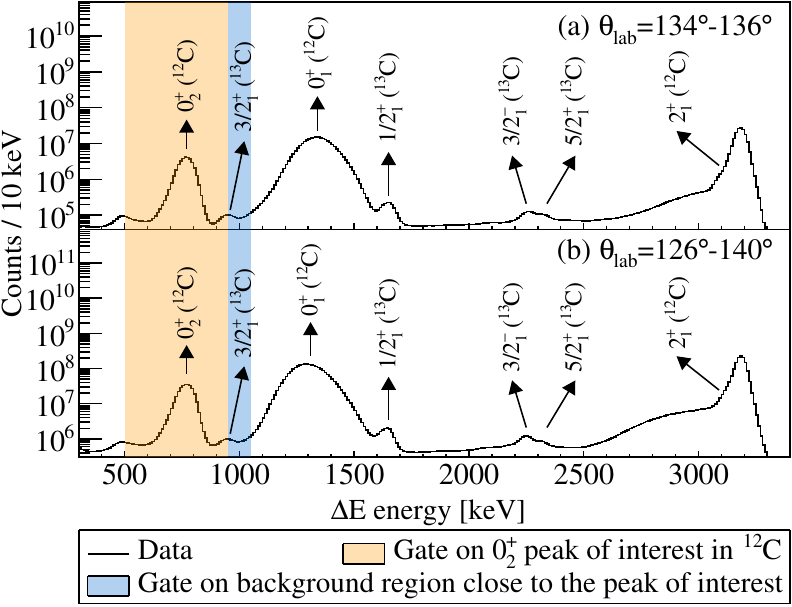}
\caption{\label{fig:InclusiveProtonSpectrum_2014}
(a) $\Delta E$ energy spectrum of inelastically scattered proton ejectiles, detected at $\theta_{\textrm{lab}} = 134^{\circ}$--$136^{\circ}$ with SiRi from the measurement performed in 2014.
The orange shaded area denotes the energy gate employed for the $0_{2}^{+}$ Hoyle state.
The blue shaded area denotes the energy gate employed for the background gated analysis, as explained in Appendix~\ref{subsec:investigationOfBackgroundInclusiveSpectra}.
The total projection of all angles of SiRi ($\theta_{\textrm{lab}} = 126^{\circ}$--$140^{\circ}$) is shown in panel (b).
The energy calibration for this data was conducted under the assumption of no dead layer in the particle detector (see text for details).
}
\end{figure}

\begin{figure}[htbp]
\centering
\includegraphics[width=\columnwidth]{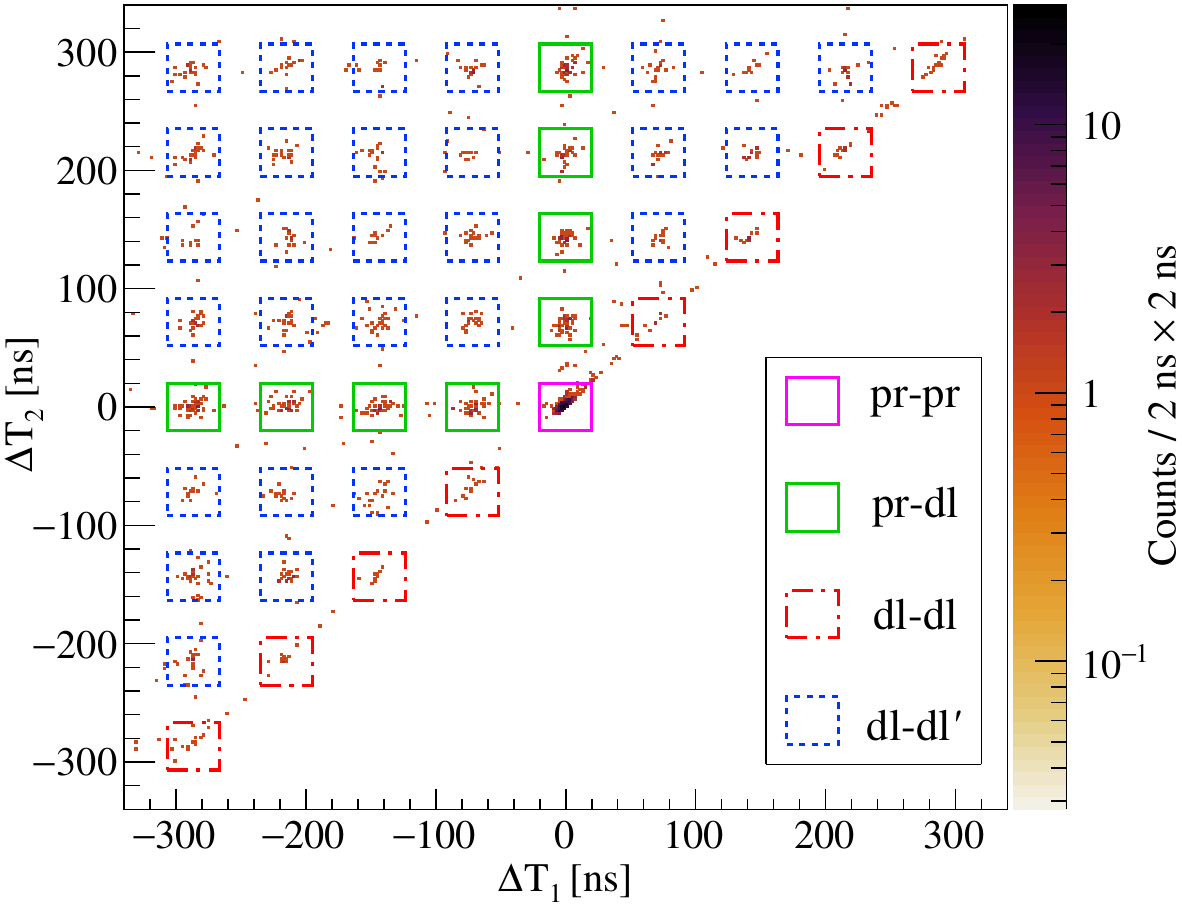}
\caption{\label{fig:NaI_time_matrix_2014}
Time differences between protons populating the Hoyle state and two $\gamma$ rays with $E_{\gamma1}=3.21$ MeV and $E_{\gamma2}=4.44$ MeV, from the experiment performed in 2014.
The matrix is sorted so the largest of the two timing values is along the y-axis.
The colored boxes denote the different background gates used for the analysis. See text for explanation of coincidence events.}
\end{figure}

Similarly to the experiment performed in 2019 (which had a beam energy of $E_{p}=10.8$ MeV, see Sec.~\ref{subsec:Data_Analysis_main_measurement_12C_2019}), the beam energy of $E_{p}=10.7$ MeV for this 2014 experiment resulted in the ejectile protons from the Hoyle state being stopped in the $\Delta$E layer of SiRi.
The same effect on the ordering of the $0_{1}^{+}$, $2_{1}^{+}$ and $0_{2}^{+}$ peaks (to not follow the corresponding excitation energy) was therefore also present.
The energy calibration was performed under the assumption of no dead layer, as for the 2019 measurement (see Sec.~\ref{subsec:Data_Analysis_main_measurement_12C_2019}).
The energy depositions from scattered protons in the $\Delta$E layer of SiRi are shown in Fig.~\ref{fig:InclusiveProtonSpectrum_2014}.
The orange shaded area shows the gate employed to obtain the triple-coincidence yield from the Hoyle state.
The blue shaded area shows the gate employed for the background-gated analysis, as explained in Appendix~\ref{subsec:results_28Si_2012}.
Fig.~\ref{fig:NaI_time_matrix_2014} presents the time difference between the detected ($\Delta$E) proton ejectile corresponding to the $0_{2}^{+}$ Hoyle state, and the NaI(Tl) timing signals corresponding to $E_{\gamma}=3.21$ MeV and $E_{\gamma}=4.44$ MeV.

\begin{figure}[htbp]
\centering
\includegraphics[width=\columnwidth]{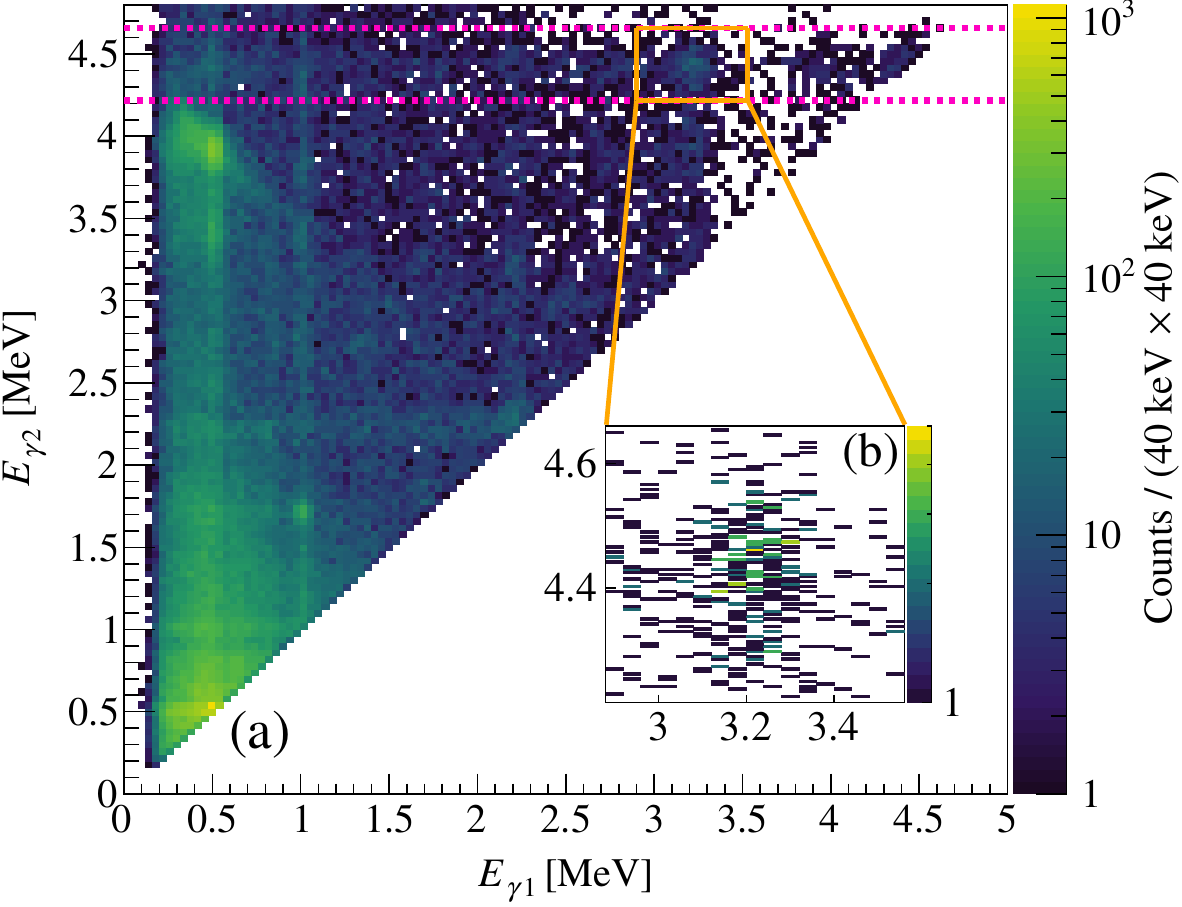}\\
\vspace{4pt}
\includegraphics[width=\columnwidth]{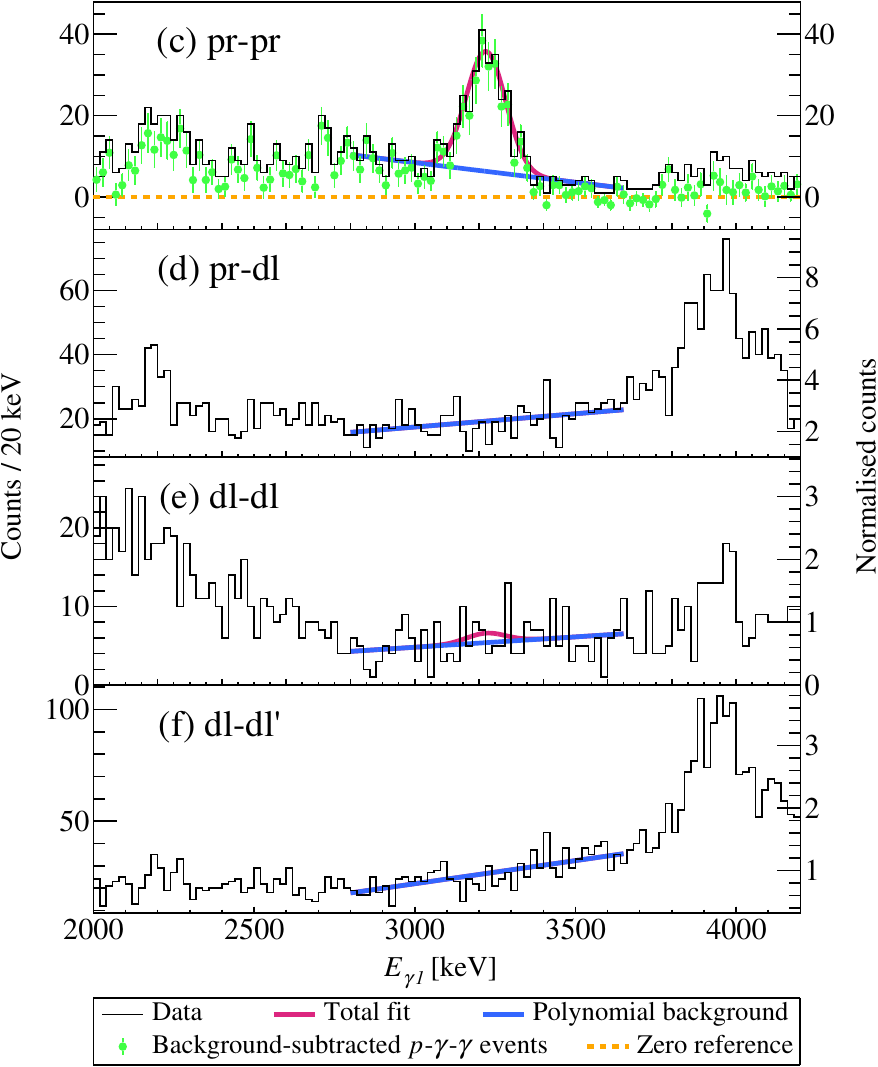}
\caption{\label{fig:gammaGamma_2014}
(a) The $\gamma$--$\gamma$ matrix (largest energy on the y-axis) gated on the \mbox{pr-pr} time locus (see Fig.~\ref{fig:NaI_time_matrix_2014}) from the $^{12}\textrm{C}(p,p')$ measurement performed in 2014.
A zoomed-in region, indicated with solid orange lines, is shown in panel (b).
The horizontal, violet dashed lines correspond to the $3\sigma$ gate on the $E_{\gamma}=4.44$ MeV $\gamma$ ray from the Hoyle state cascade.
The events of interest are those within the violet dashed region, with the projected spectra shown in panels (c)--(f) for the different time loci \mbox{pr-pr}, \mbox{pr-dl}, \mbox{dl-dl} and \mbox{dl-dl'} (see Fig.~\ref{fig:NaI_time_matrix_2014} for details on time loci), respectively.
The y-axis on the right side shows the normalized counts, accounting for the number of time loci for each spectrum.
The solid violet line shows the simultaneous peak fit in panels (c)--(f), where the peak shape has been constrained by the data in panel (c).
Individual first-order polynomials are fitted in each panel (c)--(f).
This simultaneous fit was performed with maximum likelihood estimation due to the very low statistics of the various background components.
}
\end{figure}

In Fig.~\ref{fig:gammaGamma_2014}(a) the $\gamma$--$\gamma$ matrix gated on the \mbox{pr-pr} time locus is shown. The violet, dashed horizontal lines correspond to the $3\sigma$ gate employed on the $E_{\gamma}=4.44$ MeV $\gamma$-ray in the $0_{2}^{+}$ Hoyle state cascade, with a zoomed-in region focused on the $E_{\gamma}=3.21$ MeV and $E_{\gamma}=4.44$ MeV coincidence locus shown in Fig.~\ref{fig:gammaGamma_2014}(b).
For the $\gamma$-ray spectra corresponding to these gated $\gamma$--$\gamma$ matrices (for the various time loci \mbox{pr-pr}, \mbox{pr-dl}, \mbox{dl-dl} and \mbox{dl-dl'}), the simultaneous fit is shown in Fig.~\ref{fig:gammaGamma_2014}(c)--(f).
To accurately fit the data, the spectra for the \mbox{pr-pr}, \mbox{pr-dl}, \mbox{dl-dl}, \mbox{dl-dl'} and the spectrum employed to extract the efficiency of the $E_{\gamma}=3.21\approx 3.20$ MeV $\gamma$-ray were simultaneously fitted with the mean and experimental width of the peaks being shared parameters.
This simultaneous fit was performed with maximum likelihood estimation due to the very low statistics of some of the components.
The independent backgrounds for each spectrum were simultaneously optimized in the fit analysis with a first-order polynomial for the spectra for the \mbox{pr-pr}, \mbox{pr-dl}, \mbox{dl-dl}, \mbox{dl-dl'} and a first-order polynomial and an exponential for the spectrum employed to extract the efficiency of the $E_{\gamma}=3.20$ MeV $\gamma$-ray.

\begin{figure}[htbp]
\centering
\includegraphics[width=\columnwidth]{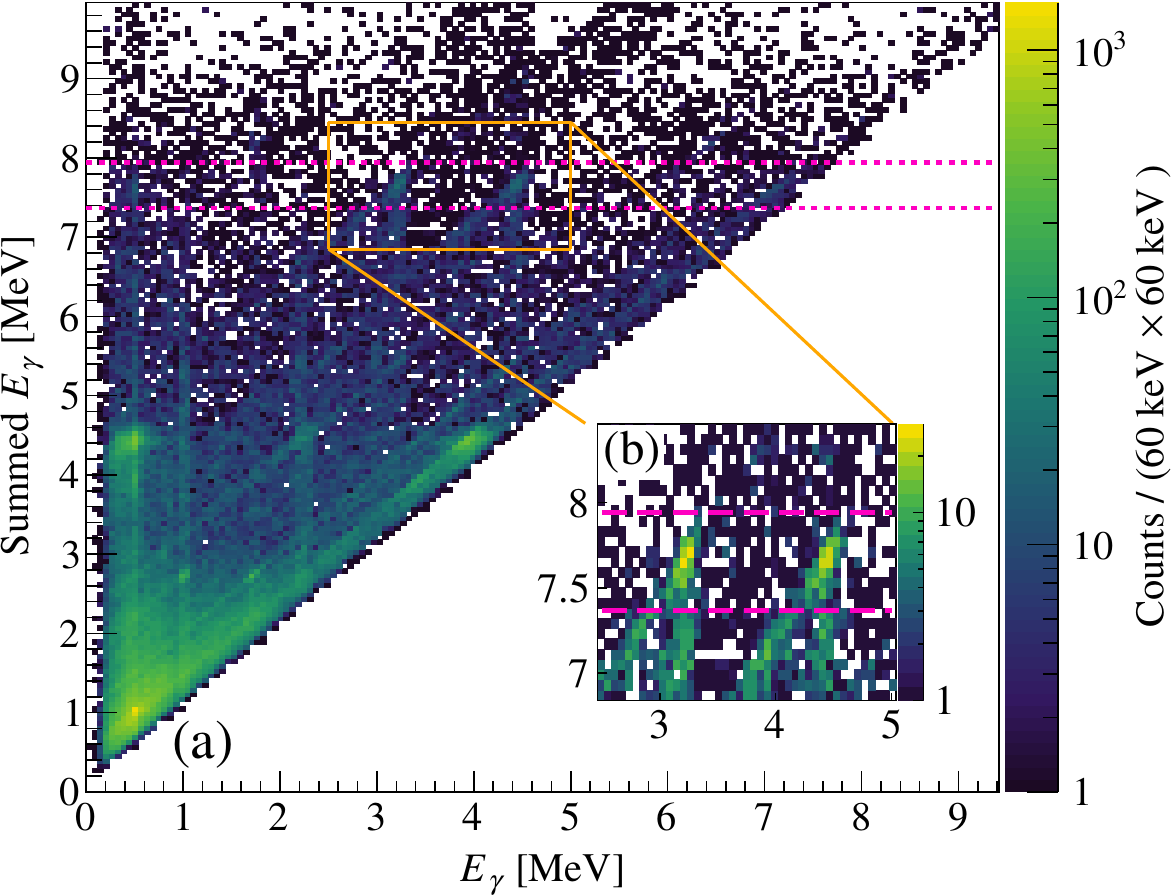}\\
\vspace{4pt}
\includegraphics[width=\columnwidth]{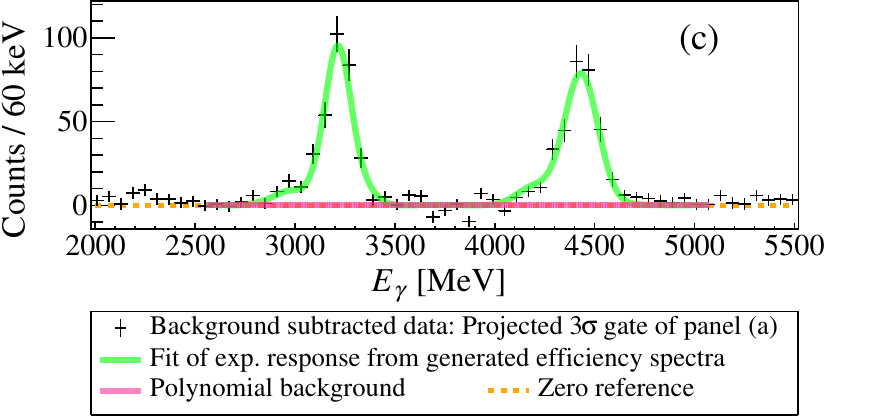}
\caption{\label{fig:summedGamma_2014}
(a) The summed-$\gamma$ matrix background subtracted with respect to time (see Fig.~\ref{fig:NaI_time_matrix_2014}) according to Eq.~\ref{eq:TripleCoincidence_backgroundSubtraction}) from the $^{12}\textrm{C}(p,p')$ measurement performed in 2014.
A zoomed-in region, indicated with solid orange lines, is shown in panel (b).
The horizontal, violet dashed lines correspond to the $3\sigma$ gate on the summed-$\gamma$ energies from the Hoyle state in $^{12}\textrm{C}$, similar to that employed in Ref.~\cite{PhysRevLett.125.182701}.
In panel (c) the background subtracted (according to Eq.~\ref{eq:TripleCoincidence_backgroundSubtraction}) projection of the violet dashed lines in panel (a) is presented.
The triple-coincidence yield extraction is performed on a single spectrum, this is to utilize the extracted non-trivial peak shape from the generated summed $\gamma$-ray detector response in Fig.~\ref{fig:summedGammaGatingCorrectionFactor_12C_2014}.
The fit is done with two parameters, one for the shared amplitude of the peaks and one for a zeroth-order polynomial of negligible size.
}
\end{figure}

The summed-$\gamma$ matrix is presented in Fig.~\ref{fig:summedGamma_2014} (a)-(c), background subtracted according to Eq.~\ref{eq:TripleCoincidence_backgroundSubtraction}.
In panel (c) the projection of the $3\sigma$ gate on the Hoyle state, indicated by the violet lines in panel (a), is shown together with the fit obtained from the $E_{\gamma}=7.65$ MeV generated summed $\gamma$-ray efficiency (see Fig.~\ref{fig:summedGammaGatingCorrectionFactor_12C_2014}).
The non-trivial peak-shape seen in panel (c) originating from performing a gate on the summed $\gamma$-rays from the Hoyle state is easily observable in this measurement.
The results of this analysis is presented in Sec.~\ref{subsec:results_12C_2014}.
\subsection{$\mathbf{^{28}\textrm{Si}(p,p')}$ with $\mathbf{E_{p}=10.7}$ MeV performed in 2014}\label{subsec:Data_Analysis_reanalysis_28Si_2014}
Similarly to the other $^{28}\textrm{Si}$ measurements presented in Secs.~\ref{subsec:Data_Analysis_main_measurement_28Si_2019} and \ref{subsec:Data_Analysis_main_measurement_28Si_2020}, this $^{28}\textrm{Si}(p,p')$ reaction with a beam energy of $E_{p}=10.7$ MeV from 2014 was analyzed with the intention of validating the analysis method and obtaining the Obst ratio (see Secs.~\ref{subsec:Data_Analysis_main_measurement_28Si_2020} and \ref{subsec:Data_Analysis_main_measurement_28Si_2019}, respectively).
In this measurement the $0_{2}^{+}$ state at $E_{\textrm{x}}=4.98$ MeV in $^{28}\textrm{Si}(p,p')$ was accessible using both layers of the particle telescope, whilst the $3_{1}^{+}$ state at $E_{\textrm{x}}=6.28$ MeV was only accessible through the first layer of the particle telescope.
Due to kinematic smearing across each subtended angle, the data of each individual $\Delta$E-detector were fitted separately for both the $0_{2}^{+}$ state at $E_{\textrm{x}}=4.98$ MeV and the $3_{1}^{+}$ state at $E_{\textrm{x}}=6.28$ MeV (see Sec.~\ref{subsec:Data_Analysis_main_measurement_12C_2019} for more in-depth explanation).
The extraction of the triple coincidences for both states were performed using an identical analysis method as in Secs.~\ref{subsec:Data_Analysis_main_measurement_12C_2019}, \ref{subsec:Data_Analysis_main_measurement_28Si_2020}, \ref{subsec:Data_Analysis_main_measurement_28Si_2019} and \ref{subsec:Data_Analysis_reanalysis_12C_2014}.
The results of this analysis is presented in Sec.~\ref{subsec:results_28Si_2014}.
\section{Data analysis: Angular-correlation correction factors and efficiencies}
\label{sec:angularCorrelation_efficiency}
This section explains and presents the angular correlation correction factors and efficiencies for the detectors used in the measurements of this work.
\subsection{Angular-correlation correction factors and efficiencies for OSCAR}
\label{subsec:angularCorrelation_efficiency_OSCAR}
This section explains the procedure to obtain the angular-correlation correction factors and $\gamma$-ray efficiencies for OSCAR, which are used in the measurements performed in 2019 and 2020 (see Table~\ref{tab:Experimental_apparatus_OSCAR} for summary of the experimental apparatus). 

\begin{figure}[tbp]
\centering
\includegraphics[width=\columnwidth]{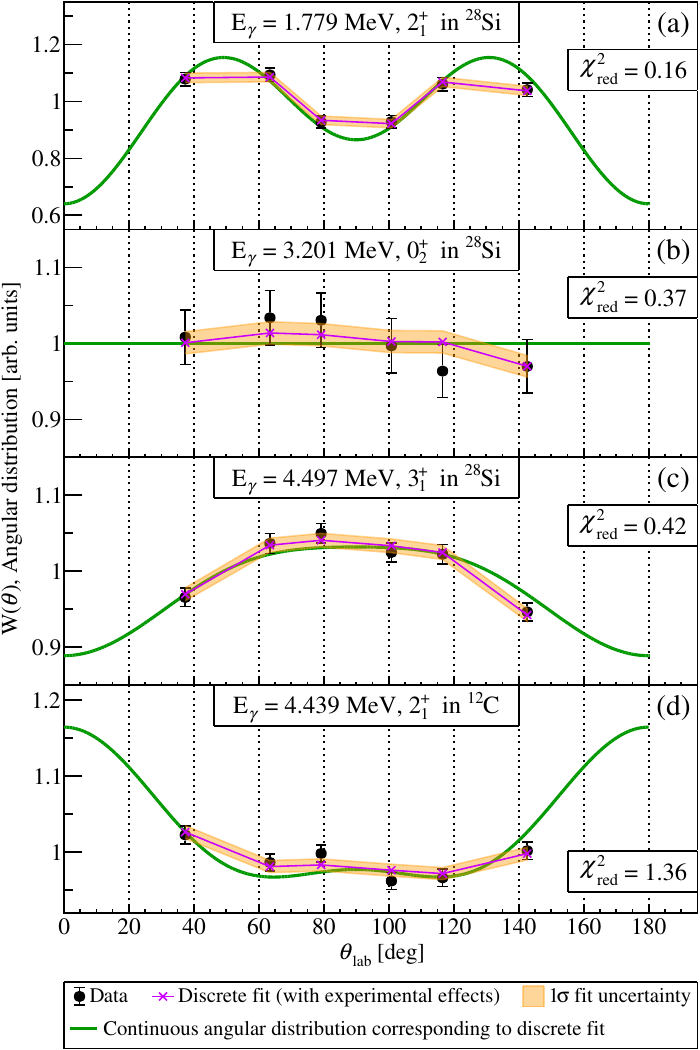}
\caption{\label{fig:angularDistributions}
The angular distributions of $\gamma$ rays with respect to the beam axis. 
See text for more information.
}
\end{figure}

\begin{figure}[tbp]
\centering
\includegraphics[width=\columnwidth]{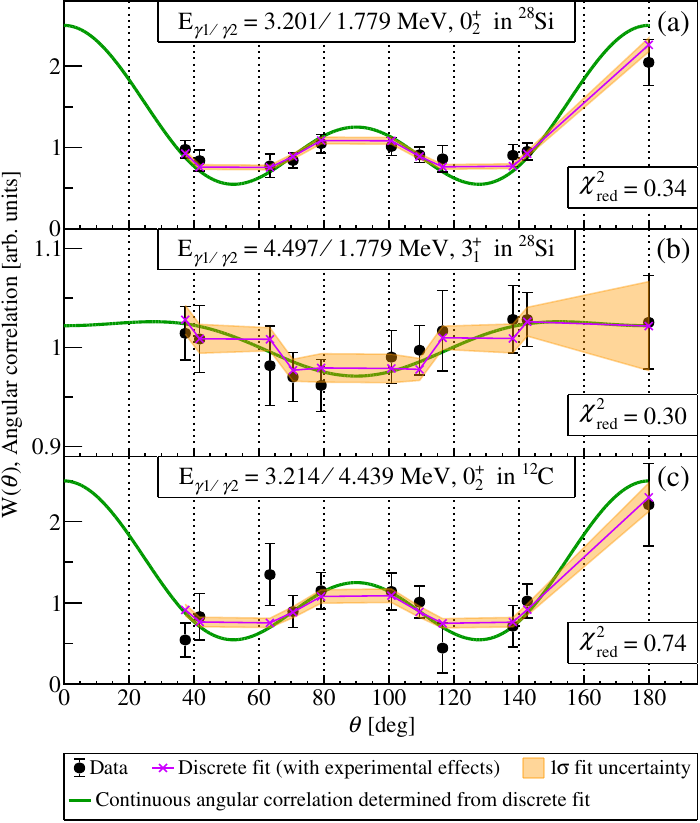}
\caption{\label{fig:angularCorrelations}
Angular correlations between two successive $\gamma$ rays from states of interest.
}
\end{figure}

For OSCAR in-beam data was employed together with a \textsc{Geant4} simulation developed by Zeiser \textit{et al.}~\cite{ZEISER2021164678} to determine the absolute photopeak efficiencies ($\epsilon_{3.21}$ and $\epsilon_{4.44}$) and the angular correlation correction factor ($W_{020}^{7.65}$) in Eq.~\ref{eq: gamma_gamma/gamma}.
The simulation enabled the response of the entire apparatus to be accounted for, e.g., attenuation from the spherically asymmetric scattering chamber.
The corresponding $\gamma$-ray angular correlations in Figs.~\ref{fig:angularDistributions} and \ref{fig:angularCorrelations} were extracted with a simultaneous fitting procedure (analogous to Fig~\ref{fig:gammaGamma_2019}(c)--(f)) to reduce the fit uncertainties.
These correlations were analyzed with fits of the form \cite{RevModPhys.25.729}
\begin{equation}\label{eq:W_angularCorrelationFit}
    W(\theta)=\sum\limits_{k=\textrm{even}}^{2L}A_{k}P_{k}\left(\cos\theta\right),
\end{equation}
\noindent where $P_{k}$ are the Legendre polynomials and $L$ is the angular momentum of the $\gamma$-ray transition.
To accurately account for the finite-angle effects of the LaBr$_{3}$(Ce) detectors, the total response of the apparatus was simulated to produce energy-dependent, discrete values for the Legendre polynomials for each LaBr$_{3}$(Ce) detector of OSCAR.
For all discrete fits, the systematic uncertainty from the geometry of OSCAR was accounted for by repeating the simulations and fits with the LaBr$_{3}$(Ce) detectors positioned at $\pm1$ mm from their nominal 16.3 cm displacement from the target.
These resultant efficiencies and correction factors are summarized in Table~\ref{tab:Efficiency_OSCAR}.
The associated uncertainty is conservatively estimated by taking a weighted average of the relative uncertainty for each discretely fitted point in Fig.~\ref{fig:angularDistributions}(a)-(d).
This was then added in quadrature to that originating from the aforementioned $\pm1$ mm geometrical uncertainty.
The $E_{\gamma} = 3.21$ MeV efficiency ($\epsilon_{3.21}$) for $\gamma$ decay from the Hoyle state was determined by using the $E_{\gamma} = 3.20$ MeV $0_{2}^{+} \rightarrow 2_{1}^{+}$ transition in $^{28}\textrm{Si}$ as a proxy, corresponding to $\epsilon_{3.20}$.
This approximation is justified as \textsc{Geant4} simulations demonstrate that $\epsilon_{3.21} \approx \epsilon_{3.20} = 0.00337(3)$.
For the cases of $^{28}\textrm{Si}$ exclusively, a dataset with a proton beam energy of $E_{p} = 16.0$ MeV was employed (see Sec.~\ref{subsec:Data_Analysis_main_measurement_28Si_2020}.
A measurement utilising SiO$_2$ was also performed in 2019, but the $0_{2}^{+}$ in $^{28}\textrm{Si}$ was inaccessible (see Secs.~\ref{subsec:Data_Analysis_main_measurement_28Si_2020} and \ref{subsec:Data_Analysis_main_measurement_28Si_2019} for more information).
The measurement with higher beam energy was also employed to avoid contamination in the inclusive spectra of the $0_{2}^{+}$ state in $^{28}\textrm{Si}$ with the $E_{\textrm{x}} = 3.68$ MeV and 3.85 MeV states from $^{13}\textrm{C}$ in the backing foil, as these states are separated at higher beam energies.
By gating on the $0_{2}^{+}$ state in $^{28}\textrm{Si}$, the associated $E_{\gamma} = 3.20$ MeV $\gamma$-ray decay is isotropic.
The corresponding $E_{\gamma} = 3.20$ MeV efficiency required a relative $3\%$ correction as the subsequent $2_{1}^{+} \rightarrow 0_{1}^{+}$ $\gamma$ ray may cause summing effects, thus reducing the experimentally measured efficiency.
The experimental value of $\epsilon_{3.20} = 0.0035(1)$ agrees well with the simulated value of 0.00337(3).
The $E_{\gamma} = 4.44$ MeV efficiency ($\epsilon_{4.44}$) for $\gamma$ decay from the $2_{1}^{+}$ state in $^{12}\textrm{C}$ was determined by gating on $2_{1}^{+}$ events detected with SiRi (see Fig.~\ref{fig:InclusiveProtonSpectrum_2019} for inclusive spectra).
The anisotropy of this $E2$ $\gamma$-ray was fitted to determine the (relative $1\%$) angular-correlation correction factor to yield $\epsilon_{4.44} = 0.00268(4)$ and $\epsilon_{4.44} = 0.00258(4)$ for the $^{12}\textrm{C}(p,p')$ measurements performed in 2019 and 2020, respectively.
Both of these $\epsilon_{4.44}$ values agree well with the simulated value of 0.00277(3).
The reason why both the data and fits of the angular correlations (relative to the beam direction) in Figs.~\ref{fig:angularDistributions}(a) and \ref{fig:angularDistributions}(b) are slightly asymmetric across $\theta = 90^{\circ}$ is the increased attenuation from SiRi, which was configured at backward angles.
As mentioned in Sec.~\ref{subsec:Data_Analysis_main_measurement_28Si_2019}, we have also analyzed the $3_{1}^{+}$ state in $^{28}\textrm{Si}$ at $E_{\textrm{x}}=6.28$ MeV as the cascade from this state was employed as a monitor transition in Ref.~\cite{PhysRevC.13.2033} and analyzed in Ref.~\cite{PhysRevLett.125.182701}.
The corresponding $E_{\gamma} = 4.50$ MeV efficiency from the $3_{1}^{+}$ state at $E_{\textrm{x}}=6.28$ MeV required a relative $3\%$ correction as the subsequent $3_{1}^{+} \rightarrow 2_{1}^{+}$ $\gamma$ ray may cause summing effects, thus reducing the experimentally measured efficiency.
In Fig.~\ref{fig:angularDistributions}(c), it is observed that the angular distribution of the initial $E_{\gamma}=4.50$ MeV $\gamma$ ray was more isotropic in comparison to the $E_{\gamma}=1.79$ MeV transition in panel Fig.~\ref{fig:angularDistributions}(a).

To determine the $W_{020}^{7.65}$ angular-correction factor, the $\gamma$--$\gamma$ angular correlations for the $0_{2}^{+} \rightarrow 2_{1}^{+} \rightarrow 0_{1}^{+}$ transitions in both $^{28}\textrm{Si}$ and $^{12}\textrm{C}$ were analyzed, see Figs.~\ref{fig:angularCorrelations}(a) and \ref{fig:angularCorrelations}(c).
These associated fits were of the form \mbox{$W(\theta) =  a_{0}\left[1.0 + A_{2}P_{2}\left( \cos\theta \right) + A_{4}P_{4}\left( \cos\theta \right)\right]$}, where  $A_{2} = 0.3571$ and $A_{4} = 1.1429$ are theoretical coefficients and $a_{0}$ is the only free parameter in the fit \cite{Hamilton}.
These fits yielded $\goodchi^{2}_{\textrm{red}} = 0.34$ and $0.74$ for the $0_{2}^{+} \rightarrow 2_{1}^{+} \rightarrow 0_{1}^{+}$ cascades in $^{28}\textrm{Si}$ and $^{12}\textrm{C}$, respectively.
Using these theoretical coefficients, the corresponding angular-correlation correction factors for the $^{28}\textrm{Si}$ and $^{12}\textrm{C}$ cases were both simulated to be $W_{020}^{7.65} \approx W_{020}^{4.98} = 0.96(2)$.
The associated uncertainty is conservatively estimated by taking a weighted average of the relative uncertainty for each discretely fitted point in Fig.~\ref{fig:angularCorrelations}(a).
This was then added in quadrature to that originating from the aforementioned $\pm1$ mm geometrical uncertainty.
The use of the simulated $W_{020}^{7.65}$ value in Eq.~\ref{eq: gamma_gamma/gamma} is justified as the corresponding theoretical (simulated) angular correlations describe the data appropriately.
Fig.~\ref{fig:angularCorrelations}(b) presents the $3_{1}^{+} \rightarrow 2_{1}^{+} \rightarrow 0_{1}^{+}$ cascade in $^{28}\textrm{Si}$, which is observed to be quite isotropic in comparison to the angular correlations in Figs.~\ref{fig:angularCorrelations}(a) and \ref{fig:angularCorrelations}(c).
\subsection{Efficiency of CACTUS}
\label{subsec:Efficiency_CACTUS}
\begin{table*}[btp]
\caption{\label{tab:Efficiency_CACTUS}%
A summary of the various experimental conditions used to determine the detector response of the CACTUS array used in Eqs.~\ref{eq: gamma_gamma/gamma}, \ref{eq: gamma_gamma/gamma_28Si_4_98MeV} and \ref{eq: gamma_gamma/gamma_28Si_6_28MeV}.
The absolute photopeak efficiencies ($\epsilon_{1.78}$, $\epsilon_{3.20}$, $\epsilon_{4.44}$, $\epsilon_{4.49}$) are presented as fractions, per detector. 
The summed detector response ($\epsilon_{1.78}\epsilon_{3.20}c_{\textrm{det}}$, $\epsilon_{1.79}\epsilon_{4.49}c_{\textrm{det}}$, $\epsilon_{3.20}\epsilon_{4.44}c_{\textrm{det}}$) are presented as total efficiency of the array.
The points marked with "generated" are obtained through convoluting detector responses, as described in Sec.~\ref{subsec:Data_Analysis_main_measurement_12C_2019}.
The efficiencies marked as "fitted" are obtained through fitting the narrow photopeak above a smooth underlying contribution (e.g., the Compton continuum or background events).
The efficiencies marked as "gated" are obtained through a raw integral of the histogram.
}
\begin{ruledtabular}
\begin{tabular}{l D{,}{}{4.0} D{,}{}{4.0} D{,}{}{4.0} } 
 & \multicolumn{1}{c}{$^{12}\textrm{C}(p,p')$ / $^{28}\textrm{Si}(p,p')$ with $E_p=16.0$ MeV} &  \multicolumn{1}{c}{$^{12}\textrm{C}(p,p')$ / $^{28}\textrm{Si}(p,p')$ with $E_p=10.7$ MeV}  \\ [0.8ex]
 & \multicolumn{1}{c}{performed in 2012 } &  \multicolumn{1}{c}{performed in 2014 \cite{PhysRevLett.125.182701}}\\ 
\midrule \\ [-2.0ex]
\multirow{1}{*}{$\epsilon_{1.78}$ (data, fitted)} & \multicolumn{1}{c}{$0.0024(2)$} & \multicolumn{1}{c}{$0.0028(1)$\footnotemark[1]}\\ 
\multirow{1}{*}{$\epsilon_{1.78}$ (data, gated)} & \multicolumn{1}{c}{$0.0024(1)$} &\multicolumn{1}{c}{$0.0031(1)$\footnotemark[1]}\\ 
\midrule \\ [-2.0ex]
\multirow{1}{*}{$\epsilon_{3.20}$ (data, fitted)} & \multicolumn{1}{c}{$0.00168(8)$}  & \multicolumn{1}{c}{$0.00169(6)$}\\ 
\midrule \\ [-2.0ex]
\multirow{1}{*}{$\epsilon_{4.44}$ (data, fitted)} & \multicolumn{1}{c}{$0.00136(4)$} & \multicolumn{1}{c}{$0.00143(4)$}\\ 
\multirow{1}{*}{$\epsilon_{4.44}$ (data, gated)} & \multicolumn{1}{c}{$0.00173(5)$} & \multicolumn{1}{c}{$0.00172(5)$}\\
\midrule \\ [-2.0ex]
\multirow{1}{*}{$\epsilon_{4.49}$ (data, fitted)} & & \multicolumn{1}{c}{$0.00152(9)$} \\ 
\midrule \\ [-2.0ex]
\multirow{1}{*}{$\epsilon_{1.78}\epsilon_{3.20}c_{\textrm{det}}$ (data, gated)} & \multicolumn{1}{c}{$0.0022(1)$} & \multicolumn{1}{c}{$0.0035(2)$}\\ 
\multirow{1}{*}{$\epsilon_{1.78}\epsilon_{3.20}c_{\textrm{det}}$ (generated, gated)} & \multicolumn{1}{c}{$0.0023(1)$} & \multicolumn{1}{c}{$0.0032(2)$}\\ 
\midrule \\ [-2.0ex]
\multirow{1}{*}{$\epsilon_{1.78}\epsilon_{4.49}c_{\textrm{det}}$ (data, gated)} &  &\multicolumn{1}{c}{$0.0030(2)$}\\ 
\multirow{1}{*}{$\epsilon_{1.78}\epsilon_{4.49}c_{\textrm{det}}$ (generated, gated)} &  &\multicolumn{1}{c}{$0.0034(2)$}\\ 
\midrule \\ [-2.0ex]
\multirow{1}{*}{$\epsilon_{3.20}\epsilon_{4.44}c_{\textrm{det}}$ (generated, gated)} &  &\multicolumn{1}{c}{$0.0025(2)$}\\ 
\end{tabular}
\end{ruledtabular}
\footnotetext[1]{
These efficiency points were not used because of contamination from the $4_{1}^{+}$ state at $E_{\textrm{x}}=4.62$ MeV in $^{28}\textrm{Si}$ in the inclusive spectrum.
}
\end{table*}
This section explains the procedure to obtain the $\gamma$-ray efficiencies for CACTUS, which are used in the measurements performed in 2012 and 2014 (see Table~\ref{tab:Experimental_apparatus_CACTUS} for summary of the experimental apparatus).

The CACTUS efficiency at $E_{\gamma}=4.44$ MeV was obtained from the reaction $^{12}\textrm{C}(p,p')$ at the $E_{\textrm{x}} = 4.44$ MeV $0_{1}^{+}$ state in $^{12}\textrm{C}$ from the measurement performed in 2014.
For Eq.~\ref{eq: gamma_gamma/gamma} and Eq.~\ref{eq: gamma_gamma/gamma_28Si_4_98MeV}, additional efficiency points were derived from two measurements in 2012.
One of these measurements also provided a second efficiency point at $E_{\gamma}=4.44$ MeV from the $0_{1}^{+}$ state at $E_{\textrm{x}} = 4.44$ MeV in $^{12}\textrm{C}$.
The experimental efficiency points in $E_{\gamma}=1.78$ MeV and $E_{\gamma}=3.20$ MeV were obtained from the $E_{\textrm{x}} = 1.78$ MeV $2_{1}^{+}$ state and the $E_{\textrm{x}} = 4.98$ MeV $0_{2}^{+}$ state in $^{28}\textrm{Si}$, respectively.
These measurements were selected because they utilized a higher beam energy of 16.0 MeV, which facilitates the use of both layers of SiRi for increased ejectile selectivity.
Table~\ref{tab:Experimental_apparatus_CACTUS} provides the experimental details and measured efficiencies for all three measurements used to determine these efficiencies.
A systematic uncertainty of $3\%$ was incorporated due to the absence of simulated correction factors for angular correlation and attenuation.
In contrast, such corrections factors were employed for OSCAR, as discussed in Sec.~\ref{subsec:angularCorrelation_efficiency_OSCAR}.
In this work the angular correlation correction factors from Kib\'edi \textit{et al.}~\cite{PhysRevLett.125.182701} for the $^{28}\textrm{Si}$ and $^{12}\textrm{C}$ measurements were utilized for all measurements using CACTUS.
\subsection{Angular correlation and efficiency of the Obst setup}
\label{subsec:AngCorrEfficiency_Obst}
To test the validity of comparing the Obst ratio between different experimental setups (see Sec.~\ref{subsec:Data_Analysis_main_measurement_28Si_2019}), a \textsc{Geant4} simulation of the experimental setup in Obst \textit{et al.}~\cite{PhysRevC.13.2033} was employed to obtain absolute-photopeak efficiencies and angular correlation correction factors required in Eq.~\ref{eq:ObstRatio_W}.
The geometry was estimated from the text and Figs. 3 and 4 of Ref.~\cite{PhysRevC.13.2033}.
Whilst this only yields an approximate geometry, this study is sufficient to test the stringent statement made in Ref.~\cite{PhysRevLett.125.182701} that despite the differences in experimental setup ``various combinations of these ratios ($A$-$E$) should agree within a few percent''.
Any slight deviations between the simulated and actual geometry of Ref.~\cite{PhysRevC.13.2033} are small in comparison to the substantial difference in experimental setups between Obst \textit{et al.}~\cite{PhysRevC.13.2033} and Kib\'edi \textit{et al.}~\cite{PhysRevLett.125.182701}.
For this investigation, accurate reproduction of the exact Obst ratio measured in Ref.~\cite{PhysRevC.13.2033} is not required to test whether the Obst ratio is as independent of experimental setup as claimed in Ref.~\cite{PhysRevLett.125.182701}.

The resulting efficiencies and correction factors are presented in Table~\ref{tab:Efficiency_Obst}.
By employing the values presented in Table~\ref{tab:Efficiency_Obst} to Eq.~\ref{eq:ObstRatio_W} an Obst ratio of $B\times D = 0.52(3)$ was obtained, assuming that the branching ratio of the $3^{+}$ state at $E_{\textrm{x}}=6.28$ MeV is equal to the current literature value.
\begin{table}[btph]
\caption{\label{tab:Efficiency_Obst}%
The simulated efficiencies and correction factors of the experimental setup employed by Obst \textit{et al.} \cite{PhysRevC.13.2033}.
}
\begin{ruledtabular}
\begin{tabular}{lll}
$\epsilon_{3.20}$ (simulation) & 0.04514(2) \\ [0.5ex]
$\epsilon_{4.49}$ (simulation) & 0.03619(1) \\ [0.5ex]
$W_{020}^{4.98}$ (simulation) & 1.303(3) \\ [0.5ex]
$W_{320}^{6.28}$ (simulation) & 1.025(2) \\ [0.5ex]
$W_{020}^{7.65}$ (simulation) & 1.319(4) \\
\end{tabular}
\end{ruledtabular}
\end{table}
\section{Results: Main measurements of this work}
\label{sec:ResultsMainMeasurements}
For the main measurements of this work performed in 2019 and 2020 (see Table~\ref{tab:Experimental_apparatus_OSCAR} for an overview), the extracted efficiencies and angular-correlation correction factors are summarized in Table~\ref{tab:Efficiency_OSCAR}.
The associated results for all of measurements performed with OSCAR (2019 and 2020) are summarized in Table~\ref{tab:Results_yields_OSCAR}.
\begin{table*}[btp]
\caption{\label{tab:Results_yields_OSCAR}%
A summary of the inclusive and \mbox{$p$-$\gamma$--$\gamma$} triple-coincidence yields employed in Eq.~\ref{eq: gamma_gamma/gamma} to determine $\Gamma_{\gamma}/\Gamma$, $\Gamma_{rad}/\Gamma$ and $\Gamma_{rad}$ for measurements performed with the OSCAR array.
$N_{020}$ represents the background-subtracted triple-coincidence yield.
These reported uncertainties combine both statistical and systematic components.
}
\begin{ruledtabular}
\begin{tabular}{l D{,}{}{4.0} D{,}{}{4.0} D{,}{}{4.0} D{,}{}{4.0} D{,}{}{4.0} D{,}{}{4.0} } 
 & \multicolumn{2}{c}{$^{12}\textrm{C}(p,p')$ with $E_p=10.8$ MeV} &  \multicolumn{2}{c}{$^{28}\textrm{Si}(p,p')$ with $E_p=10.8$ MeV} &   \multicolumn{2}{c}{$^{28}\textrm{Si}(p,p')$ with $E_p=16.0$ MeV}   \\ [0.5ex]
 & \multicolumn{2}{c}{performed in 2019 \cite{PhysRevLett.125.182701}} &  \multicolumn{2}{c}{performed in 2019} &   \multicolumn{2}{c}{performed in 2020}   \\ [0.5ex]
 \cline{2-3} \cline{4-5} \cline{6-7} \\ [-1.5ex]
& \multicolumn{1}{c}{$0_{2}^{+}$in$^{12}\textrm{C}$} & \multicolumn{1}{c}{$0_{2}^{+}$in$^{12}\textrm{C}$} & \multicolumn{1}{c}{$3_{1}^{+}$in$^{28}\textrm{Si}$} & \multicolumn{1}{c}{$3_{1}^{+}$in$^{28}\textrm{Si}$} & \multicolumn{1}{c}{$0_{2}^{+}$in$^{28}\textrm{Si}$} & \multicolumn{1}{c}{$0_{2}^{+}$in$^{28}\textrm{Si}$}\\ [0.5ex]

& \multicolumn{1}{c}{using summed-$\gamma$} & \multicolumn{1}{c}{using $\gamma\text{--}\gamma$} & \multicolumn{1}{c}{using summed-$\gamma$} & \multicolumn{1}{c}{using $\gamma\text{--}\gamma$} & \multicolumn{1}{c}{using summed-$\gamma$} & \multicolumn{1}{c}{using $\gamma\text{--}\gamma$}\\ [-2.0ex]
\\ \midrule
$N_{\textrm{inc}}$ & \multicolumn{2}{c}{$3.1922(4)\times10^8$}  & \multicolumn{2}{c}{$4.072(6)\times10^6$} & \multicolumn{2}{c}{$1.658(9)\times10^5$}   \\ [0.8ex]
$N_{\textrm{020}}$              & 1169,(62) & 1013,(68) & 41050,(285) & 38871,(196) & 2356,(34) & 2199,(68) \\ [0.8ex]
$N_{\textrm{\mbox{pr-pr}}}$     & $---$ & 1260,(50) & $---$ & 39330,(196) & $---$ & 2206,(68)  \\ [0.8ex]
$N_{\textrm{\mbox{pr-dl}}}$     & $---$ & 272,(23) & $---$ & 423,(8) & $---$ & 7,(2) \\ [0.8ex]
$N_{\textrm{\mbox{dl-dl}}}$     & $---$ & 37,(9) & $---$ & 40,(3) &  $---$ & \approx 0,(3) \\ [0.8ex]
$N_{\textrm{\mbox{dl-dl'}}}$    & $---$ & 63,(11) & $---$ & 4,(1) &  $---$ & \approx 0,(3)\\ 
\midrule
$N_{\textrm{020}}/N_{\textrm{inc}}$ & \multicolumn{1}{c}{$3.7(2)\times10^{-6}$} & \multicolumn{1}{c}{$3.2(2)\times10^{-6}$} & \multicolumn{1}{c}{$9.83(4)\times10^{-3}$} & \multicolumn{1}{c}{$9.31(5)\times10^{-3}$} & \multicolumn{1}{c}{$1.42(2)\times10^{-2}$} & \multicolumn{1}{c}{$1.33(4)\times10^{-2}$} \\[0.8ex]
$\Gamma_{\gamma}/\Gamma$ & \multicolumn{1}{c}{$4.5(4)\times10^{-4}$} & \multicolumn{1}{c}{$4.0(3)\times10^{-4}$} & \multicolumn{1}{c}{$0.82(3)$} & \multicolumn{1}{c}{$0.85(2)$} & \multicolumn{1}{c}{$1.02(5)$} & \multicolumn{1}{c}{$0.98(4)$} \\[0.8ex]
$\Gamma_{\text{rad}}/\Gamma$ & \multicolumn{1}{c}{$4.6(4)\times10^{-4}$} & \multicolumn{1}{c}{$4.1(4)\times10^{-4}$} & & & & \\[0.8ex]
$\Gamma_{\text{rad}}$ [meV] & 3.7,(4) & 3.4,(4) & & & & \\[0.8ex]

\end{tabular}
\end{ruledtabular}
\end{table*}
\begin{table*}[btp]
\caption{\label{tab:Results_yields_CACTUS}%
A summary of the inclusive and \mbox{$p$-$\gamma$--$\gamma$} triple-coincidence yields employed in Eq.~\ref{eq: gamma_gamma/gamma} to determine $\Gamma_{\gamma}/\Gamma$, $\Gamma_{rad}/\Gamma$ and $\Gamma_{rad}$ for measurements performed with the CACTUS array.
$N_{020}$ represents the background-subtracted triple-coincidence yield.
These reported uncertainties combine both statistical and systematic components.
}
\begin{ruledtabular}
\begin{tabular}{l D{,}{}{4.0} D{,}{}{4.0} D{,}{}{4.0} D{,}{}{4.0} D{,}{}{4.0} D{,}{}{4.0} } 
 & \multicolumn{2}{c}{$^{12}\textrm{C}(p,p')$ with $E_p=10.7$ MeV} &  \multicolumn{2}{c}{$^{28}\textrm{Si}(p,p')$ with $E_p=10.7$ MeV} &   \multicolumn{2}{c}{$^{28}\textrm{Si}(p,p')$ with $E_p=10.7$ MeV}   \\ [0.5ex]
 & \multicolumn{2}{c}{performed in 2014 \cite{PhysRevLett.125.182701}} &  \multicolumn{2}{c}{performed in 2014 \cite{PhysRevLett.125.182701}} &   \multicolumn{2}{c}{performed in 2014 \cite{PhysRevLett.125.182701}}   \\ [0.5ex]
 \cline{2-3} \cline{4-5} \cline{6-7} \\ [-1.5ex]
& \multicolumn{1}{c}{$0_{2}^{+}$in$^{12}\textrm{C}$} & \multicolumn{1}{c}{$0_{2}^{+}$in$^{12}\textrm{C}$} & \multicolumn{1}{c}{$3_{1}^{+}$in$^{28}\textrm{Si}$} & \multicolumn{1}{c}{$3_{1}^{+}$in$^{28}\textrm{Si}$} & \multicolumn{1}{c}{$0_{2}^{+}$in$^{28}\textrm{Si}$} & \multicolumn{1}{c}{$0_{2}^{+}$in$^{28}\textrm{Si}$}\\ [0.5ex]

& \multicolumn{1}{c}{using summed-$\gamma$} & \multicolumn{1}{c}{using $\gamma\text{--}\gamma$} & \multicolumn{1}{c}{using summed-$\gamma$} & \multicolumn{1}{c}{using $\gamma\text{--}\gamma$} & \multicolumn{1}{c}{using summed-$\gamma$} & \multicolumn{1}{c}{using $\gamma\text{--}\gamma$}\\ [-2.0ex]
\\ \midrule
$N_{\textrm{inc}}$ & \multicolumn{2}{c}{$2.95783(4)\times10^8$}  & \multicolumn{2}{c}{$9.60(1)\times10^5$} & \multicolumn{2}{c}{$2.326(7)\times10^5$}   \\ [0.8ex]
$N_{\textrm{020}}$              & 319,(22) & 239,(25) & 2809,(65) & 1805,(50) & 761,(34) & 654,(28) \\ [0.8ex]
$N_{\textrm{\mbox{pr-pr}}}$     & $---$ & 240,(25) & $---$ & 1813,(50) & $---$ & 655,(28)  \\ [0.8ex]
$N_{\textrm{\mbox{pr-dl}}}$     & $---$ & \approx 0,(7) & $---$ & 8,(2) & $---$ & 1,(1) \\ [0.8ex]
$N_{\textrm{\mbox{dl-dl}}}$     & $---$ & \approx 1,(4) & $---$ & 1,(1) & $---$ & 1,(1) \\ [0.8ex]
$N_{\textrm{\mbox{dl-dl'}}}$    & $---$ & \approx 0,(4) & $---$ & \approx 0,(1) & $---$ & \approx 0,(1)\\ 
\midrule
$N_{\textrm{020}}/N_{\textrm{inc}}$ & \multicolumn{1}{c}{$1.08(8)\times10^{-6}$} & \multicolumn{1}{c}{$8.1(9)\times10^{-7}$} & \multicolumn{1}{c}{$2.92(8)\times10^{-3}$} & \multicolumn{1}{c}{$2.20(5)\times10^{-3}$} & \multicolumn{1}{c}{$3.3(1)\times10^{-3}$} & \multicolumn{1}{c}{$2.8(1)\times10^{-3}$} \\[0.8ex]
$\Gamma_{\gamma}/\Gamma$ & \multicolumn{1}{c}{$4.7(5)\times10^{-4}$} & \multicolumn{1}{c}{$4.5(6)\times10^{-4}$} & \multicolumn{1}{c}{$0.92(6)$} & \multicolumn{1}{c}{$0.93(6)$} & \multicolumn{1}{c}{$1.09(9)$} & \multicolumn{1}{c}{$1.07(7)$} \\[0.8ex]
$\Gamma_{\text{rad}}/\Gamma$ & \multicolumn{1}{c}{$4.8(5)\times10^{-4}$} & \multicolumn{1}{c}{$4.6(6)\times10^{-4}$} &  &  &  \\[0.8ex]
$\Gamma_{\text{rad}}$ [meV] & 3.9,(5) & 3.7,(6) & & & & \\[0.8ex]

\end{tabular}
\end{ruledtabular}
\end{table*}
\begin{table*}[btp]
\caption{\label{tab:Efficiency_OSCAR}%
The experimental and simulated correction factors employed in Eq.~\ref{eq: gamma_gamma/gamma}, \ref{eq: gamma_gamma/gamma_28Si_4_98MeV} and \ref{eq: gamma_gamma/gamma_28Si_6_28MeV} for OSCAR.
The absolute photopeak efficiencies ($\epsilon_{1.78}$, $\epsilon_{3.20}$, $\epsilon_{3.21}$, $\epsilon_{4.44}$, $\epsilon_{4.49}$) are presented as fractions, per detector. 
The absolute photopeak efficiencies account for any required corrections, such as scattering/summing effects and/or anisotropic angular distributions.
The summed detector response ($\epsilon_{1.78}\epsilon_{3.20}c_{\textrm{det}}$, $\epsilon_{1.79}\epsilon_{4.49}c_{\textrm{det}}$, $\epsilon_{3.20}\epsilon_{4.44}c_{\textrm{det}}$) are presented as total efficiency of the array.
The points marked with "generated" are obtained through convoluting detector responses, as described in Sec.~\ref{subsec:Data_Analysis_main_measurement_12C_2019} and necessary corrections to the individual spectra have been included.
The efficiencies marked as "fitted" are obtained through fitting the narrow photopeak above a smooth underlying contribution (e.g., the Compton continuum or background events).
The efficiencies marked as "gated" are obtained through a raw integral of the histogram.
}
\begin{ruledtabular}
\begin{tabular}{l D{,}{}{4.0} D{,}{}{4.0} D{,}{}{4.0} } 
  & \multicolumn{1}{c}{$^{12}\textrm{C}(p,p')$ / $^{28}\textrm{Si}(p,p')$ with $E_p=10.8$ MeV} & \multicolumn{1}{c}{$^{12}\textrm{C}(p,p')$ / $^{28}\textrm{Si}(p,p')$ with $E_p=16.0$ MeV}\\
  & \multicolumn{1}{c}{performed in 2019} & \multicolumn{1}{c}{performed in 2020}\\
  \midrule \\ [-2.0ex]
\multirow{1}{*}{$\epsilon_{1.78}$ (simulation)} & \multicolumn{1}{c}{$0.00457(3)$} & \multicolumn{1}{c}{$0.00457(3)$}\\ [0.8ex] 
\multirow{1}{*}{$\epsilon_{1.78}$ (data, fitted)} &  & \multicolumn{1}{c}{$0.00472(8)_{\textrm{stat.}}\times 0.98(2)_{\textrm{syst.}}=0.0046(1)$}\\
 \multirow{1}{*}{$\epsilon_{1.78}$ (data, gated)} &  & \multicolumn{1}{c}{$0.00474(3)_{\textrm{stat.}} \times 0.98(2)_{\textrm{syst.}}=0.0046(1)$}\\
\midrule \\ [-2.0ex]
\multirow{1}{*}{$\epsilon_{3.20}$ (simulation)} & \multicolumn{1}{c}{0.00337(3)} & \multicolumn{1}{c}{0.00337(3)}\\ [0.8ex] 
\multirow{1}{*}{$\epsilon_{3.20}$ (data, fitted)} &  & \multicolumn{1}{c}{$0.003370(1)_{\textrm{stat.}} \times 1.03(2)_{\textrm{syst.}}= 0.0035(1)$}\\ [0.8ex]
\midrule \\ [-2.0ex]
\multirow{1}{*}{$\epsilon_{3.21}$ (simulation)} & \multicolumn{1}{c}{0.00337(3)} & \multicolumn{1}{c}{0.00337(3)}\\ 
\midrule \\ [-2.0ex]
\multirow{1}{*}{$\epsilon_{4.44}$ (simulation)} & \multicolumn{1}{c}{0.00277(3)} & \multicolumn{1}{c}{0.00277(3)} \\ [0.8ex]
\multirow{1}{*}{$\epsilon_{4.44}$ (data, fitted)} & \multicolumn{1}{c}{$0.002650(5)_{\textrm{stat.}} \times 1.01(2)_{\textrm{syst.}} = 0.00268(4)$} & \multicolumn{1}{c}{$0.002511(6)_{\textrm{stat.}} \times 1.03(2)_{\textrm{syst.}} = 0.00258(4)$}\\
\multirow{1}{*}{$\epsilon_{4.44}$ (data, gated)} & \multicolumn{1}{c}{$0.002717(1)_{\textrm{stat.}} \times 1.01(2)_{\textrm{syst.}}=0.00275(4)$} & \multicolumn{1}{c}{$0.002635(5)_{\textrm{stat.}} \times 1.03(2)_{\textrm{syst.}}=0.00271(5)$}\\
\midrule \\ [-2.0ex]
\multirow{1}{*}{$\epsilon_{4.49}$ (simulation)} & \multicolumn{1}{c}{0.00275(3)} & \multicolumn{1}{c}{0.00275(3)} \\[0.8ex] 
\multirow{1}{*}{$\epsilon_{4.49}$ (data, fitted)} & \multicolumn{1}{c}{$0.0024(1)_{\textrm{stat.}} \times 1.01(1)_{\textrm{syst.}}=0.0024(1)$} & \\
\midrule \\ [-2.0ex]
\multirow{1}{*}{$\epsilon_{1.78}\epsilon_{3.20}c_{\textrm{det}}$ (data, gated)} &  & \multicolumn{1}{c}{0.0141(5)}\\
\multirow{1}{*}{$\epsilon_{1.78}\epsilon_{3.20}c_{\textrm{det}}$ (generated, gated)} &  & \multicolumn{1}{c}{0.0154(7)}\\ 
\midrule \\ [-2.0ex]
\multirow{1}{*}{$\epsilon_{1.78}\epsilon_{4.49}c_{\textrm{det}}$ (data, gated)} & \multicolumn{1}{c}{0.0120(6)\footnotemark[1]} &\\ 
\multirow{1}{*}{$\epsilon_{1.78}\epsilon_{4.49}c_{\textrm{det}}$ (generated, gated)} & \multicolumn{1}{c}{0.0123(4)\hphantom{\footnotemark[1]}} &\\ 
\midrule \\ [-2.0ex]
\multirow{1}{*}{$\epsilon_{3.20}\epsilon_{4.44}c_{\textrm{det}}$ (generated, gated)} & \multicolumn{1}{c}{0.0091(4)} &\\ 
\midrule \\ [-2.0ex]
\multirow{1}{*}{$W_{020}^{4.98/7.65}$ (simulation)} & \multicolumn{1}{c}{0.96(2)} & \multicolumn{1}{c}{0.96(2)}\\ [0.9ex]
\multirow{1}{*}{$W_{020}^{4.98/7.65}$ (data, fitted)} &  & \multicolumn{1}{c}{0.96(4)}\\ [0.9ex]
\midrule \\ [-2.0ex]
\multirow{1}{*}{$W_{320}^{6.28}$ (simulation)} & \multicolumn{1}{c}{1.01(2)} &\\ [0.9ex]
\end{tabular}
\end{ruledtabular}
\footnotetext[1]{
Obtained using the literature value of $\Gamma_{\gamma}^{E2/M1}/\Gamma^{6.28}=0.88(4)$~\cite{SHAMSUZZOHABASUNIA20131189}.
}
\end{table*}

\subsection{$\mathbf{^{12}\textrm{C}(p,p')}$ with $\mathbf{E_p=10.8}$ MeV performed in 2019}
\label{subsec:results_12C_2019}
The $\gamma$-decay branching ratio of the Hoyle state was determined with Eq.~\ref{eq: gamma_gamma/gamma} and the values reported in Table~\ref{tab:Results_yields_OSCAR} and \ref{tab:Efficiency_OSCAR} to be \mbox{$\Gamma_{\gamma}^{E2}/\Gamma^{7.65}=4.0(3)\times 10^{-4}$} when using the triple-coincidence yields obtained through $\gamma\text{--}\gamma$, and \mbox{$\Gamma_{\gamma}^{E2}/\Gamma^{7.65}=4.5(4)\times 10^{-4}$} when using the triple-coincidence yields obtained through summed-$\gamma$ matrices.
In this work, the former value is used to calculate the associated radiative branching ratio and radiative width in order to implement a more conservative uncertainty.
Using the theoretical total conversion coefficient, \mbox{$\alpha_{tot}(E2, E_{\gamma} = 3.21 \textrm{ MeV}) = 8.77(13)\times10^{-4}$} \cite{KIBEDI2008202_bricc,PhysRevLett.125.182701} and the recommended value of \mbox{$\Gamma_{\pi}(E0)/\Gamma=7.6(4)\times 10^{-6}$} from Eriksen \textit{et al.}~\cite{PhysRevC.102.024320}, the radiative branching ratio of the Hoyle state was determined as \mbox{$\Gamma_{\textrm{rad}}/\Gamma=4.1(4) \times 10^{-4}$} when using the triple-coincidence yields obtained through $\gamma\text{--}\gamma$, and \mbox{$\Gamma_{\textrm{rad}}/\Gamma=4.6(4) \times 10^{-4}$} when using the triple-coincidence yields obtained through summed-$\gamma$ matrices.
Using Eq.~\ref{eq: gamma_rad}, the radiative width of the Hoyle state is determined as \mbox{$\Gamma_{\textrm{rad}}= 3.4(4)$ meV} when using the triple-coincidence yields obtained through $\gamma\text{--}\gamma$, and \mbox{$\Gamma_{\textrm{rad}}= 3.7(4)$ meV} when using the triple-coincidence yields obtained through summed-$\gamma$ matrices.
\subsection{$\mathbf{^{28}\textrm{Si}(p,p')}$ with $\mathbf{E_p=16.0}$ MeV performed in 2020}
\label{subsec:results_28Si_2020}
The $\gamma$-decay branching ratio of the $4.98$ MeV $0_{2}^{+}$ state in $^{28}\textrm{Si}$ was determined with Eq.~\ref{eq: gamma_gamma/gamma_28Si_4_98MeV} and the values reported in Table~\ref{tab:Results_yields_OSCAR} and \ref{tab:Efficiency_OSCAR} to be $\Gamma_{\gamma}^{E2}/\Gamma^{4.98}=0.98(4)$ when using the triple-coincidence yields obtained through $\gamma\text{--}\gamma$, and $\Gamma_{\gamma}^{E2}/\Gamma^{4.98}=1.02(5)$ when using the triple-coincidence yields obtained through summed-$\gamma$ matrices.
\subsection{$\mathbf{^{28}\textrm{Si}(p,p')}$ with $\mathbf{E_p=10.8}$ MeV performed in 2019}
\label{subsec:results_28Si_2019}
The $\gamma$-decay branching ratio of the $6.28$ MeV $3_{1}^{+}$ state in $^{28}\textrm{Si}$ was determined with Eq.~\ref{eq: gamma_gamma/gamma_28Si_6_28MeV} and the values reported in Table~\ref{tab:Results_yields_OSCAR} and \ref{tab:Efficiency_OSCAR} to be $\Gamma_{\gamma}^{E2/M1}/\Gamma^{6.28}=0.85(2)$ when using the triple-coincidence yields obtained through $\gamma\text{--}\gamma$ matrices, and $\Gamma_{\gamma}^{E2/M1}/\Gamma^{6.28}=0.82(3)$ when using the triple-coincidence yields obtained through summed-$\gamma$ matrices.
These results are in good agreement with the literature value of $\Gamma_{\gamma}^{E2/M1}/\Gamma^{6.28}=0.88(4)$ \cite{SHAMSUZZOHABASUNIA20131189}.
For the 2019 experiment in this work, the Obst ratio in Eq.~\ref{eq:ObstRatio} was determined to be $B\times D=0.70(2)$.
\section{Results: Reanalysis of previous work}
\label{sec:results_reanalysis}
This section presents results from measurements used for the reanalysis of the data published in Ref.~\cite{PhysRevLett.125.182701} as well as additional measurements performed with CACTUS (see Table~\ref{tab:Experimental_apparatus_CACTUS} for an overview), with the goal of independently verifying aspects of the analysis performed in Ref.~\cite{PhysRevLett.125.182701}.
For the measurements performed in 2012 and 2014 the extracted efficiencies and angular-correlation correction factors are summarized in Table~\ref{tab:Efficiency_CACTUS}.
The associated results for all measurements performed with CACTUS (2012 and 2014) are summarized in Table~\ref{tab:Results_yields_CACTUS}.
\subsection{$\mathbf{^{12}\textrm{C}(p,p')}$ with $\mathbf{E_p=10.7}$ MeV performed in 2014}
\label{subsec:results_12C_2014}
In order to understand the discrepancy between the results of this work and Ref.~\cite{PhysRevLett.125.182701}, the analysis methodology used in this work was employed to independently verify aspects of the analysis in Ref.~\cite{PhysRevLett.125.182701}.
In this work, the inclusive and triple-coincidence yields for the 2014 experiment of Ref.~\cite{PhysRevLett.125.182701} were determined, see Table~\ref{tab:Results_yields_CACTUS}.
The CACTUS efficiencies employed in this work were independently extracted from the same 2014 measurements, as well as from two measurements performed in 2012 (see Table~\ref{tab:Experimental_apparatus_CACTUS} for a summary).
Due to the lack of a \textsc{Geant4} simulation for the CACTUS array in this study, the angular correlation correction factor ($W_{020}^{7.65}$ from Eq.~\ref{eq: gamma_gamma/gamma}) presented in the work of Kib\'edi \textit{et al.}~\cite{PhysRevLett.125.182701} was used in this work.
The use of $W_{020}^{7.65}$ from Ref.~\cite{PhysRevLett.125.182701} does not play a significant role as this correction factor has been independently verified to be close to unity with OSCAR, which subtends similar angles to CACTUS (see Table~\ref{tab:Efficiency_CACTUS} and Table~\ref{tab:Efficiency_OSCAR}).
The inclusive yield from this independent analysis is $\approx 6\%$ larger than the value of $N_{\textrm{inc}}=2.78(6)\times 10^8$ reported in Ref.~\cite{PhysRevLett.125.182701} (yields presented in Table~\ref{tab:Results_yields_CACTUS}).
This slight discrepancy is mostly due to difference in fit parametrization of the inclusive peak, given that the identical $^{12}\textrm{C}(p,p')$ data was utilized in this reanalysis as in Ref.~\cite{PhysRevLett.125.182701}.

The triple-coincidence yields obtained through gated projections of summed-$\gamma$ and $\gamma\text{--}\gamma$ matrices are $319(22)$ and $239(25)$, respectively (see Table~\ref{tab:Results_yields_CACTUS}).
The significant difference between these yields is due to the difference in gates, whereby the $\gamma\text{--}\gamma$ matrix is gated on the photopeak of a single transition, whilst the summed-$\gamma$ matrix is gated on the summed photopeak of the cascade.
As described in Sec.~\ref{subsec:Data_Analysis_main_measurement_12C_2019}, the poor resolution of the NaI(Tl) detectors means that gates on the narrow photopeak contain correlated continuum events, whilst only the peak component is defined to correspond to the fitted, photopeak efficiency.
Gating on a summed-peak of the Hoyle state compounds this effect (summed-$\gamma$ matrix), as opposed to only gating on the single 4.44 MeV $\gamma$ ray in the Hoyle-state cascade ($\gamma$-$\gamma$ matrix).
By accounting for the two types of efficiency, the $\gamma$-$\gamma$ and summed-$\gamma$ methods respectively yield \mbox{$\Gamma_{\gamma}^{7.65}/\Gamma^{7.65}=4.7(5)\times 10^{-4}$} and \mbox{$\Gamma_{\gamma}^{7.65}/\Gamma^{7.65}=4.5(6)\times 10^{-4}$}, which are in excellent agreement.

In this independent analysis, the triple-coincidence yield obtained through gated projections of the $\gamma\text{--}\gamma$ matrices is $\approx 10\%$ larger than that in Ref.~\cite{PhysRevLett.125.182701} (see Table~\ref{tab:Results_yields_CACTUS}).
This is due to the differences in gates, yield extraction and background subtraction.
For completeness, the triple-coincidence yield obtained through gated projections of summed-$\gamma$ matrices of 312(22) is $\approx 30\%$ larger than the value of 217(21) reported in Ref.~\cite{PhysRevLett.125.182701} (which uses a similar method).
This discrepancy was shown to be primarily due to a combination of slightly different calibrations [for the NaI(Tl) detectors] and slightly different gates on the summed-peak of the Hoyle state.
For the 2014 reanalysis in this work, the gate on the summed-$\gamma$ peak of the Hoyle state was 7369––7939 keV.
It was verified that shifts of approximately 100 keV of this gate (this shift being a surrogate for a combined effect of different energy calibrations and gate ranges) induced relative changes of approximately 25 \% in the triple-coincidence yield.
This dependence on gates is accounted for if the summed-$\gamma$ gate is completely consistent with the corresponding summed-$\gamma$ efficiency, as is in this work.
In this work, the triple-coincidence yield obtained through $\gamma\text{--}\gamma$ is performed through the simultaneous fitting of five spectra where the peak shape is defined from the a high-statistics coincidence spectrum of the same $\gamma$-ray energy and each spectrum has individual background components.
Furthermore, the spectrum utilized to constrain the peak shape in the simultaneous fitting is also employed to measure the fitted $\gamma$-ray efficiency,  ensuring that every event included in the triple-coincidence yield is accurately factored into the efficiency extraction.
In this work the triple-coincidence yield obtained through summed-$\gamma$ matrices is done through fitting a single, background-subtracted spectrum containing the constituents of the summed $\gamma$-ray using peak shapes defined by the generated summed $\gamma$-ray response.
The substantial difference in yields between the $\gamma\text{--}\gamma$ and summed-$\gamma$ methods is linked to significant differences in $\gamma$-ray efficiency.

For the absolute photopeak efficiencies and detector combinations in Eq.~\ref{eq: gamma_gamma/gamma}, a significant discrepancy was determined between this work and the study by Kib\'edi \textit{et al.}~\cite{PhysRevLett.125.182701}.
Firstly, $\gamma$-ray photopeak efficiencies reported in Ref.~\cite{PhysRevLett.125.182701} are not absolute as was reported, but relative.
Notably, the CACTUS efficiencies in Ref.~\cite{PhysRevLett.125.182701} were determined from simulation, whereas all $\gamma$-ray efficiencies reported in this work were determined through experimental data. 
Secondly, the number of detector combinations ($c_{\text{det}}$) in Ref.~\cite{PhysRevLett.125.182701} was incorrectly reported as $c_{\text{det}} = n_{\text{det}}(n_{\text{det}} - 1)/2 = 325$, whereas the correct value is $c_{\text{det}} = n_{\text{det}}(n_{\text{det}} - 1) = 650$.
This $c_{\text{det}}$ factor was not explicitly defined in Eq.~2 of Ref.~\cite{PhysRevLett.125.182701}, which is analagous to Eqs.~\ref{eq: gamma_gamma/gamma}, ~\ref{eq: gamma_gamma/gamma_28Si_4_98MeV} and ~\ref{eq: gamma_gamma/gamma_28Si_6_28MeV} in this work.
In the independent reanalysis of the 2014 experiment in this work, the radiative branching ratio of the Hoyle state was determined to be \mbox{$\Gamma_{\textrm{rad}}/\Gamma=4.5(6)\times 10^{-4}$} and \mbox{$\Gamma_{\textrm{rad}}/\Gamma=4.8(5)\times 10^{-4}$} (using the $\gamma$--$\gamma$ and summed-$\gamma$ matrices, respectively), which are in disagreement with the result reported by Kib\'edi \textit{et al.}~\cite{PhysRevLett.125.182701}.

\subsection{$\mathbf{^{28}\textrm{Si}(p,p')}$ with $\mathbf{E_p=10.7}$ MeV performed in 2014}
\label{subsec:results_28Si_2014}
In this work, the independent analysis of the 2014 $^{12}\textrm{C}(p,p')$ measurement (employing CACTUS) was validated by determining the well-known $\gamma$-decay branching ratio of the $0_{2}^{+}$ state at $E_{\textrm{x}}=4.98$ MeV in $^{28}\textrm{Si}$.
Similarly to the $^{12}\textrm{C}(p,p')$ measurement from 2014 (see Sec.~\ref{subsec:results_12C_2014}) the the angular correlation correction factor ($W_{020}^{4.98}$ from Eq.~\ref{eq: gamma_gamma/gamma_28Si_4_98MeV}) presented in the work of Kib\'edi \textit{et al.}~\cite{PhysRevLett.125.182701} was used in this work.
In this work, independent experimental efficiencies were used, as detailed in Appendix \ref{subsec:Efficiency_CACTUS}.
The $\gamma$-decay branching ratio of the $4.98$ MeV $0_{2}^{+}$ state in $^{28}\textrm{Si}$ was determined with Eq.~\ref{eq: gamma_gamma/gamma_28Si_4_98MeV} and the values reported in Table~\ref{tab:Results_yields_CACTUS} and \ref{tab:Efficiency_CACTUS} to be $\Gamma_{\gamma}^{4.98}/\Gamma^{4.98}=1.07(7)$ when using the triple-coincidence yields obtained through $\gamma\text{--}\gamma$, and $\Gamma_{\gamma}^{4.98}/\Gamma^{4.98}=1.09(9)$ when using the triple-coincidence yields obtained through summed-$\gamma$ matrices.
These results agree with Kib\'edi \textit{et al.}~\cite{PhysRevLett.125.182701}.
The $\gamma$-decay branching ratio of the $E_{\textrm{x}} = 6.28$ MeV $3_{1}^{+}$ state in $^{28}\textrm{Si}$ was determined with Eq.~\ref{eq: gamma_gamma/gamma_28Si_6_28MeV} and the values reported in Table~\ref{tab:Results_yields_CACTUS} and \ref{tab:Efficiency_CACTUS} to be $\Gamma_{\gamma}^{6.28}/\Gamma^{6.28}=0.93(6)$ when using the triple-coincidence yields obtained through $\gamma\text{--}\gamma$, and $\Gamma_{\gamma}^{6.28}/\Gamma^{6.28}=0.92(6)$ when using the triple-coincidence yields obtained through summed-$\gamma$ matrices.
These results agree with Kib\'edi \textit{et al.}~\cite{PhysRevLett.125.182701} and with the literature value of $\Gamma_{\gamma}^{6.28}/\Gamma^{6.28}=0.88(4)$ \cite{SHAMSUZZOHABASUNIA20131189}.
The Obst ratio in Eq.~\ref{eq:ObstRatio} was determined to be $B\times D=0.82(4)$, which agrees well with the value obtained by Kib\'edi \textit{et al.}~\cite{PhysRevLett.125.182701}.
\subsection{$\mathbf{^{28}\textrm{Si}(p,p')}$ with $\mathbf{E_p=16.0}$ MeV performed in 2012}
\label{subsec:results_28Si_2012}
In this work, the independent reanalysis of the 2014 $^{28}\textrm{Si}(p,p')$ measurement (employing CACTUS), where the $\gamma$-decay branching ratio of the $E_{\textrm{x}}=4.98$ MeV $0^{+}_{2}$ state was determined, was validated by determining the same quantity from a similar measurement using CACTUS from 2012.
Details of this $^{28}\textrm{Si}(p,p')$ measurement are presented in column three of Table~\ref{tab:Experimental_apparatus_CACTUS}.
This $^{28}\textrm{Si}(p,p')$ measurement performed in 2012 with $E_p=16.0$ MeV enabled both layers of SiRi to be used for increased ejectile selectivity.
In this work, independent experimental efficiencies were used, as detailed in Appendix~\ref{subsec:Efficiency_CACTUS}.
Similarly to the $^{28}\textrm{Si}(p,p')$ measurement from 2014 (see Secs.~\ref{subsec:results_28Si_2014}) the angular correlation correction factor ($W_{020}^{4.98}$ from Eq.~\ref{eq: gamma_gamma/gamma_28Si_4_98MeV}) presented in the work of Kib\'edi \textit{et al.}~\cite{PhysRevLett.125.182701} was used in this work.
The $\gamma$-decay branching ratio of the $0_{2}^{+}$ state at $E_{\textrm{x}}=4.98$ MeV in $^{28}\textrm{Si}$ was determined to be $\Gamma_{\gamma}^{4.98}/\Gamma^{4.98}=1.07(8)$ when using the triple-coincidence yields obtained through $\gamma$--$\gamma$, and $\Gamma_{\gamma}^{4.98}/\Gamma^{4.98}=1.09(7)$ when using the triple-coincidence yields obtained through summed-$\gamma$.
These results thus validate the analysis method for the 2014 $^{28}\textrm{Si}(p,p')$ and $^{12}\textrm{C}(p,p')$ measurements employing CACTUS.
\section{Discussion}
\label{sec:discussion}
A summary of previous results for the radiative branching of the Hoyle state in $^{12}\textrm{C}$ is presented in Table~\ref{tab:Results_radiativeBranchingRatio}.
\begin{table}[btp]
\caption{\label{tab:Results_radiativeBranchingRatio}%
Summarised results of recent measurements for the radiative branching ratio of the Hoyle state.
}
\begin{ruledtabular}
\begin{tabular}{lll}
Reference                   & $\Gamma_{\textrm{rad}}/\Gamma \times 10^{4}$ & Methodology                    \\ \midrule
Kelley \cite{KELLEY201771}                  & $4.16(11)$                              & Adopted value                  \\
Kib\'edi \cite{PhysRevLett.125.182701}                  & $6.2(6)$                        & $\gamma$-ray spectroscopy \\
Tsumura \cite{TSUMURA2021136283}                 & $4.3(8)$                       & Charged-particle spectroscopy  \\
Luo \cite{PhysRevC.109.025801}                  & $4.0(3)$                       & Charged-particle spectroscopy  \\
Dell'Aquila \cite{dellAquila_hoyleState}                  &  4.2(6)                       & Charged-particle spectroscopy  \\
Rana \cite{RANA2024139083}                  &  4.01(30)                       & $\gamma$-ray spectroscopy  \\
Rana \cite{RANA2024139083}                  &  4.04(30)                       & Charged-particle spectroscopy  \\
This work                  & $4.1(4)$                       & $\gamma$-ray spectroscopy  \\
\end{tabular}
\end{ruledtabular}
\end{table}
The radiative branching ratio reported in this work of \mbox{$\Gamma_{\textrm{rad}}/\Gamma=4.1(4) \times 10^{-4}$} for the Hoyle state (from the 2019 measurement, see Table~\ref{tab:Results_yields_OSCAR}) agrees well with the adopted value of \mbox{$\Gamma_{\textrm{rad}}/\Gamma=4.16(11)\times 10^{-4}$} given by Kelley \textit{et al.}~\cite{KELLEY201771} as well as all recent measurements, with the lone exception being the result by Kib\'edi \textit{et al.}~\cite{PhysRevLett.125.182701}, which is significantly larger.

To investigate the discrepantly large result by Kib\'edi \textit{et al.}~\cite{PhysRevLett.125.182701}, the analysis methodology used in this work was employed to independently verify the analysis in Ref.~\cite{PhysRevLett.125.182701}.
Two different methods were employed in Ref.~\cite{PhysRevLett.125.182701} to determine the $\gamma$-ray branching ratio of the Hoyle state.
The first method employed a gate on the summed-$\gamma$ matrix to determine triple-coincidence events, using Eq.~2 in Ref.~\cite{PhysRevLett.125.182701} which is analogous to 
Eq.~\ref{eq: gamma_gamma/gamma} in this work.
As detailed in Sec.~\ref{subsec:results_12C_2014}, several issues were discovered: firstly, the $\gamma$-ray photopeak efficiencies reported in Ref.~\cite{PhysRevLett.125.182701} are not absolute as was reported, but relative.
Secondly, the number of detector combinations for the NaI(Tl) $\gamma$-ray detectors in Ref.~\cite{PhysRevLett.125.182701} was incorrectly reported.
Thirdly, it was found that due to the relatively poor resolution of the NaI(Tl) detectors in the 2014 experiment of Ref.~\cite{PhysRevLett.125.182701}, a gate on the summed-$\gamma$ peak contained a large ``smooth'' contribution of events beneath the narrow photopeak, where only events in the photopeak were considered in the corresponding efficiency.
This smooth contribution beneath the narrow photopeak also resulted in non-trivial peak shapes to be fitted for the triple-coincidence yield obtained through summed-$\gamma$ matrices (see Sec.~\ref{subsec:Data_Analysis_main_measurement_12C_2019} for a detailed discussion).
In Sec.~\ref{subsec:Data_Analysis_reanalysis_12C_2014}, a reanalysis of the 2014 experiment of Ref.~\cite{PhysRevLett.125.182701} was performed, and after correcting for the three issues listed above, we obtain \mbox{$\Gamma_{\gamma}^{7.65}/\Gamma^{7.65}=4.7(5)\times 10^{-4}$} and \mbox{$\Gamma_{\textrm{rad}}^{7.65}/\Gamma^{7.65}=4.8(5)\times 10^{-4}$} in this work, which is significantly lower than the discrepant result of \mbox{$\Gamma_{\gamma}^{7.65}/\Gamma^{7.65}=6.1(6)\times 10^{-4}$}~\cite{PhysRevLett.125.182701}, whilst being in reasonable agreement with the adopted value of \mbox{$\Gamma_{\textrm{rad}}/\Gamma=4.16(11)\times 10^{-4}$} given by Kelley \textit{et al.}~\cite{KELLEY201771}.
Furthermore, in this work, an additional method was performed using a $\gamma$--$\gamma$ coincidence matrix to select for triple-coincidence events (see Sec.~\ref{subsec:Data_Analysis_main_measurement_12C_2019}).
As with the previously described gate on the summed-$\gamma$ matrix, a gate on the $\gamma$--$\gamma$ matrix is also affected by detector resolution.
By accounting for these effects in this additional method using $\gamma$--$\gamma$ matrices, we obtain \mbox{$\Gamma_{\gamma}^{7.65}/\Gamma^{7.65}=4.5(6)\times 10^{-4}$} and \mbox{$\Gamma_{\textrm{rad}}^{7.65}/\Gamma^{7.65}=4.6(6)\times 10^{-4}$} for the 2014 experiment.

In the second method employed in Ref.~\cite{PhysRevLett.125.182701}, the branching ratio of the $0_{2}^{+}$ Hoyle state was determined relative to that of the $0_{2}^{+}$ state in $^{28}\textrm{Si}$ at $E_{\textrm{x}}=4.98$ MeV using the following equation:
\begin{equation}\label{eq: gamma_gamma/gamma_7_65_4_98_MeV}
\frac{\Gamma_\gamma^{7.65}}{\Gamma^{7.65}}=\frac{N_{020}^{7.65}}{N_{020}^{4.98}} \times \frac{N_{\text {singles }}^{4.98}}{N_{\text {singles }}^{7.65}} \times \frac{\epsilon_\gamma^{1.78}}{\epsilon_\gamma^{4.44}} \times \frac{\epsilon_\gamma^{3.20}}{\epsilon_\gamma^{3.21}} \times \frac{W_{020}^{4.98}}{W_{020}^{7.65}}.
\end{equation}
The $\gamma$-decay branching ratio determined with this relative method was not affected by the fact that relative efficiencies (which were reported as absolute) were employed in Ref.~\cite{PhysRevLett.125.182701}.
The issue with how this equation was employed in Ref.~\cite{PhysRevLett.125.182701} is once again linked to the resolution of the $\gamma$-ray detectors.
Specifically, since the triple-coincidence yields for this equation were determined by gating on the summed-$\gamma$ matrix, a gated efficiency (instead of a fitted photopeak efficiency) must be employed for the sum peak (see Sec.~\ref{subsec:Data_Analysis_main_measurement_12C_2019} for a detailed explanation).
Whilst the ratio of $\epsilon_\gamma^{3.20}/\epsilon_\gamma^{3.21}$ is the same whether one uses fitted or gated efficiencies (since the energies are similar), the ratio of $\epsilon_\gamma^{1.78}/\epsilon_\gamma^{4.44}$ is strongly dependent due to the energy difference, see panel (a) and (e) in Fig.~\ref{fig:gammaGatingCorrectionFactor_all}.
With the reanalysis of the 2014 experiment in this work, calculating Eq.~\ref{eq: gamma_gamma/gamma_7_65_4_98_MeV} using fitted photopeak efficiencies (whether they be absolute, or relative as in Ref.~\cite{PhysRevLett.125.182701}) yields a $\gamma$-decay branching ratio of \mbox{$\Gamma_{\gamma}/\Gamma=5.6(6)\times 10^{-4}$}, which agrees with the value of \mbox{$\Gamma_{\gamma}/\Gamma=6.1(6)\times 10^{-4}$} by Kib\'edi \textit{et al.}~\cite{PhysRevLett.125.182701}.
Using the correct gated efficiency to account for the gate on the summed-$\gamma$ matrix, corresponding to $\epsilon_\gamma^{1.78}\epsilon_\gamma^{3.20} = 0.0034(2)$ and $\epsilon_\gamma^{4.44}\epsilon_\gamma^{3.21} = 0.0025(2)$ (see Table~\ref{tab:Efficiency_CACTUS}), Eq.~\ref{eq: gamma_gamma/gamma_7_65_4_98_MeV} yields a $\gamma$-decay branching ratio of \mbox{$\Gamma_{\gamma}^{7.65}/\Gamma^{7.65}=4.3(7) \times 10^{-4}$}, corresponding to \mbox{$\Gamma_{\textrm{rad}}/\Gamma=4.4(8)\times 10^{-4}$}.
As with the first method, our correction applied to this second method yields branching ratios which are significantly lower than that of Ref.~\cite{PhysRevLett.125.182701}, whilst being in good agreement with the adopted value of \mbox{$\Gamma_{\textrm{rad}}/\Gamma=4.16(11)\times 10^{-4}$} \cite{KELLEY201771}.

The same analysis methods using $\gamma$-$\gamma$ and summed-$\gamma$ matrices were applied to the new 2019 and 2020 measurements which employed the OSCAR array of LaBr$_{3}$(Ce) detectors (see Secs.~\ref{subsec:Data_Analysis_main_measurement_12C_2019}, \ref{subsec:Data_Analysis_main_measurement_28Si_2019} and \ref{subsec:Data_Analysis_main_measurement_28Si_2020} for full details).
For the Hoyle state, we obtain \mbox{$\Gamma_{\gamma}^{7.65}/\Gamma^{7.65}=4.0(3)\times 10^{-4}$} and \mbox{$\Gamma_{\textrm{rad}}^{7.65}/\Gamma^{7.65}=4.1(4)\times 10^{-4}$} by gating on $\gamma$--$\gamma$ matrices.
By gating on summed-$\gamma$ matrices, we obtain \mbox{$\Gamma_{\gamma}^{7.65}/\Gamma^{7.65}=4.5(4)\times 10^{-4}$} and \mbox{$\Gamma_{\textrm{rad}}^{7.65}/\Gamma^{7.65}=4.6(4)\times 10^{-4}$}.
Whilst both sets of results are in good agreement with each other and the adopted value of \mbox{$\Gamma_{\textrm{rad}}/\Gamma=4.16(11)\times 10^{-4}$} \cite{KELLEY201771}, in this work, we do not recommend this latter result using summed-$\gamma$ matrices as the $\gamma$--$\gamma$ analysis is simpler and does not require the summed-$\gamma$ response to be generated (which can introduce hidden systematic errors).
These results were verified by analyzing the $0_{2}^{+}~\rightarrow~2_{1}^{+}~\rightarrow~0_{1}^{+}$ cascade in $^{28}\textrm{Si}$, which is a surrogate for the analogous cascade from the Hoyle state (with identical angular correlation).
For the $0_{2}^{+}$ state in $^{28}\textrm{Si}$, the analysis methods in this work yielded \mbox{$\Gamma_{\gamma}/\Gamma=0.98(4)$} from the triple coincidences obtained through $\gamma$--$\gamma$ and \mbox{$\Gamma_{\gamma}/\Gamma=1.02(5)$} for the triple coincidences obtained through summed-$\gamma$ for the $^{28}\textrm{Si}(p,p')$ measurement in 2020.
This is in excellent agreement with reference value for the $0_{2}^{+}$ state in $^{28}\textrm{Si}$ of \mbox{$\Gamma_{\gamma}/\Gamma=1$} \cite{SHAMSUZZOHABASUNIA20131189}.

Finally, Ref.~\cite{PhysRevLett.125.182701} makes a claim regarding Eq.~14 in Obst \textit{et al.}~\cite{PhysRevC.13.2033} [shown in Eq.~\ref{eq:ObstEquationGDBR} of this work]: ``Despite some differences between their experiment and ours, various combinations of these ratios should agree within a few percent.''
This specifically relates to the ``Obst ratio'' defined in Eqs.~\ref{eq:ObstRatio} and \ref{eq:ObstRatio_W}.
In this work, we simulated the efficiencies and angular-correlation correction factors for the experimental setup for Obst \textit{et al.}~\cite{PhysRevC.13.2033} (constrained by the angular-correlation fits in Sec.~\ref{sec:angularCorrelation_efficiency}), see Table~\ref{tab:Efficiency_Obst}.
In Table~\ref{tab:results_Obst} the measured Obst ratios from the $^{28}\textrm{Si}(p,p')$ measurements from 2014 and 2019 (see Secs.~\ref{subsec:Data_Analysis_reanalysis_28Si_2014} and \ref{subsec:Data_Analysis_main_measurement_28Si_2019}) and the simulated Obst setup are presented together with the Obst ratio measured by Ref.~\cite{PhysRevLett.125.182701} and Ref.~\cite{PhysRevC.13.2033}.
The Obst ratio from the reanalysis of the $^{28}\textrm{Si}(p,p')$ measurement from 2014 is in excellent agreement with Ref.~\cite{PhysRevLett.125.182701}.
The Obst ratio obtained from the $^{28}\textrm{Si}(p,p')$ measurement in 2019 utilizing OSCAR is slightly lower due to the difference in the efficiency between the LaBr$_3$ detectors of OSCAR and NaI(Tl) detectors of CACTUS, as the angular correlation correction factors of the two arrays are very similar.
The simulation of the Obst setup yielded an Obst ratio which is $\approx 3\sigma$ away from the result published by Obst \textit{et al.}~\cite{PhysRevC.13.2033}.
This level of agreement is reasonable given the approximate nature of the simulation, with the geometry based on figures and text in Ref.~\cite{PhysRevC.13.2033}.
Another factor that may contribute to this slight discrepancy is systematic differences in how the $\gamma$-decay branching ratios were extracted between the work of Obst \textit{et al.}~\cite{PhysRevC.13.2033} and this work (e.g., different gates to obtain efficiencies and triple coincidences).
Despite this discrepancy, the simulation reproduces the significantly lower Obst ratio for the setup used in Ref.~\cite{PhysRevC.13.2033}.
Due to the limited $\gamma$-ray detection angles employed by Obst \textit{et al.}~\cite{PhysRevC.13.2033}, the ratio of angular-correlation correction factors ($W_{320}^{6.28}/W_{020}^{4.98}$) for Obst \textit{et al.}~\cite{PhysRevC.13.2033} of $0.787(2)$ is significantly different from the ratio of $1.057(2)$ for the CACTUS setup in Kib\'edi \textit{et al.}~\cite{PhysRevLett.125.182701} and the ratio of $1.047(1)$ for the OSCAR setup in this work.
The latter two ratios are very similar since the CACTUS and OSCAR arrays share similar detector angles.
To verify this analysis, the $\gamma$-decay branching ratio of the $E_{\textrm{x}}=6.28$ MeV $3_{1}^{+}$ state was measured to be in good agreement with the literature value with both CACTUS and OSCAR arrays (see Secs.~\ref{subsec:Data_Analysis_reanalysis_28Si_2014} and \ref{subsec:Data_Analysis_main_measurement_28Si_2019}).
From the above arguments, it is clear that various combinations of the ratios in Eq.~14 of Obst \textit{et al.}~\cite{PhysRevC.13.2033} can vary more than a few percent, which is consistent with the limited detector coverage in Ref.~\cite{PhysRevC.13.2033}.
\begin{table}[btph]
\caption{\label{tab:results_Obst}%
Summarised results of the Obst ratio (see Sec.~\ref{subsec:Data_Analysis_main_measurement_28Si_2019}).
}
\begin{ruledtabular}
\begin{tabular}{lll}
Reference                   & Obst ratio (Eq.~\ref{eq:ObstRatio}) & \\ \midrule
Obst \textit{et al.}~\cite{PhysRevC.13.2033}  & 0.409(15) \\
Kib\'edi \textit{et al.}~\cite{PhysRevLett.125.182701}  & 0.80(4) \\
$^{28}\textrm{Si}(p,p')$ data from 2014 (this work) & 0.82(4) \\
$^{28}\textrm{Si}(p,p')$ data from 2019 (this work) & 0.70(2) \\
Simulation of Obst setup (this work) & 0.52(3) \\
\end{tabular}
\end{ruledtabular}
\end{table}

\begin{figure}[htbp]
\centering
\includegraphics[width=\columnwidth]{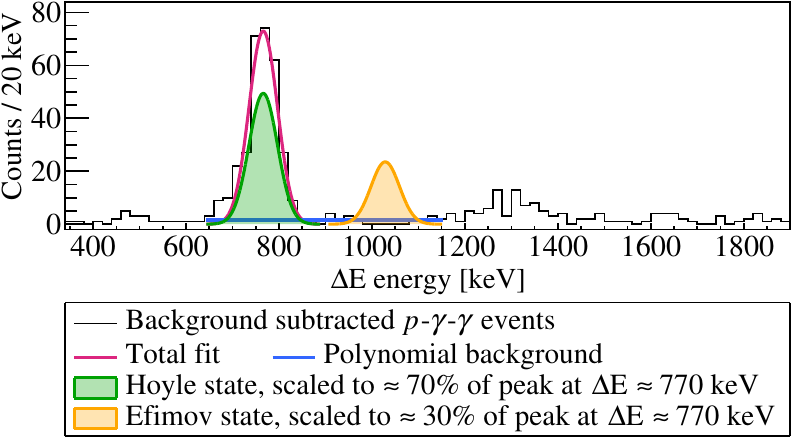}
\caption{\label{fig:inclusiveProtons_deltaE_2014_efimov_combinedFigure_cropped}
$\Delta E$ spectrum for the 2014 $^{12}\textrm{C}(p,p')$ measurement, gated on the $E_{\gamma}=3.21$ MeV and $E_{\gamma}=4.44$ MeV $\gamma$-ray photopeaks from the Hoyle state (see text for details).
}
\end{figure}

A physical justification for the discrepant $\gamma$-decay branching ratio reported in Ref.~\cite{PhysRevLett.125.182701} was suggested by Cardella \textit{et al.}~\cite{PhysRevC.104.064315, Cardella_EFJ_conferencePreceding} as originating from the hypothetical Efimov state in $^{12}\textrm{C}$.
If the Efimov state manifests at the predicted energy of $E_{\textrm{x}} = 7.485$ MeV \cite{Naidon_2017}, then such a contamination to the result by Kib\'edi \textit{et al.}~
\cite{PhysRevLett.125.182701} and the corresponding reanalysis in this work can be excluded.
Firstly, the photopeak of the $E_{\gamma}=3.018$ MeV $E2$ $\gamma$ decay from the Efimov state would be clearly resolved from the $E_{\gamma}=3.214$ MeV photopeak in Fig.~\ref{fig:gammaGamma_2014}(c).
Secondly, the ejectiles corresponding to the Hoyle state and Efimov state at $E_{\textrm{x}} = 7.485$ would be separated by 150 or 250 keV, corresponding to a 0 or 25 $\upmu$m deadlayer in SiRi, respectively (an $\approx 25$ $\upmu$m deadlayer is required to reproduce the ordering of the peaks in Figs.~\ref{fig:InclusiveProtonSpectrum_2019} and \ref{fig:InclusiveProtonSpectrum_2014}).
If the Efimov state was sufficiently populated at exactly $E_{\textrm{x}} = 7.485$ MeV, the $\approx 90$ keV FWHM resolution of the $\Delta E$ detectors in SiRi would enable such a contamination to be observable.
Fig.~\ref{fig:inclusiveProtons_deltaE_2014_efimov_combinedFigure_cropped} presents a selective test for such a contamination in the $\Delta E$ spectrum by gating on the $E_{\gamma}=3.21$ MeV and $E_{\gamma}=4.44$ MeV $\gamma$-ray photopeaks from the Hoyle state (with the same background subtraction method as applied in Sec.~\ref{subsec:Data_Analysis_reanalysis_12C_2014}).
If the discrepancy between the $\gamma$-decay branching ratio reported by Ref.~\cite{PhysRevLett.125.182701} ($\Gamma_{\textrm{rad}}/\Gamma=6.2(6)\times 10^{-4}$) and the ENSDF average (\mbox{$\Gamma_{\textrm{rad}}/\Gamma=4.16(11)\times 10^{-4}$}) \cite{KELLEY201771} is assumed to be purely due to the Efimov state, the corresponding Efimov contribution would be $\approx 30\%$ of the triple-coincidence yield reported by Kib\'edi \textit{et al.}~\cite{PhysRevLett.125.182701} and in the independent analysis of the 2014 data performed in this work (see Table~\ref{tab:Results_yields_OSCAR}).
Such a division of the triple-coincidence yield is completely excluded by the data as no such peak is observed at $\Delta E\approx 1.030$ MeV (corresponding to $E_{\textrm{x}} = 7.485$ MeV).
Furthermore, a possible contaminant Efimov state should affect the results of both the 2014 and 2019 experiments since these experimental conditions were very similar in both beam energy and detector configurations.
However, only the 2014 measurement (Ref.~\cite{PhysRevLett.125.182701} and the independent reanalysis in this work) yields a discrepantly large $\gamma$-decay branching ratio for the Hoyle state (see Table~\ref{tab:Results_yields_OSCAR}).
Finally, a study by Bishop \textit{et al.}, which employed the $\beta$-decay population of $^{12}\textrm{C}$ states and astrophysical-rate calculations, has yielded strong evidence \textit{against} the very existence of the Efimov state at $E_{\textrm{x}} = 7.485$ MeV in $^{12}\textrm{C}$ \cite{PhysRevC.103.L051303}.
From these arguments, we conclude that contamination from the hypothetical Efimov state is unlikely to explain why the radiative branching ratio of the Hoyle state reported by Ref.~\cite{PhysRevLett.125.182701} is discrepant.

\section{Conclusion}
\label{sec:conclusion}
In summary, a new measurement of the $\gamma$-decay and radiative branching ratio of the Hoyle state has been performed using OSCAR at the OCL.
In this work, the reported radiative branching ratio of \mbox{$\Gamma_{\textrm{rad}}/\Gamma=4.1(4) \times 10^{-4}$} is in good agreement with the previously adopted value of $\Gamma_{\textrm{rad}}/\Gamma=4.19(11)\times 10^{-4}$ from Kelley \textit{et al.}~\cite{KELLEY201771}, as well as recent measurements by Tsumura \textit{et al.}~\cite{TSUMURA2021136283}, Luo \textit{et al.}~\cite{PhysRevC.109.025801}, Dell'Aquila \textit{et al.}~\cite{dellAquila_hoyleState} and Rana \textit{et al.}~\cite{RANA2024139083}.
The analysis method employed in this study was verified with the extraction of the radiative branching ratio of the $0_{2}^{+}$ state in $^{28}\textrm{Si}$ at $E_{\textrm{x}}=4.98$ MeV.

In this work, the 2014 experiment published by Kib\'edi \textit{et al.}~\cite{PhysRevLett.125.182701} was reanalyzed with the newly developed analysis method.
Several issues were discovered, and the corrected values for this 2014 experiment are also in good agreement with the previously adopted value \cite{KELLEY201771} and aforementioned recent studies \cite{TSUMURA2021136283, PhysRevC.109.025801, dellAquila_hoyleState, RANA2024139083}.



\begin{acknowledgments}
This study has been funded by the Research Council of Norway through its grant to the Norwegian Nuclear Research Centre (Project No. 341985, 245882 and 325714).
The authors would like to thank J.~C. M\"uller, P.~A. Sobas, and J.~C. Wikne at the Oslo Cyclotron Laboratory (OCL) for the accelerated proton beams.
We would like to thank F. Zeiser for his contributions through the development of the \textsc{Geant4} simulation.
We would like to express our gratitude to T.~Kib\'{e}di, A.~E~Stuchbery and B.~J.~Coombes for their invaluable efforts and in-depth discussions that greatly contributed to the advancement of this work.
We would like to thank P.~Adsley for useful discussions during this work.
We would like to thank F.~W. Furmyr, M.~M. Bjørøen and J.T.H.~Dowie for participating in taking shifts during the main measurement of this work.
The computations were performed on resources provided by UNINETT Sigma2 - the National Infrastructure for High Performance Computing and Data Storage in Norway (using ``SAGA'' on Project No. NN9895K and NN9464K).
A. C. Larsen gratefully acknowledges funding of this research by the European Research Council through ERC-STG-2014 under Grant Agreement No. 637686 and from the Research Council of Norway, Project Grant No. 316116.
\end{acknowledgments}

\section{Appendix}
\subsection{Investigation of background in inclusive particle spectra}
\label{subsec:investigationOfBackgroundInclusiveSpectra}
\begin{figure*}[htbp]
\centering
\includegraphics[width=2\columnwidth]{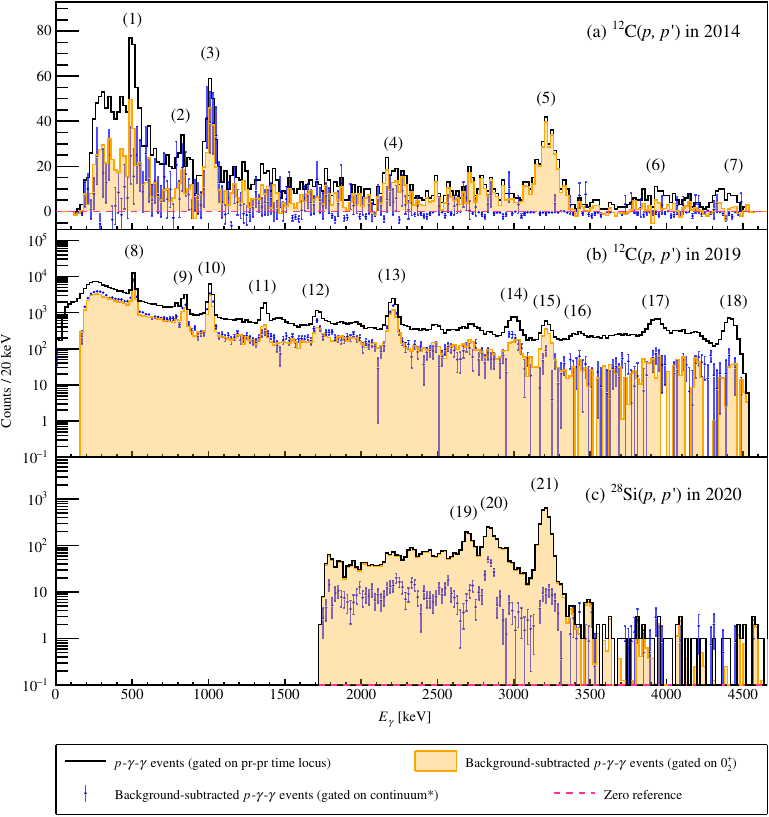}
\caption{\label{fig:simultaneousFit_allExperiments_gammaGamma}
A comparison of $p$-$\gamma$--$\gamma$ events obtained from the $\gamma$--$\gamma$ matrix with different conditions on the gating on the particle spectrum. Events gated on the $0_{2}^{+}$ state in the nucleus of interest and the \mbox{"pr-pr"} time locus are indicated in black in all panels.
By subtracting the background using Eq.~\ref{eq:TripleCoincidence_backgroundSubtraction}, the background-subtracted $p$-$\gamma$--$\gamma$ events gated on the $0_{2}^{+}$ state in the nucleus of interest are shown in orange.
Additionally, the background-subtracted $p$-$\gamma$--$\gamma$ events when gating on the continuum in the nucleus of interest are illustrated in blue in all panels.
A pink horizontal dashed line shows the zero reference in all panels.
Panel (a) shows the comparison for the $^{12}\textrm{C}$(\textit{p},\textit{p}') reaction populated in 2014 (see Sec.~\ref{subsec:Data_Analysis_reanalysis_12C_2014} for more information).
Panel (b) shows the comparison for the $^{12}\textrm{C}$(\textit{p},\textit{p}') reaction populated in 2019 (see Sec.~\ref{subsec:Data_Analysis_main_measurement_12C_2019} for more information).
Panel (c) shows the comparison for the $^{28}\textrm{Si}$(\textit{p},\textit{p}') reaction populated in 2020 (see Sec.~\ref{subsec:Data_Analysis_main_measurement_28Si_2020} for more information).}
\end{figure*}

\begin{figure*}[htbp]
\centering
\includegraphics[width=2\columnwidth]{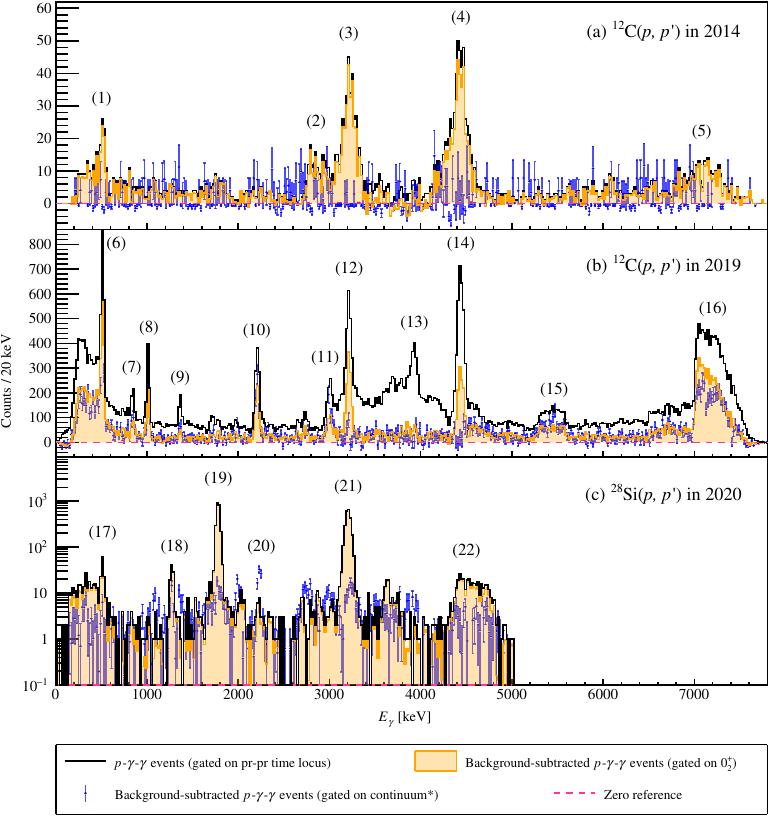}
\caption{\label{fig:simultaneousFit_allExperiments_summedGamma}
A comparison of $p$-$\gamma$--$\gamma$ events obtained from the summed-$\gamma$ matrix with different conditions on the gating on the particle spectrum. Events gated on the $0_{2}^{+}$ state in the nucleus of interest and the \mbox{"pr-pr"} time locus are indicated in black in all panels.
The gate on the summed-$\gamma$ matrix in these spectra consists of only a $3\sigma$ band around the summed $\gamma$-ray energy of interest, in contrast to the modified gate used to extract p-$\gamma$--$\gamma$ events (see Sec.~\ref{subsec:Data_Analysis_main_measurement_12C_2019} for more information).
By subtracting the background using Eq.~\ref{eq:TripleCoincidence_backgroundSubtraction}, the background-subtracted $p$-$\gamma$--$\gamma$ events gated on the $0_{2}^{+}$ state in the nucleus of interest are shown in purple.
Additionally, the background-subtracted $p$-$\gamma$--$\gamma$ events when gating on the continuum in the nucleus of interest are illustrated in green in all panels.
A pink horizontal dashed line shows the zero reference in all panels.
Panel (a) shows the comparison for the $^{12}\textrm{C}$(\textit{p},\textit{p}') reaction populated in 2014 (see Sec.~\ref{subsec:Data_Analysis_reanalysis_12C_2014} for more information).
Panel (b) shows the comparison for the $^{12}\textrm{C}$(\textit{p},\textit{p}') reaction populated in 2019 (see Sec.~\ref{subsec:Data_Analysis_main_measurement_12C_2019} for more information).
Panel (c) shows the comparison for the $^{28}\textrm{Si}$(\textit{p},\textit{p}') reaction populated in 2020 (see Sec.~\ref{subsec:Data_Analysis_main_measurement_28Si_2020} for more information).}
\end{figure*}

For some of the measurements presented in Tables~\ref{tab:Experimental_apparatus_OSCAR} and \ref{tab:Experimental_apparatus_CACTUS}, the nature of the background events beneath the inclusive peak of interest (in SiRi) was investigated.
Three cases comparing the effect on the triple-coincidence yields are each presented in Figs.~\ref{fig:simultaneousFit_allExperiments_gammaGamma} and \ref{fig:simultaneousFit_allExperiments_summedGamma}.
The first case corresponding to a gate on the inclusive peak of interest and the \mbox{pr-pr} time locus. The second case corresponding to a gate on the inclusive peak of interest with background subtraction according to Eq.~\ref{eq:TripleCoincidence_backgroundSubtraction}.
The third case corresponds to a gate on the smooth region close to the inclusive peak of interest, with background subtraction according to Eq.~\ref{eq:TripleCoincidence_backgroundSubtraction}.
In some cases, these background gates were of a different size than for the peak of interest and for such cases, the scaling of the corresponding triple-coincidence spectra was commensurate.

Good selectivity with SiRi was observed as the correlated $\gamma$-ray peaks are negligible for the background-regions in the inclusive spectra.
However, several uncorrelated peaks persist even after background subtraction by gating on different time loci.
The relative strength of these uncorrelated $\gamma$-ray peaks are reproduced in the triple-coincidence spectra gated on the smooth region near the peak of interest.
This indicates that the uncorrelated $\gamma$-ray peaks which persist through timing-related background subtraction (in Figs.~\ref{fig:gammaGamma_2019}, \ref{fig:summedGamma_2019}, \ref{fig:gammaGamma_2014} and \ref{fig:summedGamma_2014}) are due to scattered ejectiles which lie beneath the inclusive peak of interest.
Tables~\ref{tab:table_simfit_allExperiments_gammaGamma} and \ref{tab:table_simfit_allExperiments_summedGamma} summarizes the various transitions in Figs.~\ref{fig:simultaneousFit_allExperiments_gammaGamma} and ~\ref{fig:simultaneousFit_allExperiments_summedGamma}.

\begin{table}[btp]
\caption{\label{tab:table_simfit_allExperiments_gammaGamma}%
Summary of the labeled features in Fig.~\ref{fig:simultaneousFit_allExperiments_gammaGamma}.
}
\begin{ruledtabular}
\begin{tabular}{l r r}
Label & $\textrm{Energy}$ & Source  \\
& $\textrm{[keV]}$ &  \\ \midrule
(1) & 511 &  \\ \midrule
(2) & 844 & $^{27}\textrm{Al}$ \\ \midrule
(3) & 1015 & $^{27}\textrm{Al}$ \\ \midrule
(4) & 2212 & $^{27}\textrm{Al}$ \\ \midrule
(5) & 3214 & $^{12}\textrm{C}$ \\ \midrule
(6) & 3927 & First-escape of 4440 keV in $^{12}\textrm{C}$ \\ \midrule
(7) & 4440 & $^{12}\textrm{C}$ \\ \midrule
(8) & 511 &  \\ \midrule
(9) & 844 & $^{27}\textrm{Al}$ \\ \midrule
(10) & 1015 & $^{27}\textrm{Al}$ \\ \midrule
(11) & 1436 & $^{138}\textrm{La}$ \\ \midrule
(12) & 1720 & $^{27}\textrm{Al}$ \\ \midrule
(13) & 2212 & $^{27}\textrm{Al}$ \\ \midrule
(14) & 3004 & $^{27}\textrm{Al}$ \\ \midrule
(15) & 3214 & $^{12}\textrm{C}$ \\ \midrule
(16) & 3416 & Second-escape of 4440 keV in $^{12}\textrm{C}$ \\ \midrule
(17) & 3927 & First-escape of 4440 keV in $^{12}\textrm{C}$ \\ \midrule
(18) & 4440 & $^{12}\textrm{C}$ \\ \midrule
(19) & 2713 & $^{29}\textrm{Si}$ \\ \midrule
(20) & 2838 & $^{28}\textrm{Si}$ \\ \midrule
(21) & 3201 & $^{28}\textrm{Si}$ \\
\end{tabular}
\end{ruledtabular}
\end{table}

\begin{table}[btp]
\caption{\label{tab:table_simfit_allExperiments_summedGamma}%
Summary of the labeled features in Fig.~\ref{fig:simultaneousFit_allExperiments_summedGamma}.
}
\begin{ruledtabular}
\begin{tabular}{l r r}
Label & $\textrm{Energy}$ & Source  \\
& $\textrm{[keV]}$ &  \\ \midrule
(1) & 511 &  \\ \midrule
(2) & $\approx$2900 & Compton scattered 3004 keV from $^{27}\textrm{Al}$\\ \midrule
(3) & 3214 & $^{12}\textrm{C}$ \\ \midrule
(4) & 4440 & $^{12}\textrm{C}$ \\ \midrule
(5) & $\approx$7100 & Compton scattered 4440 keV from $^{12}\textrm{C}$ \\ \midrule
(6) & 511 &  \\ \midrule
(7) & 844 & $^{27}\textrm{Al}$ \\ \midrule
(8) & 1015 & $^{27}\textrm{Al}$ \\ \midrule
(9) & 1436 & $^{138}\textrm{La}$ \\ \midrule
(10) & 2212 & $^{27}\textrm{Al}$ \\ \midrule
(11) & 3004 & $^{27}\textrm{Al}$ \\ \midrule
(12) & 3214 & $^{12}\textrm{C}$ \\ \midrule
(13) & 3927 & First-escape of 4440 keV in $^{12}\textrm{C}$ \\ \midrule
(14) & 4440 & $^{12}\textrm{C}$ \\ \midrule
(15) & $\approx$5400 & Compton background in coincidence \\ 
     &  & with 2212 keV (10) from $^{27}\textrm{Al}$ \\ \midrule
(16) & $\approx$7200 & Compton background in coincidence \\ 
     &  & with low-energy below 511 keV (6) \\ \midrule
(17) & 511 &  \\ \midrule
(18) & 1273/1263 & $^{29}\textrm{Si}$/$^{30}\textrm{Si}$ \\ \midrule
(19) & 1779 & $^{28}\textrm{Si}$ \\ \midrule
(20) & 2212/2235 & $^{27}\textrm{Al}$/$^{30}\textrm{Si}$ \\ \midrule
(21) & 3201 & $^{28}\textrm{Si}$ \\ \midrule
(22) & $\approx$4500 & Compton background in coincidence \\
     &  & with low-energy below 511 keV (17) \\ 
\end{tabular}
\end{ruledtabular}
\end{table}

\bibliography{apssamp}

\end{document}